\documentclass[twoside,12pt]{article}
\usepackage{epsfig}

\usepackage{url}
\usepackage[bookmarks, pagebackref=false]{hyperref}
\usepackage{color}
	\definecolor{rossoCP3}{cmyk}{0,.88,.77,.40}
		\definecolor{graa}{rgb}{0.8,0.8,0.8}
		\definecolor{blaa}{rgb}{0.2,0.2,0.6}
		\definecolor{gron}{RGB}{0,150,0}
		\hypersetup{
			colorlinks,
			bookmarksopen,
			bookmarksnumbered,
			citecolor=blaa, 		
			linkcolor=black, 
			urlcolor=blaa,			
			}

\def\Journal#1#2#3#4{{#1} {#2} (#4) #3 }

\def\NC{\em Nuovo Cimento}

\def\NPB{{\em Nucl. Phys.} B}

\def\PLB{{\em Phys. Lett.} B}

\def\PRL{\em Phys. Rev. Lett.}

\def\PREP{\em Phys. Rep.}

\def\PRD{{\em Phys. Rev.} D}

\def\ZPC{{\em Z. Phys.} C}
\def\ZPA{{\em Z. Phys.} A}

\def\RMP{\em Rev. Mod. Phys.}

\def\HPA{\em Helv. Phys. Acta}
\def\DANS{\em Dok. Akad. Nauk SSSR}
\def\CMP{\em Commun. Math. Phys.}
\def\JPG{\em J. Phys. G}

\def\CPC{\em Comput. Phys. Commun.}
\def\JHEP{\em JHEP}
\def\TMF{\em Teor. Mat. Fiz.}
\def\PPNP{\em Prog. Part. Nucl. Phys.}
\def\EPJC{{\em Eur. Phys. J.} C}
\def\TMP{\em Theor. Math. Phys.}
\def\IJMPA{{\em Int. J. Mod. Phys.} A}
\def\MPLA{{\em Mod. Phys. lett.} A}
\def\JETPL{\em JETP Lett.}
\def\NPPS{\em Nucl. Phys. Proc. Suppl.}
\def\SJNP{\em Sov. J. Nucl. Phys.}
\def\JETP{\em Sov. Phys. JETP}
\def\PR{\em Phys. Rev.}

\def\PZETF{\em Pisma Zh. Eksp. Teor. Fiz.}
\def\JCP{\em J. Comp. Phys.}
\def\ACM{\em Adv. Comput. Math.}
\def\ZP0{\em Zeit. Phys.}
\def\YF{\em Yad. Fiz.}

\def\AP{\em Annals Phys.}

\newcommand{\dRRNA}{\frac{d_F^{abcd}d_F^{abcd}}{N_A}}
\newcommand{\dRANA}{\frac{d_F^{abcd}d_A^{abcd}}{N_A}}
\newcommand{\dAANA}{\frac{d_A^{abcd}d_A^{abcd}}{N_A}}

\begin{document}

\title{ \vspace{1cm} The Renormalization Scale-Setting Problem in QCD}

\author{Xing-Gang Wu$^{1}$, Stanley J. Brodsky$^2$ and Matin Mojaza$^{2,3}$\\
$^1$Department of Physics, Chongqing University, Chongqing 401331, P.R. China\\
$^2$SLAC National Accelerator Laboratory, Stanford University, CA 94039, USA \\
$^3$CP3-Origins, Danish Institute for Advanced Studies, \\
University of Southern Denmark, DK-5230}

\maketitle

\begin{abstract}

A key problem in making precise perturbative QCD predictions is to set the proper renormalization scale of the running coupling. The conventional scale-setting procedure assigns an arbitrary range and an arbitrary systematic error to fixed-order pQCD predictions. In fact, this {\it ad hoc} procedure gives results which depend on the choice of the renormalization scheme, and it is in conflict with the standard scale-setting procedure used in QED. Predictions for physical results should be independent of the choice of scheme or other theoretical conventions. We review current ideas and points of view on how to deal with the renormalization scale ambiguity and show how to obtain renormalization scheme- and scale- independent estimates. We begin by introducing the renormalization group (RG) equation and an extended version, which expresses the invariance of physical observables under both the renormalization scheme and scale-parameter transformations. The RG equation provides a convenient way for estimating the scheme- and scale- dependence of a physical process. We then discuss self-consistency requirements of the RG equations, such as reflexivity, symmetry, and transitivity, which must be satisfied by a scale-setting method. Four typical scale setting methods suggested in the literature, {\it i.e.,} the Fastest Apparent Convergence (FAC) criterion, the Principle of Minimum Sensitivity (PMS), the Brodsky-Lepage-Mackenzie method (BLM), and the Principle of Maximum Conformality (PMC), are introduced. Basic properties and their applications are discussed. We pay particular attention to the PMC, which satisfies all of the requirements of RG invariance. Using the PMC, all non-conformal terms associated with the $\beta$-function in the perturbative series are summed into the running coupling, and one obtains a unique, scale-fixed, scheme-independent prediction at any finite order. The PMC provides the principle underlying the BLM method, since it gives the general rule for extending BLM up to any perturbative order; in fact, they are equivalent to each other through the PMC - BLM correspondence principle. Thus, all the features previously observed in the BLM literature are also adaptable to the PMC. The PMC scales and the resulting finite-order PMC predictions are to high accuracy independent of the choice of initial renormalization scale, and thus consistent with RG invariance. The PMC is also consistent with the renormalization scale-setting procedure for QED in the zero-color limit. The use of the PMC thus eliminates a serious systematic scale error in perturbative QCD predictions, greatly improving the precision of empirical tests of the Standard Model and their sensitivity to new physics.

\end{abstract}

\begin{description}
\item[PACS numbers] 12.38.Bx, 11.15.Bt, 11.10.GH
\item[Keywords] Renormalization Group, Renormalization Scale, BLM/PMC, QCD
\end{description}

\eject

\tableofcontents

\section{Introduction}

Quantum chromodynamics (QCD) is believed to be the field theory of hadronic strong interactions. Due to its asymptotic freedom property~\cite{qcd1,qcd2}, the QCD running coupling becomes numerically small at short distances, allowing perturbative calculations of cross sections for high momentum transfer physical processes. In the perturbative QCD (pQCD) framework, a physical quantity ($\rho$) is expanded to $n$-th order in the QCD coupling $\alpha_s(\mu_r)$; i.e.,
\begin{equation}\label{phyvalue}
\rho_n = {\cal C}_0 \; \alpha_s^p(\mu_r) + \sum_{i=1}^{n}{\cal C}_i(\mu_r) \; \alpha_s^{p+i}(\mu_r), \;\; (p\geq 0)
\end{equation}
where ${\cal C}_0$ is the tree-level term, ${\cal C}_1$ the one-loop correction, ${\cal C}_2$ the two-loop correction, etc., and $p$ is the power of the coupling associated with the tree-level term. The renormalization scale $\mu_r$ must be specified in order to obtain a definite prediction. The calculation of the coefficients ${\cal C}_i(\mu_r)$ involves ultraviolet divergences which must be regulated and removed by a renormalization procedure. The infinite series $\rho_{n\to\infty}$ is in principle renormalization scheme and renormalization scale independent because of renormalization group (RG) invariance~\cite{peter1,gml,bogo,peter2,callan,symanzik}; i.e., the physical predictions of a theory, calculated up to all orders, are formally independent of the choice of renormalization scale and renormalization scheme. However, at any finite order, the scale/scheme dependence from $\alpha_s(\mu_r)$ and ${\cal C}_i(\mu_r)$ do not exactly cancel, leading to renormalization-scheme and renormalization-scale ambiguities. Such ambiguities are well-known~\cite{ambi1,ambi2,ambi3,ambi4,fac1,fac2,fac3,fac4,pms1,pms2,pms3,pms4,blm}. A guiding principle for resolving such problems is that physical results must be independent of theoretical conventions.

It should be recalled that there is no ambiguity in setting the renormalization scale in quantum electrodynamics (QED) at any finite order. Mass renormalization is straightforward in QED. Due to the Ward-Takahashi identity~\cite{WTidentity}, the divergences in the vertex and fermion wavefunction corrections exactly cancel, and the remaining ultraviolet divergence $Z_3$ associated with the vacuum polarization insertions defines a natural scale for the running QED coupling $\alpha_{em}(q^2)$. For example, the renormalization scale for the electron-muon elastic scattering due to the one-photon exchange skeleton graph in the conventional Gell Mann-Low ({\rm GM-L}) scheme~\cite{gml} is simply equal to the momentum transfer squared $t= q^2$ carried by the photon propagator. The renormalization scale $\mu^2_r = t = q^2 $ is independent of the choice of initial renormalization scale $t_0$ since in QED
\begin{equation}\label{tt0}
\alpha_{em}(t) = \frac{\alpha_{em}(t_0)}{1 - \Pi(t,t_0)} \;\;, \label{qed}
\end{equation}
where
\begin{displaymath}
\Pi(t,t_0) = \frac{\Pi(t,0) -\Pi(t_0,0)}{1-\Pi(t_0,0)} ,
\end{displaymath}
which sums all vacuum polarization contributions, both proper and improper, to the dressed photon propagator. Equation (\ref{qed}) shows explicitly that although the initial renormalization scale $t_0$ is arbitrary, the final scale $t$ is unique and unambiguous, in agreement with the RG invariance. With any other choice of initial scale, one will recover the same result, but only after summing an infinite number of vacuum polarization corrections. In the case of muonic atoms $(\mu^{-}Z)$, the modified muon-nucleus Coulomb potential is precisely $-Z\alpha(-{\vec q}^{2})/ \vec{q}^{2}$; i.e., $\mu^2=-{\vec q}^2.$ Again, the renormalization scale is also unique.

The renormalization scale in QED can be determined unambiguously in any scheme, including dimensional regularization; the scale for different schemes are connected to the {\rm GM-L} scale by commensurate scale relations (CSRs)~\cite{scale1}, a topic which we will discuss below. The resulting perturbative prediction is then scheme-independent. The computation of higher-order $\{\beta_i\}$-functions for the RG equation is important for perturbative calculations at high orders~\cite{fiveqed0,fiveqed1,fiveqed2,fiveqed3}.

The scale-setting question is much more complicated in QCD due to its non-Abelian nature. Unlike QED, where there is a preferred ({\rm GM-L}) scheme and a precisely known value of the coupling at zero momentum scale (the fine-structure constant $\alpha_{em}(0)\simeq 1/137.036 \cdots$~\cite{pdg}); in the case of QCD we do not have a uniquely preferred scheme and well-determined value for the coupling in the perturbative region. Consequently, in QCD, the uncertainty from the choice of renormalization scheme and scale must be treated with great care. It should be noted, however, that pQCD reduces to Abelian theory in the zero-color $N_C \to 0$ limit~\cite{qed1}. This analytic limit provides an important constraint on the renormalization scale problem in QCD.

In the standard procedure for a first estimate of the physical observable, one chooses a renormalization scheme with an initial renormalization scale $\mu_r=\mu_r^{\rm init}$ in Eq.(\ref{phyvalue}), and then applies some scale-setting method to improve the pQCD estimate. After scale setting, the perturbative series for the physical observable (\ref{phyvalue}) can be rewritten as
\begin{equation}\label{phyvalue2}
\rho_n = {\cal C}_0 \; \alpha_s^p(\tilde\mu^{0}_r) + \sum_{i=1}^{n} \tilde{\cal C}_i(\tilde\mu^{i}_r) \; \alpha_s^{p+i}(\tilde\mu^{i}_r), \;\; (p\geq 0)
\end{equation}
where the new leading-order (LO) and higher-order scales $\tilde\mu^{0}_r$ and $\tilde\mu^{i}_r$ are functions of the initial renormalization $\mu^{\rm init}_r$, depending on the choice of the scale-setting method. At the same time, the new coefficients $\tilde{\cal C}_i(\tilde\mu^{i}_r)$ are changed accordingly so as to obtain a consistent result.

A common practice adopted in the literature is to directly deal with Eq.(\ref{phyvalue}), which is a very special case of Eq.(\ref{phyvalue2}) by simply taking $\{\tilde\mu^{i}_{r}\} \equiv \mu^{\rm init}_r = Q$ and $\tilde{\cal C}_i(\tilde\mu^{i}_r)\equiv {\cal C}_i(Q)$. Here $Q$ is usually taken as the typical momentum transfer of the process or a value which minimizes the contributions of the loop diagrams. As compensation, one varies the value of $Q$ over a certain range, such as the typical range $[Q/2, 2\,Q]$, to ascertain the renormalization scale uncertainty. This is the simplest scale setting method. It is often argued that by setting and varying the renormalization scale in this way, one can estimate contributions from higher-order terms; i.e. a change in the renormalization scale will affect how much of a result comes from Feynman diagrams without loops, and how much comes from the leftover finite parts of loop diagrams. Because of its perturbative nature, it is a common belief that those scheme and scale uncertainties will be reduced after finishing a higher-and-higher order calculation. Especially, because of the improvement of loop calculation technologies developed in recently years, many high-energy processes involving heavy particles have been calculated up to next-to-next-to-leading-order (NNLO) or even higher, which greatly improves our theoretical estimations in comparison with the experimental data. However, this {\it ad hoc} assignment of scale and its range leads to an important systematic error in the present theoretical and experimental analysis.

Besides the complexity of higher-and-higher order calculations, there are many weak points of this conventional scale-setting method:

\begin{enumerate}
\item Following the discussion below Eq.(\ref{phyvalue}), the fixed-order estimation is renormalization scheme dependent: different choice of renormalization schemes will lead to different theoretical results. In addition to the ad hoc dependence on the choice of $Q$,

 \begin{itemize}
  \item It is clearly artificial to guess a renormalization scale and to study its uncertainty by simply varying $\mu_{r}\in [Q/2, 2\,Q]$. Why is the scale uncertainty estimated only by varying a factor of $1/2$ or $2$, and not, say, $3$ times of Q ? For example, Ref.~\cite{Q6} argues that after including the first and second order corrections to several deep inelastic sum rules which are due to heavy flavor contributions, the renormalization scale $\mu_{r}$ should be taken as $\mu_r \sim 6.5 \,m$, if taking the typical scale $Q$ to be the corresponding heavy quark mass $m$. The variation of $Q$ allows one to estimate some of the contributions from higher-order terms, however, this only exposes the $\{\beta_i\}$-dependent non-conformal terms, not the entire perturbative series. It also should be emphasized that the renormalization scale for the heavy-quark loop that appears in the three-gluon coupling depends nontrivially on the virtualities of the three gluons entering the three-gluon vertex~\cite{binger}.

  \item Sometimes, there are several choices for the typical momentum transfer of the process, all of which, according to the arguments of the conventional scale setting, can be taken as the renormalization scale, such as the heavy-quark mass $m$, the collision energy of the subprocess $\sqrt{s}$, etc. Which one provides the correct theoretical estimate ?  Taking the $B_c$-meson hadroproduction as an example, different choices of typical momentum flow will lead to about $30\%$ error to the total cross-section~\cite{typicalscale}. Moreover, the idea of the typical momentum transfer as the renormalization scale only gives us the order of magnitude for the scale and we do not know which one is `optimal', $Q/2$, $Q$, $2Q$ or any others.
 \end{itemize}

\item There are uncancelled large logarithms as well as the ``renormalon" terms in higher orders which diverge as ($n!\beta_i^{n}\alpha_s^n$). The renormalon divergence was discovered in 1970s~\cite{renormalon1,renormalon2,renormalon3}. It has been found that those renormalon terms can give sizable contributions to the theoretical estimates, such as $e^+e^-$ annihilation, $\tau$ decays, deep inelastic scattering, hard processes involving heavy quark, etc.; a detailed discussion on the renormalon problem can be found in the review~\cite{renormalon4}. As a recent example, for the case of $W$-boson plus three-jet production at the hadronic colliders, due to the renormalization terms and the uncanceled large logs, Ref.~\cite{wjet} shows that a poor choice of the scale using the conventional scale setting method can manifest itself as a strong dependence on the ratio of next-to-leading-order (NLO) cross section to LO cross section (the so-called $K$ factor), which can even predict unreasonable negative NLO QCD differential cross-sections in certain kinematical regions.

\item By taking the Abelian limit $N_c \to 0$ at fixed $\alpha_{em}=C_F \alpha_s$ with $C_F=(N_c^2-1)/2N_c$, we can transform the QCD case effectively to the QED case \cite{qed1,qed2}. A self-consistent scale-setting method should be adaptable to both QCD and QED. This fact can be treated as a criterion on whether a suggested scale setting is correct or not. Conventional scale setting gives wrong results when applied to QED processes: As shown above, there is no ambiguity in setting the renormalization scale in QED. In the {\rm GM-L} scheme, the renormalization scale is the virtuality of the virtual photon, which naturally sums all vacuum polarization contributions into the coupling. There is thus no reason to vary the renormalization scale $\mu^{\rm GM-L}_r$ by a factor of $1/2$ or $2$, since it is already the optimized scale.

\item As more and more data appear, especially because of the running of the high collision energy and high luminosity Large Hadronic Collider (LHC), we need more accurate theoretical estimates to suit the needs of those forthcoming high precision data. It would be helpful to know to what fixed order we can achieve the desired precision. However, the perturbative series does not appear to converge, when using conventional scale setting. The conventional scale setting appears as a lucky guess, and we have no strict criteria to ensure the perturbative convergence, which is especially because of the renormalon terms or uncanceled large logarithms. Taking the top quark pair production as an example, it is found that the total cross-section at NNLO level for the $(q\bar{q})$-channel, $q\bar{q}\to t+\bar{t}$, by taking the conventional renormalization scale choice of $m_t$, is about $50\%$ of the NLO cross-section~\cite{pmc3,nnlo}. On the other hand, the experimental result on the $t\bar{t}$ total cross section has been measured with a precision $\Delta \sigma_{t \bar{t}}/\sigma_{t \bar{t}}\sim\pm 7\%$ at the Tevatron~\cite{cdft,d0t} and $\sim \pm 10\%$ at the LHC~\cite{atlas,cms}. Thus, to derive a more precise perturbative estimation, one would need to do even higher order calculations, at least at the NNNLO level, which however is not expected to be available in the near future.

\end{enumerate}

In summary, the conventional scale-setting method assigns an arbitrary range and an arbitrary systematic error to fixed-order pQCD prediction. One may argue that the correct renormalization scale for the fixed-order prediction can be decided by comparing with the experimental data. But this surely is process dependent and greatly depresses the predictive power of the pQCD theory.

\begin{figure}[tb]
\begin{center}
\begin{minipage}[t]{8 cm}
\epsfig{file=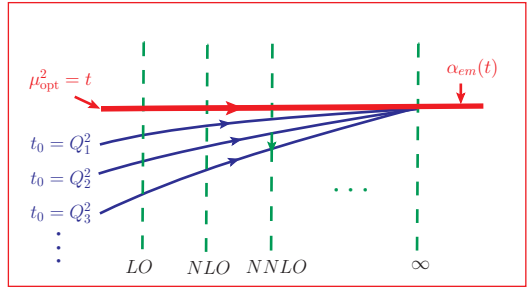,scale=0.5}
\end{minipage}
\begin{minipage}[t]{16.5 cm}
\caption{Pictorial representation of the optimized renormalization scale $\mu_{\rm opt}$. Taking electron-muon elastic scattering through one-photon exchange as an example: In the {\rm GM-L} scheme, the optimized scale is $\mu^2_{\rm opt}=t$ which corresponds to the scale-invariant value $\alpha_{em}(t)$. As a comparison, the values of $\alpha$ at fixed-orders for different choice of $t_0=Q^2_{i}$ under the conventional scale setting ($i=1,2,3,\cdots$) are shown by thin-and-solid curves. \label{optimizalscale}}
\end{minipage}
\end{center}
\end{figure}

For a general fixed-order calculation, what is the correct ``physical" scale or optimized scale? To our understanding, it should provide a prediction independent of the renormalization scheme and the choice of initial scale $\mu^{\rm init}_r$. In fact, this is a criteria of the renormalization group. A pictorial representation of what is the optimized renormalization scale is shown in Fig.(\ref{optimizalscale}), where the electron-muon elastic scattering through one-photon exchange is taken as an illustration. In the {\rm GM-L} scheme, the optimized scale $\mu^2_{\rm opt}=t$ which corresponds to the initial scale-invariant value $\alpha_{em}(t)$. This optimal scale $t$ is independent to the choice of initial scale $t_0$; i.e., any choice of $t_0$ will lead to the same scale $t$ (and then same $\alpha_{em}$) as shown by Eq.(\ref{tt0}). Moreover, by using the proper scale setting method, such as the Brodsky-Lepage-Mackenzie (BLM)~\cite{blm} method and the Principle of Maximum Conformality (PMC)~\cite{pmc3,pmc1,pmc2,pmc5,pmc6}, the prediction is also scheme independent and the argument of the coupling in different schemes have the correct displacement. For example, by using the BLM/PMC procedure, one can obtain the well-known one-loop displacement between the argument of the coupling in the ${\overline{MS}}$ scheme relative to the {\rm GM-L} scheme~\cite{conLam1}, $\alpha^{GM-L}_{em}(t)=\alpha^{\overline{MS}}_{em}(e^{-5/3}t)$.

As a comparison, the values of $\alpha_{em}$ at fixed order for different choice of $t_0=Q^2_{i}$ ($i=1,2,3,\cdots$) are shown by thin-and-solid curves in Fig.(\ref{optimizalscale}). The value of $\alpha_{em}$ strongly depends on the value of (initial) scale $t_0$ under the conventional scale-setting method. Thus, even if a particular choice of $t_0$ may lead to a value of $\alpha_{em}$ close to $\alpha_{em}(t)$ using conventional scale setting, this would not be the correct answer. As one includes higher-and-higher orders, the guessed scale will lead to a better estimate; when doing the perturbative calculation up to infinite order, any choice of $t_0$ will lead to the correct value $\alpha_{em}(t)$ as required by the RG invariance. However, if one chooses $t_0=t$, the complete all-orders result is obtained from the onset.

Does there exist such an optimized renormalization scale for a general high-energy process in non-Abelian QCD? If it does exist, how can one set it in a systematic and process-independent way?

The attempt to solve the renormalization scale and renormalization scheme ambiguities has a long history. Many authors have presented their views on how to find such an optimized scale, which have served to help clarify the issues involved. In this report, we summarize the principal ideas and results for each topic which can be served as a guide to the original literature. We first collect the improvements from a general point of view, not following their development in time but follow the sequence of how the renormalization scheme and scale questions are understood. Then, we present a detailed introduction on the BLM~\cite{blm} and its underlying principle, the PMC~\cite{pmc3,pmc1,pmc2,pmc5,pmc6,BMW}. We shall show that the PMC provides the solution for solving the renormalization scale and renormalization scheme ambiguities.

In Sec.\ref{secII} we begin with the RG equation which governs the running (scale) behavior of the QCD coupling $\alpha_s(\mu_r)$. For convenience, we extend the RG equation also to know the evolution of the renormalization scheme parameters; i.e. the extended RG equations, which was first suggested by Stevenson~\cite{pms1,pms2,pms3,pms4}, and later improved by Brodsky and Lu~\cite{HJLu}. The extended RG equations provide a convenient way for estimating both the scheme- and scale- dependence of the QCD predictions for a physical process. Any physical observable is independent of the renormalization scale and renormalization scheme; this is the main property of RG invariance~\cite{peter1,bogo,peter2,callan,symanzik}. We utilize the extended RG equations for a general discussion on this point. The solution for a special case in which all scheme parameters are set to zero, i.e. the 't Hooft scheme~\cite{tH}, is also discussed. The advantage of the 't Hooft scheme is that its coupling is scheme-independent and it gives a precise definition for the QCD asymptotic scale under a possible renormalization scheme ${\cal R}$; i.e., the scale for the 't Hooft scheme associated with the ${\cal R}$-scheme $\Lambda^{'tH-{\cal R}}_{QCD}$~\cite{HJLu}.

As a natural deduction of RG invariance, in Sec.\ref{secIII} we discuss the self-consistency requirements, such as reflexivity, symmetry and transitivity, which must be satisfied by a scale setting method~\cite{pmc6,self1,self2}. The transitivity property is especially important for self-consistent scale setting. The fact that the renormalization group is called a ``group" is mainly because of such transitivity property \cite{peter1,bogo,peter2}. These self-consistency theoretical requirements can shed light on the reliability of the scale setting method suggested in the literature.

In Sec.\ref{secIV} we present a brief summary of some typical scale setting methods, such as the Fastest Apparent Convergence (FAC) or more strictly the RG-improved effective coupling/charge method~\cite{fac1,fac2,fac3,fac4}, the Principle of Minimum Sensitivity (PMS)~\cite{pms1,pms2,pms3,pms4}, the BLM~\cite{blm} and the PMC~\cite{pmc3,pmc1,pmc2,pmc5}. The FAC and the PMS are designed to improve the perturbative series either by requiring all higher-order terms vanish~\cite{fac1,fac2,fac3,fac4} or by forcing the fixed-order series to satisfy the RG invariance at the renormalization point~\cite{pms1,pms2,pms3,pms4}. The BLM and the PMC instead improve the perturbative series by absorbing only the $n_f$-terms or the $\{\beta_i\}$-terms of the series into the argument of the coupling. Thus, these four scale-setting methods have quite different consequences. It has been found that the PMS does not satisfy the RG-properties symmetry, reflexivity, and transitivity, so that the relations among different observables depend on the choice of the intermediate renormalization scheme~\cite{pmc6,self1,self2}. Furthermore, the predicted PMS scale for the jet production from $e^+e^-$-annihilation does not yield the correct physical behavior; it anomalously rises without bound for small jet energy~\cite{Kramer1,Kramer2}. At present, the BLM is widely adopted in the literature and we will present its features observed and developed in recent years. The PMC provides the underlying principle for BLM, since it provides a rule to set the BLM scales up to all orders, and they are equivalent to each other through the PMC - BLM correspondence principle~\cite{pmc2}. Thus, all features observed in the BLM-literature are inherited by PMC.

The main idea of the PMC is that after proper procedures, all non-conformal $\{\beta_i\}$-terms in the perturbative expansion are summed into the running coupling so that the remaining terms in the perturbative series are identical to that of a conformal theory; i.e. the corresponding theory with $\{\beta_i\}=\{0\}$. The QCD predictions from PMC are then independent of renormalization scheme, because the proper displacement of the scales are included. In fact, this can be shown explicitly by considering a generalization of the conventional $\overline{MS}$-scheme for dimensional regularization, the ${\cal R}_\delta$-scheme, where a further constant $\delta$ from the ${1}/{\epsilon}$ poles is subtracted; i.e. $\frac{1}{\bar{\epsilon}} =\frac{1}{\epsilon} +\ln(4 \pi) - \gamma_E-\delta$. The $\delta$-terms in the perturbative series will always accompany $\{\beta_i\}$-terms, and thus the elimination of $\delta$-terms is equivalent to the elimination of $\{\beta_i\}$-terms. Therefore the PMC estimate can be achieved directly through a proper treatment of $\delta$-terms. This leads to a systematic prescription of setting the scales to all-orders, and opens the opportunity to start a program for automatically setting the PMC scales~\cite{BMW,BMW2}.

It has been found that PMC satisfies all self-consistency conditions. After PMC scale setting, the divergent ``renormalon" series does not appear in the conformal series; thus as in QED, the scale can be unambiguously set by PMC. The scheme independence can be adopted to derive commensurate scale relations among different observables and to find the displacements among the effective BLM/PMC scales which are derived under different schemes or conventions. The PMC renormalization scale and the resulting finite-order PMC prediction are both to high accuracy independent of the choice of the initial renormalization scale $\mu^{\rm init}_r$, consistent with the RG invariance. Even the residual scale-dependence at fixed order due to unknown higher-order $\{\beta_i\}$-terms is substantially suppressed.  Since the PMC eliminates a serious systematic scale-error in pQCD predictions, it greatly improves the precision of tests of the Standard Model (SM) and the sensitivity to new physics at the colliders. Surely, it is necessary to compute the higher-order terms of the conformal theory to estimate the true accuracy, a better understanding of the $\{\beta_i\}$-series will lead to a more accurate estimation. It is the main task of this report to present a detailed introduction to PMC by including all its developments and useful features, its detailed technologies, and its potential phenomenological applications.

In Sec.\ref{secV} we present some applications of PMC, such as the total cross-section or the forward-backward asymmetry for the top-quark pair-hadronic production at the NNLO level. In which, we show much more subtle points in applying PMC to high energy processes.

In Sec.\ref{secVI} we summarize and present an outlook.

\section{Renormalization Group Equations}
\label{secII}

In addition to the purpose of solving the renormalization scheme and scale dependence of the pQCD process, another important goal of a scale setting method is to improve the convergence of the pQCD series. A recent review on the development of the QCD coupling is presented in Ref.~\cite{prosperi}. The infrared behavior of the coupling, in the space-like and time-like regions are discussed using dispersion theory and analytic perturbation theory in Refs.~\cite{apt1,apt2,apt3}. Here we will concentrate on the behavior of the coupling and how to deal with its renormalization scheme dependence, based on the RG equation and its extended version.

\subsection{\it Renormalization Group Equation and Its Extended Version}

Predictions for observables in pQCD are expressed in terms of the renormalized coupling whose values at any fixed order depends on which renormalization scheme we choose. Conventionally, the scale dependence of the coupling is controlled by the $\beta^{\cal R}$-function,
\begin{equation} \label{basic-RG}
\beta^{\cal R}=\mu_r^2\frac{\partial}{\partial\mu_r^2} \left(\frac{\alpha^{\cal R}_s(\mu_r)}{4\pi}\right) =-\sum_{i=0}^{\infty}\beta^{\cal R}_{i}\left(\frac{\alpha^{\cal R}_s(\mu_r)}{4\pi}\right)^{i+2} ,
\end{equation}
where the superscript ${\cal R}$ stands for an arbitrary renormalization scheme, such as $MS$ scheme~\cite{MS}, $\overline{MS}$ scheme~\cite{conLam1}, $MOM$ scheme~\cite{MOM}, etc.. The $\beta^{\cal R}_i$-functions for any $MS$-like scheme are the same~\cite{kataev-beta}. The various terms in $\beta^{\cal R}_0$, $\beta^{\cal R}_1$, $\ldots$, correspond to one-loop, two-loop, $\ldots$ contributions respectively. In general, the $\{\beta^{\cal R}_i\}$ are scheme-dependent and depend on the quark mass $m_f^2$. According to the decoupling theorem~\cite{decouple}, a quark with mass $m_{f}^2\gg\mu_r^2$ can be ignored, and we can usually neglect $m_f^2$-terms when $m_{f}^2\ll\mu_r^2$. Then, for every renormalization scale $\mu_r$, one can divide the quarks into active ones with $m_f =0$ and inactive ones that can be ignored. Within this framework, it is well-known that the first two coefficients $\beta^{\cal R}_{0,1}$ are universal, which have been calculated in Refs.\cite{qcd1,qcd2,beta101,beta102,beta11,beta12,beta13}. Hereafter, we simply write them as $\beta_0$ and $\beta_1$. The $\{\beta^{\overline{MS}}_i\}_{i\geq2}$ functions for $\overline{MS}$-scheme up to three and four loops can be found in the literature~\cite{beta1,beta2,beta3,beta4}. For convenience, we present the results for any semi-simple Lie gauge group with $n_f$ fermions and $N$ colors~\cite{beta2}:
\begin{eqnarray}
\beta_0 &=& \frac{11}{3} C_A - \frac{4}{3} T_F n_f ,\\
\beta_1 &=& \frac{34}{3} C_A^2 - \frac{20}{3} C_A T_F n_f- 4 C_F T_F n_f, \nonumber \\
\beta^{\overline{MS}}_2 &=& \frac{2857}{54} C_A^3 - \frac{1415}{27} C_A^2 T_F n_f + \frac{158}{27} C_A T_F^2 n_f^2+ \frac{44}{9} C_F T_F^2 n_f^2 - \frac{205}{9} C_F C_A T_F n_f+ 2 C_F^2 T_F n_f, \nonumber \\
\beta^{\overline{MS}}_3 &=& C_A C_F T_F^2 n_f^2 \left( \frac{17152}{243} + \frac{448}{9} \zeta_3 \right) + C_A C_F^2 T_F n_f \left( - \frac{4204}{27} + \frac{352}{9} \zeta_3 \right) + \frac{424}{243} C_A T_F^3 n_f^3 \nonumber \\
&&+ C_A^2 C_F T_F n_f \left( \frac{7073}{243} - \frac{656}{9} \zeta_3 \right)
+ C_A^2 T_F^2 n_f^2 \left( \frac{7930}{81} + \frac{224}{9} \zeta_3
\right) + \frac{1232}{243} C_F T_F^3 n_f^3 \nonumber \\
&&+ C_A^3 T_F n_f \left( - \frac{39143}{81} + \frac{136}{3} \zeta_3 \right)
+ C_A^4 \left( \frac{150653}{486} - \frac{44}{9} \zeta_3 \right)
+ C_F^2 T_F^2 n_f^2 \left( \frac{1352}{27} - \frac{704}{9} \zeta_3
\right) \nonumber \\
&&+ 46 C_F^3 T_F n_f + n_f \dRANA \left( \frac{512}{9} - \frac{1664}{3} \zeta_3 \right) + n_f^2 \dRRNA \left( - \frac{704}{9} + \frac{512}{3} \zeta_3
\right) \nonumber \\
&&+ \dAANA \left( - \frac{80}{9} + \frac{704}{3} \zeta_3 \right) ,
\end{eqnarray}
where $C_A$, $N_A$ and $C_F$, $N_F$ are the quadratic Casimir invariants~\cite{casimir} and dimensions of the adjoint and fermion representation, respectively, $T_F$ is the trace normalization of the generators of the fermions, $\zeta$ is Riemann zeta function and $d_{A/F}^{abcd}$ are the invariant quartic tensors. The expressions for the latter in any semi-simple Lie group can be found in~\cite{HCGT}. For the $SU_{C}(N)$-color-group with fundamental fermions the invariants read:
\begin{equation}
T_F = \frac{1}{2}, \;\;\;\; C_F = \frac{N^2-1}{2N}, \;\;\;\; C_A =N,
\;\;\;\; \dRRNA = \frac{N^4-6N^2+18}{96N^2},
\end{equation}
\begin{equation}
\dRANA = \frac{N(N^2+6)}{48}, \;\;\;\; \dAANA =
\frac{N^2(N^2+36)}{24}, \;\;\;\; N_A = N^2-1,
\end{equation}
In particular, for the $SU_{C}(3)$-color-group, we have~\cite{beta2}
\begin{eqnarray}
\beta_0 &=& 11-\frac{2}{3}n_f \label{beta00}\\
\beta_1 &=& 102-\frac{38}{3}n_f \label{beta01}\\
\beta^{\overline{MS}}_2 &=& \frac{2857}{2}-\frac{5033}{18}n_f+\frac{325}{54}n_f^2 \label{beta02}\\
\beta^{\overline{MS}}_3 &\simeq& 29243.0-6964.30~n_f +405.089~n_f^2 +1.49931~ n_f^3 \label{beta03}
\end{eqnarray}

We first present the solution of Eq.(\ref{basic-RG}) at the one-loop level, i.e. the solution with only the first term keeping in the right-hand-side. The solution is independent of any renormalization scheme ${\cal R}$ and it takes the form
\begin{equation}
\alpha_s(\mu_r)=\frac{\alpha_s(\mu^{\rm init}_{r})} {1-\frac{\beta_0}{4\pi} \ln\left(\frac{(\mu^{\rm init}_{r})^2}{\mu^{2}_{r}}\right) \alpha_s(\mu^{\rm init}_{r})} = \alpha_s(\mu^{\rm init}_{r})\sum_{n=0}^{\infty} \left[\frac{\beta_0}{4\pi} \ln\left(\frac{(\mu^{\rm init}_{r})^2}{\mu^{2}_{r}}\right) \alpha_s(\mu^{\rm init}_{r})\right]^n ,
\end{equation}
where $\mu^{\rm init}_{r}$ stands for an arbitrary initial renormalization scale. This equation governs the one-loop behavior of the coupling. It implies that one can first adopt any initial renormalization scale $\mu^{\rm init}_{r}$ to measure the coupling; however after we have summed all $\{\beta_0\}$-terms into $\alpha_s(\mu^{\rm init}_{r})$, its final value will not depend on the choice of $\mu^{\rm init}_{r}$. This is a crucial point, which can be extended to any loops; i.e. if summing all types of $\{\beta^{\cal R}_i\}$-terms into the $\alpha^{\cal R}_s$-running through the RG equation, the behavior of $\alpha^{\cal R}_s(\mu_r)$ will be uniquely fixed and is independent of the choice of $\mu^{\rm init}_{r}$. The final summed result will be more accurate. As will be shown later, this fact agrees with the RG invariance and will be a useful guide for setting optimal renormalization scales for any fixed-order calculation. On the other hand, if setting
\begin{equation}
\Lambda_{QCD}^2=(\mu^{\rm init}_{r})^2 \exp\left(-\frac{4\pi}{\beta_0\alpha_s(\mu^{\rm init}_{r})}\right),
\end{equation}
one can rewrite $\alpha_s(\mu_r)$ in terms of an overall (universal) scale $\Lambda_{QCD}$, without any reference to a specific initial scale $\mu^{\rm init}_{r}$,
\begin{equation}\label{lambdaQCD0}
\alpha_s(\mu_r)=\frac{4\pi}{\beta_{0} \ln\left(\frac{\mu_r^2}{\Lambda_{QCD}^2}\right)}.
\end{equation}
The value of the dimensional scale $\Lambda_{QCD}$ keeps track of the initial parametrization $(\mu^{\rm init}_{r},\alpha_s(\mu^{\rm init}_{r}))$ and is universal and scale invariant; its value is not predicted by the theory but must be extracted from the measurement of $\alpha_s$ at a given reference scale. The value of $\Lambda_{QCD}$ is commonly believed to be associated with the typical hadron size; i.e. to the energy range where confinement effects set in. In effect, $\Lambda_{QCD}$ is the scale at which the coupling approximated by Eq.(\ref{lambdaQCD0}) diverges (Landau ghost~\cite{landau}).

As suggested by Stevenson~\cite{pms1,pms2,pms3,pms4}, it is convenient to use the first two universal coefficients $\beta_0$ and $\beta_1$ to rescale the coupling and the scale-parameter in Eq.(\ref{basic-RG}); i.e., by rescaling the coupling and the scale parameters as~\cite{HJLu}
\begin{displaymath}
a^{\cal R}=\frac{\beta_1}{4\pi\beta_0}\alpha^{\cal R}_s \;\;{\rm and}\;\; \tau_{\cal R}=\frac{\beta^2_0}{\beta_1} \ln\mu^2_r ,
\end{displaymath}
one can express the RG equation (\ref{basic-RG}) into a simpler canonical form
\begin{equation}\label{scale0}
\beta^{\cal R}(a)= \frac{{\rm d} a^{\cal R}}{{\rm d} \tau_{\cal R}} = -(a^{\cal R})^2 \left[1+ a^{\cal R} +c^{\cal R}_2 (a^{\cal R})^2+c^{\cal R}_3 (a^{\cal R})^3 +\cdots \right] ,
\end{equation}
where $c^{\cal R}_i = {\beta^{\cal R}_i \beta_0^{i-1}} / {\beta^i_1}$ for $i=2, 3, \cdots$, respectively \footnote{Another way to rescale the coupling, which is consistent with the large $\beta_0$-approximation~\cite{blma2,blma3,blma4}, has also been suggested in the literature~\cite{Mikhailov1}: i.e. setting $A^{\cal R}= \frac{\beta_0}{4\pi} \alpha^{\cal R}_s$, the RG equation (\ref{basic-RG}) changes to
\begin{displaymath}
\frac{d A^{\cal R}}{d \tau_{\cal R}} = -\left[(A^{\cal R})^2 + d^{\cal R}_1 (A^{\cal R})^3 + d^{\cal R}_2 (A^{\cal R})^4 + d^{\cal R}_3 (A^{\cal R})^5 +\cdots \right] ,
\end{displaymath}
where $d^{\cal R}_i=\beta^{\cal R}_{i}/\beta^{i+1}_{0}$ for $i=1, 2, \cdots$.}.

\begin{figure}[tb]
\begin{center}
\begin{minipage}[t]{8 cm}
\epsfig{file=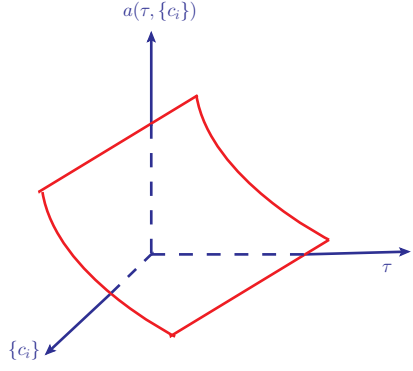, scale=0.5}
\end{minipage}
\begin{minipage}[t]{16.5 cm}
\caption{Pictorial representation of the universal coupling $a(\tau,\{c_i\})$, where $\tau$ and $\{c_i\}$ are independent scheme and scale parameters respectively. \label{unicoupling}}
\end{minipage}
\end{center}
\end{figure}

As an extension of the ordinary coupling, one can further define a universal coupling $a(\tau,\{c_i\})$ to include its dependence on both the scale parameter $\tau$ and the scheme parameters $\{c_i\}$. A pictorial representation of the universal coupling $a(\tau,\{c_i\})$ is shown in Fig.(\ref{unicoupling}). The universal coupling satisfies both the scheme and scale evolution equations~\cite{pms1,pms2,pms3,pms4,HJLu}.

Following Eq.(\ref{scale0}), the scale evolution equation can be rewritten as
\begin{equation}
\beta(a,\{c_i\}) = \frac{\partial a}{\partial \tau} = -a^2 \left[1+ a +c_2 a^2+c_3 a^3 +\cdots \right] . \label{scale}
\end{equation}
The scale-equation (\ref{scale}), similar to Eq.(\ref{scale0}), can be used to evolve the universal coupling from one scale to another. By comparing Eq.(\ref{scale0}) with Eq.(\ref{scale}), setting $\{c_i\}=\{c^{\cal R}_i\}$ , there exists a value of $\tau=\tau_{\cal R}$ for which
\begin{equation}\label{asrelation}
a^{\cal R}(\tau_{\cal R})=a(\tau_{\cal R},\{c^{\cal R}_i\}) .
\end{equation}
This shows that any coupling $a^{\cal R}(\tau_{\cal R})$ can be expressed in terms of a universal coupling $a(\tau,\{c_i\})$. Notice that the evolution equation (\ref{scale}) contains no explicit reference to QCD parameters such as the number of colors or the number of flavors. Therefore, aside from its infinite dimensional character, $a(\tau,\{c_i\})$ is just a mathematical function. Truncation of the $\{c_i\}$-terms (or $\{\beta_i\}$-functions) simply corresponds to the evaluation of $a(\tau,\{c_i\})$ in a subspace where higher-order $\{c_i\}$-terms are zero.

The scheme evolution equation is defined as
\begin{equation}
\beta_n(a(\tau,\{c_i\}),\{c_i\}) = \frac{\partial}{\partial c_n} a(\tau,\{c_i\}) .
\end{equation}
The computation of second partial derivative
\begin{equation}
\frac{\partial^2}{\partial\tau\partial c_n} a(\tau,\{c_i\})= \frac{\partial^2 }{\partial c_n \partial \tau} a(\tau,\{c_i\})
\end{equation}
implies
\begin{equation}
\frac{\partial \beta_{n}(a(\tau,\{c_i\}),\{c_i\})}{\partial\tau}=\frac{\partial \beta(a(\tau,\{c_i\}),\{c_i\})}{\partial c_n} ,
\end{equation}
which leads to
\begin{eqnarray}
\beta^2(a(\tau,\{c_i\}),\{c_i\}) \frac{\partial}{\partial a(\tau,\{c_i\})} \left(\frac{\beta_{n}(a(\tau,\{c_i\}),\{c_i\})} {\beta(a(\tau,\{c_i\}),\{c_i\})}\right) =-a^{n+2}(\tau,\{c_i\}) .
\end{eqnarray}
Finally, we obtain
\begin{equation}
\beta_n(a(\tau,\{c_i\}),\{c_i\}) = -\beta(a(\tau,\{c_i\}),\{c_i\}) \int_0^{a(\tau,\{c_i\})} \frac{ x^{n+2} dx}{\beta^2(x,\{c_i\})}, \label{scheme}
\end{equation}
where $a(0,\{0\})=\infty$ and $\beta(0,\{0\})=0$ stand for the boundary conditions. The lower limit of the integral has been set to satisfy the boundary condition $\beta_{n}(a(\tau,\{c_i\}),\{c_i\})={\cal O}(a^{n+1})$, i.e. a change in $c_n$ can only affect terms of order $a^{n+1}$ or higher~\cite{pms1,pms2,pms3,pms4}. The scheme-equation (\ref{scheme}) can be used to relate the couplings under different schemes by changing $\{c_i\}$. Equation (\ref{scheme}) can be solved perturbatively with the help of the scale-equation (\ref{scale}), which can be used to estimate how the uncalculated higher-order terms contribute to the final result. An explicit example for this point will be presented in Sec.\ref{anni}, where the value of $R(e^+ e^-)$ at the four-loop level together with its scheme error analysis will be discussed.

\subsection{\it Solution of the Scale Equation up to Four-Loop Level}

Since any coupling under any renormalization scheme can be related to a universal coupling $a(\tau,\{c_i\})$, the scale-equation (\ref{scale}) can be solved in a conventional way; i.e. the evolution of the universal running coupling can be obtained by integrating Eq.(\ref{scale}), which can be rewritten as
\begin{equation}
\left(\frac{\beta^2_0}{\beta_1}\ln\frac{\mu^2_r}{(\mu^{\rm init}_{r})^2}\right) = \int^{a(\tau,\{c_i\})}_{a(\tau_0,\{c_i\})}\frac{{\rm d} a}{\beta(a,\{c_i\})} ,
\end{equation}
where $\tau_0=({\beta^2_0}/{\beta_1}) \ln(\mu^{\rm init}_{r})^2$. Here $\mu^{\rm init}_{r}$ stands for an initial renormalization scale. Up to four-loop level, it leads to
\begin{equation}
L ={\cal C} +\frac{1}{a} + \ln a +\left(c_2-1 \right) a + \frac{c_3-2 c_2+1}{2}a^2 + {\cal O}(a^3), \label{scale2}
\end{equation}
where ${\cal C}$ is an arbitrary integration constant and
\begin{equation}
L =\frac{\beta^2_0}{\beta_1}\ln\left(\frac{\mu_r^2}{\Lambda_{QCD}^2}\right) .
\end{equation}
The value of $\Lambda_{QCD}$ can be extracted from a measurement of the QCD coupling at a given reference scale or a QCD measure with mass dimensions such as the pion decay constant $f_\pi$.

Eq.(\ref{scale2}) may be solved iteratively. In fact, up to four loops, it has the form
\begin{equation}
a(L,\{c_i\})= \frac{\kappa_1}{L}+ \frac{\kappa_2}{L^2}+ \frac{\kappa_3}{L^3}+ \frac{\kappa_4}{L^4} +{\cal O}\left(\frac{1}{L^5}\right) .
\end{equation}
Substituting it into Eq.(\ref{scale2}), these coefficients $\kappa_j$ ($j=1,\cdots,4$) can be determined by requiring all the terms with $\left(\frac{1}{L^j}\right)$ ($j<5$) vanish. We finally obtain
\begin{eqnarray}\label{alphas}
a &=& \frac{1}{L}+ \frac{1}{L^2}\left({\cal C}- \ln L\right) + \frac{1}{L^3}\left[{\cal C}^2 +{\cal C} +c_2 -(2{\cal C}-\ln L +1)\ln L -1\right] + \frac{1}{L^4}\left\{ {\cal C}\left({\cal C}^2 +\frac{5}{2}{\cal C} + 3 c_2 -2\right) \right.\nonumber\\
&& \left. -\frac{1-c_3}{2} -\left[3{\cal C}^2 +5{\cal C} +3c_2-2 -\left(3{\cal C} -\ln L +\frac{5}{2}\right)\ln L \right]\ln L\right\} +{\cal O}\left(\frac{1}{L^5}\right) .
\end{eqnarray}
One will find that the above four-loop solution agrees with Ref.~\cite{fourloopa} after proper parameter transformations. In fact, the integrated RG equation in Ref.~\cite{fourloopa} takes the form
\begin{equation}\label{alphasold}
L^* =\frac{1}{\beta^*_0}\left[\frac{1}{(a^*)}+b_1 \ln (a^*) +(b_2 -b_1^2) (a^*) +\left(\frac{b_3}{2}- b_1 b_2 +\frac{b_1^3}{2}\right) (a^*)^2\right] +C^* ,
\end{equation}
where the following definitions are adopted
\begin{eqnarray}
a^* = \frac{\alpha_s}{\pi} \;,\;
b_i = \frac{\beta^*_i}{\beta^*_0} \;,\;
L^* = \ln\left(\frac{\mu_r^2}{\Lambda_{QCD}^2}\right) ,
\end{eqnarray}
which are related to our present adopted definitions through the following relations
\begin{eqnarray}
\beta_j = {4^{j+1}} \beta^*_j\;(j=0,1,\cdots) \;,\; a=\frac{\beta_1^*}{\beta_0^*} a^* \;,\;
L =\frac{\beta^{*2}_0}{\beta^*_1}L^{*} \;,\; c^R_2 =\left(\frac{\beta_0^*}{\beta_1^*}\right)^2 b_2 \;,\; c^R_3 =\left(\frac{\beta_0^*}{\beta_1^*}\right)^3 b_3 \;.
\end{eqnarray}
It is found that by identifying the integration constant ${\cal C}^* = \frac{\beta_1}{\beta_0^2}\left({\cal C}-\ln\frac{4\beta_0}{\beta_1}\right)$, the above four-loop solution (\ref{alphas}) agrees with that of Ref.~\cite{fourloopa}.

The universal coupling has a particularly simple form when all the scheme parameters $\{c_i\}$ are set to zero (the 't Hooft scheme~\cite{tH}). The 't Hooft scheme is free of higher-order corrections and its running coupling $a^{'tH}$ is governed by the simpler RG equation
\begin{equation}
L^{'tH}= \frac{1}{a^{'tH}} + \ln\left(\frac{a^{'tH}}{1+a^{'tH}}\right) ,
\end{equation}
where
\begin{equation}
L^{'tH} =\frac{\beta^2_0}{\beta_1}\ln\left(\frac{\mu_r^2} {(\Lambda^{'tH}_{QCD})^2}\right)
\end{equation}
and the integration constant ${\cal C}$ has been absorbed into the asymptotic scale $\Lambda^{'tH}_{QCD}$ for convenience. It can be solved perturbatively as described above, being a special case of the solution (\ref{alphas}). At the two-loop level, it however has an analytic solution which can be written as a function of the scale in terms of the Lambert function $W(z)$~\cite{lambert}, which is defined through the equation, $z=W(z)\exp W(z)$.

The 't Hooft coupling presents a formal singularity at $L^{'tH}=0$; i.e. $a^{'tH}\equiv a(0,\{0\})=\infty$. Inversely, it provides a precise definition for the asymptotic scale; i.e., the 't Hooft scale $\Lambda^{'tH}_{QCD}$, which is defined to be the pole of the coupling in the 't Hooft scheme, $a^{'tH}\equiv a({\beta^2_0}/{\beta_1} \ln(\mu_r^2/(\Lambda^{'tH}_{QCD})^2),\{0\})$. Note that since the absorbed integration constant ${\cal C}$ is arbitrary, the value of $\Lambda^{'tH}_{QCD}$ is not unique, and there are infinite number of 't Hooft schemes, differing only by the value of $\Lambda^{'tH}_{QCD}$. However, under a specific renormalization scheme (${\cal R}$-scheme), its asymptotic scale can be fixed to be the 't Hooft scale associated with the ${\cal R}$-scheme $\Lambda^{'tH-{\cal R}}_{QCD}$~\cite{HJLu}, which enters into both
\begin{displaymath}
a^{\cal R}= a\left({\beta^2_0}/{\beta_1} \ln(\mu_r^2/(\Lambda^{'tH-{\cal R}}_{QCD})^2),\{c^{\cal R}_i\}\right) \;\;{\rm and}\;\; a^{'tH}= a\left({\beta^2_0}/{\beta_1} \ln(\mu_r^2/(\Lambda^{'tH-{\cal R}}_{QCD})^2),\{0\}\right) \;.
\end{displaymath}
Here the word ``associated" means we are choosing the particular 't Hooft scheme that shares the same 't Hooft scale with any given ${\cal R}$-scheme. In practice, the 't Hooft scale associated with the ${\cal R}$-scheme $\Lambda^{'tH-{\cal R}}_{QCD}$ can be fixed by setting the integration constant to be ${\cal C}_{\cal R}$. In fact, by taking the same integration constant ${\cal C}_{\cal R}$ for both the 't Hooft scheme and the chosen ${\cal R}$-scheme, one can obtain a relation between $\Lambda^{'tH-{\cal R}}_{QCD}$ and the asymptotic scale $\Lambda^{\cal R}_{QCD}$ for the ${\cal R}$-scheme; i.e.
\begin{equation}
\Lambda^{'tH-{\cal R}}_{QCD}=\exp\left(\frac{\beta_1}{2\beta_0^2}{\cal C}_{\cal R}\right)\Lambda^{\cal R}_{QCD} .
\end{equation}
As a special case, by choosing ${\cal C}_{\overline{MS}}=\ln{\beta_0^2}/{\beta_1}$ \cite{conLam1,conLam2}, we obtain
\begin{equation}\label{relation}
\Lambda^{'tH-\overline{MS}}_{QCD}=\left(\frac{\beta_1}{\beta_0^2}\right)^{-\beta_1/2\beta_0^2} \Lambda^{\overline{MS}}_{QCD} .
\end{equation}
Such a relation is consistent with the observation shown in Refs.~\cite{pms1,pms2} and has lately been observed in Refs.~\cite{pmc2,HJLu}. The present definition of $\Lambda^{\overline{MS}}_{QCD}$ is the conventional one, which is associated with the choice of ${\cal C}_{\overline{MS}}=\ln{\beta_0^2}/{\beta_1}$ and is originally suggested by Refs.~\cite{conLam1,conLam2}. There are other choices for ${\cal C}_{\overline{MS}}$ together with the choice of $\Lambda^{\overline{MS}}_{QCD}$~\cite{lams1,lams2,lams3}, which would be helpful in certain cases.

\subsection{\it Renormalization Group Invariance}
\label{RGInvariance}

Grunberg has pointed out that any perturbatively calculable physical quantity can be used to define an effective coupling, or ``effective charge", by incorporating the entire radiative corrections into its definition~\cite{fac1,fac2,fac3,fac4}. The effective coupling satisfies the same RG equation as the usual coupling. Thus, the running behavior for both the effective coupling and the usual coupling are the same if their RG equations are calculated under the same choice of scheme parameters. This idea has been discussed in more detail in Refs.~\cite{gruta1,gruta2}. Such an effective coupling can be used as a reference to define the renormalization procedure. For example, the effective coupling $\alpha_R$ from the total hadronic cross section in $e^+e^-$ annihilation can be defined as~\cite{eekataev}
\begin{equation}\label{alphaRQ}
R_{e^+e^-}(Q^2) \equiv R^0_{e^+e^-}(Q^2) \left[1+{\alpha_R(Q)\over \pi}\right] \ ,
\end{equation}
where $R^0_{e^+e^-}(Q^2)$ is the Born result and $s= Q^2$ is the squared $e^+e^-$ annihilation energy; the effective coupling $\alpha_{g_1}$ from the Bjorken sum rule for polarized electro-production can be defined as~\cite{epbj}
\begin{equation}\label{alphag1Q}
\int_0^1 {\rm d} x \left[ g_1^{ep}(x,Q^2) - g_1^{en}(x,Q^2) \right] \equiv \frac{1}{3} \left|\frac{g_A}{g_V}\right|\left[ 1- \frac{\alpha_{g_1}(Q)}{\pi} \right] ,
\end{equation}
where $Q^2 = -q^2$ and $q^2$ is the momentum transfer squared. An important suggestion is that all effective couplings must satisfy the RG equation~\cite{fac1,fac2,fac3,fac4}. Different schemes or effective couplings will differ through the third and higher coefficients of the $\{\beta^{\cal R}_i\}$-functions, which are scheme ${\cal R}$ dependent. Thus, any effective coupling can be used as a reference to define the renormalization procedure.

Physical results should be independent of theoretical conventions. The RG invariance states that a physical quantity should be independent of the renormalization scale and renormalization scheme~\cite{peter1,bogo,peter2,callan,symanzik}. Thus it is helpful to use the extended coupling Eq.(\ref{asrelation}) which contains both the scheme and scale parameters for the discussion.

\subsubsection{\it Demonstration of Renormalization Group Invariance}

The RG invariance shows that if the effective coupling $a(\tau_{\cal R},\{c^{\cal R}_i\})$ corresponds to a physical observable, the result from calculating in any scheme should be independent of any other scale $\tau_{\cal S}$ and any other scheme parameters $\{c^{\cal S}_j\}$; i.e.
\begin{eqnarray}
\frac{\partial a(\tau_{\cal R},\{c^{\cal R}_i\})}{\partial \tau_{\cal S}} &\equiv& 0 \label{inv-scale} \;, \;\;\;\;{\rm [scale\; invariance]} \\
\frac{\partial a(\tau_{\cal R},\{c^{\cal R}_i\})}{\partial c^{\cal S}_j} &\equiv& 0 \; . \;\;\;{\rm [scheme\; invariance]}\label{inv-sch}
\end{eqnarray}

\noindent{\it Demonstration}: We provide an intuitive demonstration for the RG invariance from the extended RG equations. Given two effective couplings $a(\tau_{\cal R},\{c^{\cal R}_i\})$ and $a(\tau_{\cal S},\{c^{\cal S}_i\})$ defined under two different schemes ${\cal R}$ and ${\cal S}$, one can expand $a(\tau_{\cal R},\{c^{\cal R}_i\})$ in a power series of $a(\tau_{\cal S},\{c^{\cal S}_i\})$ through a Taylor expansion:
\begin{eqnarray}
a(\tau_{\cal R},\{c^{\cal R}_i\})&=& a(\tau_{\cal S}+\bar{\tau},\{c^{\cal S}_i +\bar{c}_i\})\nonumber \\
&=& a(\tau_{\cal S},\{c^{\cal S}_i \})+ \left(\frac{\partial a}{\partial \tau}\right)_{\cal S} \bar{\tau} +\sum_{i} \left(\frac{\partial a}{\partial c_i}\right)_{\cal S} \bar{c_i} +\frac{1}{2!}\left[\left(\frac{\partial^2 a}{\partial \tau^2}\right)_{\cal S} \bar{\tau}^2 + \right.\nonumber\\
&& \left. 2\left(\frac{\partial^2 a}{\partial\tau \partial c_i}\right)_{\cal S} \bar{\tau}\bar{c}_{i} +\sum_{i,j}\left(\frac{\partial^2 a}{\partial c_{i} \partial c_j}\right)_{\cal S} \bar{c}_{i}\bar{c}_{j} \right] +\frac{1}{3!}\left[\left(\frac{\partial^3 a}{\partial \tau^3}\right)_{\cal S} \bar{\tau}^3 +\cdots \right]+\cdots , \label{asexpand}
\end{eqnarray}
where $\bar{\tau}=\tau_{\cal R} -\tau_{\cal S}$, $\bar{c}_i =c^{\cal R}_i - c^{\cal S}_i$ and the subscript ${\cal S}$ next to the partial derivatives means they are evaluated at the point $(\tau_{\cal S},\{c^{\cal S}_i \})$.

The right-hand-side of Eq.(\ref{asexpand}) can be regrouped according to the different orders of scheme-parameters $\{\bar{c}_i\}$. After differentiating both side of Eq.(\ref{asexpand}) over $\tau_{\cal S}$, we obtain
\begin{eqnarray}
\frac{\partial a(\tau_{\cal R},\{c^{\cal R}_i\})}{\partial \tau_{\cal S}} =\frac{\partial^{(n+1)} a(\tau_{\cal S},\{c^{\cal S}_i \})} {\partial\tau_{\cal S}^{(n+1)}} \frac{\bar{\tau}^{n}}{n!} + \sum_{i} \frac{\partial^{(n+1)}a(\tau_{\cal S},\{c^{\cal S}_i \})} {{\partial c^{\cal S}_i}\partial\tau_{\cal S}^{(n)}} \frac{\bar{\tau}^{n-1}\bar{c}_{i}}{(n-1)!} + \cdots , \label{rgi-use}
\end{eqnarray}
where $n$ stands for the highest perturbative order for a fixed-order calculation. It is noted that Eq.(\ref{rgi-use}) can be further simplified with the help of RG equations (\ref{scale},\ref{scheme}). If setting $n\to\infty$, the right-hand-side of Eq.(\ref{rgi-use}) tends to zero, and we obtain the scale-invariance equation (\ref{inv-scale}). This shows that if $a(\tau_{\cal R},\{c^{\cal R}_i\})$ corresponds to a physical observable (corresponding to the case of infinite perturbative series, $n\to\infty$), it will be independent of any other scale $\tau_{\cal S}$. Similarly, doing the first derivative of $a(\tau_{\cal R},\{c^{\cal R}_i\})$ with respect to the scheme-parameter $c^{\cal S}_j$, one can obtain the scheme-invariance equation (\ref{inv-sch}).

In other words, if one uses an effective coupling $a(\tau_{\cal S},\{c^{\cal S}_i\})$ under the renormalization scheme ${\cal S}$ and with an initial renormalization scale $\{\tau_{\cal S}\}$ to predict the value of another effective coupling $a(\tau_{\cal R},\{c^{\cal R}_i\})$, the RG invariance (\ref{inv-scale},\ref{inv-sch}) tell us that

\begin{itemize}
\item if we have summed all types of $c^{\cal S}_i$-terms (or equivalently the $\{\beta^{\cal S}_i\}$-terms) into the effective coupling, as is the case of an infinite-order calculation, then our final prediction of $a(\tau_{\cal R},\{c^{\cal R}_i\})$ will be independent of any choice of initial scale $\tau_{\cal S}$ and any renormalization-scheme ${\cal S}$.

\item In any case, one needs to set an initial renormalization scale to initiate a calculation, and the actual scale may or may not be equal to such initial scale, depending on which scale setting method we choose. According to Eq.(\ref{rgi-use}), for a fixed-order estimation (i.e. $n\neq \infty $), there is some residual initial-scale dependence. This is reasonable: as shown by Eq.(\ref{asexpand}), for a fixed-order calculation, the $\{\beta^{S}_i\}$-terms in even higher orders are unknown which however are necessary to cancel the scale dependence from the one-lower-order terms. Those unknown-terms provide the scale-error source for the fixed-order estimate under the conventional scale setting method. In this method, by varying the scale to be within several times of the typical momentum transfer of the process, one can estimate some of the contributions from the higher-order terms, which however only exposes the $\{\beta^{\cal S}_i\}$-dependent non-conformal terms, not the entire perturbative series.

    If one can find a proper way to sum up all the known-type of $\{\beta^{\cal S}_i\}$-terms into the coupling, and at the same time effectively suppress the contributions from those unknown-type of $\{\beta^{\cal S}_i\}$-terms at higher orders, leading to highly convergent perturbative series, such residual initial scale dependence can be greatly suppressed. Then, even for a fixed-order calculation, one can eliminate the scale error and get the right estimate for a physical observable.

\item If setting all the differences of the renormalization scheme parameters to zero, $\bar{c}_i \equiv 0$ ($i=1,2,\cdots$), Eq.(\ref{asexpand}) returns to a scale-expansion series for the coupling expanding over itself but specified at another scale; i.e.
\begin{equation}
a(\tau_{\cal R},\{c^{\cal R}_i\}) = a(\tau_{\cal S},\{c^{\cal R}_i\})+\frac{\partial a(\tau_{\cal S},\{c^{\cal R}_i\})}{\partial \tau_{\cal S}} \bar{\tau} + \frac{1}{2!}\frac{\partial^2 a(\tau_{\cal S},\{c^{\cal R}_i\})}{\partial \tau_S^2} \bar{\tau}^2 +\frac{1}{3!}\frac{\partial^3 a(\tau_{\cal S},\{c^{\cal R}_i\})}{\partial \tau_{\cal S}^3} \bar{\tau}^3 +\cdots . \label{alphasbeta}
\end{equation}
  Using the RG scale-equation (\ref{scale}), the right-hand-side of the above equation can be rewritten as perturbative series of $a(\tau_{\cal S},\{c^{\cal R}_i \})$, whose coefficient at each order is a $\{\beta^{\cal R}_i\}$-series. This, inversely, tells us which $\{\beta^{\cal R}_i\}$-series controls the running coupling at each perturbative order.

\end{itemize}

The above discussions are general, which are also suitable for Abelian QED; i.e. by taking the Abelian limit $N_c \to 0$ at fixed $\alpha_{em}=C_F \alpha_s$ with $C_F=(N_c^2-1)/2N_c$ and $N_c$ the quark's color number, we effectively return to the QED case~\cite{qed1,qed2}.

\subsubsection{\it A Combined Evolution of the Coupling in Scheme and Scale}

One can use Eq.(\ref{asexpand}) together with the scheme and scale evolution equations (\ref{scale},\ref{scheme}) to evolve any coupling $a(\tau_{\cal S},\{c^{\cal S}_i\})$, either the usual one or the effective one, ``adiabatically" into another coupling $a(\tau_{\cal R},\{c^{\cal R}_i\})$, not only in scale but also in scheme. Following the idea of Ref.~\cite{HJLu}, we show how this can be achieved. This can be used to relate any two effective couplings.

First we expand the coupling $a(\tau_{\cal R},\{c^{\cal R}_i\})$ as a perturbative series of $a(\tau_{\cal S},\{c^{\cal S}_i\})$:
\begin{equation}\label{f2f3}
a(\tau_{\cal R},\{c^{\cal R}_i\})= a(\tau_{\cal S},\{c^{\cal S}_i\}) + f_2 a^2(\tau_{\cal S},\{c^{\cal S}_i\}) + f_3 a^3(\tau_{\cal S},\{c^{\cal S}_i\}) + f_4 a^4(\tau_{\cal S},\{c^{\cal S}_i\}) +\cdots .
\end{equation}
This expansion series itself is not accurate if we truncate the series to a fixed perturbative order; i.e., if these two schemes ${\cal R}$ and ${\cal S}$ are quite different and the two scales $\tau_{\cal R}$ and $\tau_{\cal S}$ are also quite different, then the series might not be convergent. However, it can give us some RG equation improved relations for the scheme- and scale- dependent parameters among different schemes.

From the scheme and scale evolution equations (\ref{scale},\ref{scheme}), up to order ${\cal O}(a^5)$, we have
\begin{eqnarray}
\left(\frac{\partial a}{\partial \tau}\right)_{\cal S} &=& -a^2(\tau_{\cal S},\{c^S_i\}) -a^3(\tau_{\cal S},\{c^{\cal S}_i\}) - c^{\cal S}_2 a^4(\tau_{\cal S},\{c^{\cal S}_i\}) +{\cal O}(a^5) , \nonumber\\
\left(\frac{\partial^2 a}{\partial \tau^2}\right)_{\cal S} &=& 2a^3(\tau_{\cal S},\{c^{\cal S}_i\}) +5a^4(\tau_{\cal S},\{c^{\cal S}_i\}) +{\cal O}(a^5) , \nonumber\\
\left(\frac{\partial^3 a}{\partial \tau^3}\right)_{\cal S} &=& -6a^4(\tau_{\cal S},\{c^{\cal S}_i\}) +{\cal O}(a^5) , \;\;\;
\left(\frac{\partial a}{\partial c_2}\right)_{\cal S} = a^3(\tau_{\cal S},\{c^{\cal S}_i\}) + {\cal O}(a^5) , \nonumber\\
\left(\frac{\partial a}{\partial c_3}\right)_{\cal S} &=& \frac{1}{2}a^4(\tau_{\cal S},\{c^{\cal S}_i\}) + {\cal O}(a^5) , \;\;\;
\left(\frac{\partial^2 a}{\partial\tau \partial c_2}\right)_{\cal S} = -3a^4(\tau_{\cal S},\{c^{\cal S}_i\}) +{\cal O}(a^5) \nonumber .
\end{eqnarray}
Then, the Taylor expansion of $a(\tau_{\cal R},\{c^{\cal R}_i\})$ over $a(\tau_{\cal S},\{c^{\cal S}_i\})$ as shown by Eq.(\ref{asexpand}) can be simplified as
\begin{eqnarray}
a(\tau_{\cal R},\{c^{\cal R}_i\}) &=& a(\tau_{\cal S},\{c^{\cal S}_i\}) -\bar{\tau} a^2(\tau_{\cal S},\{c^{\cal S}_i\}) + \left(\bar{c}_2 -\bar{\tau}+\bar{\tau}^2\right) a^3(\tau_{\cal S},\{c^{\cal S}_i\}) + \nonumber\\
&& \left[\frac{1}{2}\bar{c}_3 -\left(c_2^{\cal S} +3\bar{c}_2\right) \bar{\tau} +\frac{5}{2}\bar{\tau}^2- \bar{\tau}^3\right]a^4(\tau_{\cal S},\{c^{\cal S}_i\}) +{\cal O}(a^5) . \label{expfourier2}
\end{eqnarray}
After an order-by-order matching of Eq.(\ref{f2f3}) and Eq.(\ref{expfourier2}), we obtain
\begin{equation}
\tau_{\cal R} = \tau_{\cal S} -f_2 ,\;\;
c^{\cal R}_2 = c_2^{\cal S} -f_2-f_2^2+f_3 ,\;\;
c^{\cal R}_3 = C^{\cal S}_3-2f_2 c^{\cal S}_2 +f_2^2 +4f_2^3 -6f_2 f_3 +2f_4 . \label{diffschrel-abc}
\end{equation}
Finally, substituting these parameters into the extended RG equations for $a(\tau_{\cal R},\{c^{\cal R}_i\})$, we can obtain the required accurate scheme and scale behaviors of $a(\tau_{\cal R},\{c^{\cal R}_i\})$.

This finishes the process of deriving the coupling $a(\tau_{\cal R},\{c^{\cal R}_i\})$ at any scale and any scheme from an initial or known $a(\tau_{\cal S},\{c^{\cal S}_i\})$, which can be summarized in the following :
\begin{itemize}
\item Derive the initial scheme parameters of $a_{\cal S}$ ($\tau_{\cal S}$, $c^{\cal S}_2$, $c^{\cal S}_3$, ...) by calculating the coefficients of its fundamental $\beta(a_{\cal S},\{c^{\cal S}_i\})$-functions.

\item From Feynman diagram calculation, obtain the expansion series of $a(\tau_{\cal R},\{c^{\cal R}_i\})$ in terms of $a(\tau_{\cal S},\{c^{\cal S}_i\})$, e.g. to derive the expansion coefficients $f_i$ for Eq.(\ref{f2f3}).

\item Use the relations as Eq.(\ref{diffschrel-abc}) to identify the parameters $\tau_{\cal R}$, $c^{\cal R}_2$, $c^{\cal R}_3$, ...

\item Use the extended scale evolution equation (\ref{scale}), to run $a_{\cal R}$ to any required scale. As a byproduct, one can adopt the extended scheme evolution equation (\ref{scheme}) for an error analysis of $a_{\cal R}$ due to unknown scheme parameters.

\end{itemize}

\section{Self-Consistency Conditions for a Scale-Setting Method}
\label{secIII}

As has been discussed above, the goal of a scale-setting method is to find an optimal renormalization scale which can be systematically set in a process independent way. This universality can provide a renormalization-scheme independent, initial-scale independent, and also highly convergent perturbative series. In the literature, it has been suggested that some self-consistent requirements, such as the reflexivity, the symmetry and the transitivity, can shed light on the reliability of the scale setting method~\cite{pmc6,self1,self2}. These self-consistency requirements have a solid background, which are natural requirements of the renormalization group (RG) equation and the RG invariance.

If one knows how to set the optimal scale, then one can translate the result freely from one scheme to another scheme through scale relations~\cite{blm-ske1,blm-ske2}. This observation has been emphasized in Ref.~\cite{scale1}, where the scale transformation among different schemes are called ``commensurate scale relations" (CSRs). It shows that even though the expansion coefficients could be different under different renormalization schemes, after a proper scale setting, one can find a relation between the effective renormalization scales which ensures that the total result remains the same under any renormalization scheme. For simplicity, following the suggestion of Refs.~\cite{pmc6,self1,self2}, we also omit the scheme parameters in the coupling in the following discussions for the self-consistent requirements of a scale setting method, but will retrieve them when necessary.

In the following four self-consistency requirements are listed, which follow from the RG equation and RG invariance:

\begin{enumerate}
\item { {\it Existence} and {\it uniqueness} of the renormalization scale $\mu_r$.} Any scale setting method must satisfy these two requirements. This agrees with our common belief that there does exist an unique and optimal renormalization scale for a fixed-order estimation. A pictorial representation of the optimized renormalization scale is shown in Fig.(\ref{optimizalscale}). For example, the optimal scale for the Abelian QED case is set by the {\rm GM-L} scheme~\cite{gml}.

\item {\it Reflexivity}. Given an effective coupling $\alpha_s(\mu_r)$ specified at a renormalization scale $\mu_r$, we can express it in terms of itself but specified at another renormalization scale $\mu'_r$,
    \begin{equation}\label{asexp}
    \quad\quad\quad \alpha_s(\mu_r)=\alpha_s(\mu'_r)+f_1(\mu_r,\mu'_r)\alpha_s^2(\mu'_r)+ \cdots,
    \end{equation}
    where $f_1(\mu_r,\mu'_r)\propto\ln(\mu_r^2/\mu_r^{'2})$. Up to infinite orders due the scale-invariance (\ref{inv-scale}), we have ${\partial \alpha_s(\mu_r)} / {\partial\ln{\mu_r^{'2}}} \equiv 0$. This, inversely, means that if $\alpha_s(\mu_r)$ is known (say, a experimentally measured effective coupling), and we try to use the above perturbative equation to ``predict" $\alpha_s(\mu_r)$ from itself, then any deviation of $\mu'_r$ from $\mu_r$ would lead to an inaccurate result due to the truncation of expansion series. More explicitly, for a fixed-order expansion with the highest perturbative-order $n$, from Eq.(\ref{rgi-use}), we obtain
    \begin{displaymath}
    \quad\quad \frac{\partial\alpha_s(\mu_r)}{\partial\ln{\mu_r^{'2}}} \propto \frac{\left(\ln{\mu_r^2}/{\mu_r^{'2}}\right)^{n}}{n!} \frac{\partial^{(n+1)}\alpha_s(\mu'_r) } {\partial(\ln\mu^{'2}_r)^{(n+1)}} .
    \end{displaymath}
    This shows, generally, the right-hand-side of Eq.(\ref{asexp}) depends on $\mu'_r$ at any fixed order. Thus to get a correct fixed-order estimate for $\alpha_s(\mu_r)$, a self-consistency scale setting must take the unique value ${\mu'_r} = \mu_r$ on the right-hand-side of Eq.(\ref{asexp}). If a scale setting satisfies this property, we say it is {\it reflexive}.

\item {\it Symmetry}. Given two different effective couplings $\alpha_{s1}(\mu_1)$ and $\alpha_{s2}(\mu_2)$ under two different renormalization schemes and at the two renormalization scales $\mu_1$ and $\mu_2$ respectively, we can expand any one of them in terms of the other:
    \begin{eqnarray}
    \quad\alpha_{s1}(\mu_1)&=&\alpha_{s2}(\mu_2)+ r_{12}(\mu_1,\mu_2)\alpha_{s2}^2(\mu_2)+\cdots,\nonumber\\
    \quad\alpha_{s2}(\mu_2)&=&\alpha_{s1}(\mu_1)+ r_{21}(\mu_2,\mu_1)\alpha_{s1}^2(\mu_1)+\cdots.\nonumber
    \end{eqnarray}
    After a general scale setting, we have
    \begin{eqnarray}
    \quad\alpha_{s1}(\mu_1)&=&\alpha_{s2}(\mu^{*}_2)+ \tilde{r}_{12}(\mu_1,\mu^{*}_2) \alpha_{s2}^2(\mu^{*}_2)+\cdots \,, \label{defr12}\\
    \quad\alpha_{s2}(\mu_2)&=&\alpha_{s1}(\mu^{*}_1)+ \tilde{r}_{21}(\mu_2,\mu^{*}_1) \alpha_{s1}^2(\mu^{*}_1)+\cdots \,. \label{defr21}
    \end{eqnarray}
    Here as a general choice, we have implicitly set the effective scales at NLO-level to be equal to the LO ones; i.e., the effective scales for the highest-order terms are usually taken as the same effective scales at the one-lower-order, since they are the scales strictly set by using the known-terms.

    Setting $\mu^{*}_{2}=\lambda_{21}\mu_{1}$ and $\mu^{*}_1= \lambda_{12}\mu_{2}$, if
    \begin{equation}\label{sysbas}
    \lambda_{12}\lambda_{21}=1 \ ,
    \end{equation}
    we say that the scale setting is {\it symmetric}. \\

 {\it Explanation}: \\

 If $\mu^{*}_{2}=\lambda_{21}\mu_{1}$ and $\mu^{*}_1= \lambda_{12}\mu_{2}$, we obtain
 \begin{eqnarray}
  &&\alpha_{s1}(\mu_1)=\alpha_{s2}(\lambda_{21}\mu_{1})+
  \tilde{r}_{12}(\mu_1,\lambda_{21}\mu_{1})\alpha_{s2}^2(\lambda_{21}\mu_{1})+\cdots
  \label{eq15} \\
  &&\alpha_{s2}(\mu_2)=\alpha_{s1}(\lambda_{12}\mu_{2})+
  \tilde{r}_{21}(\mu_2,\lambda_{12}\mu_{2})\alpha_{s1}^2(\lambda_{12}\mu_{2})+\cdots . \label{eq16}
 \end{eqnarray}
 As a combination of Eqs.(\ref{eq15},\ref{eq16}), we obtain
 \begin{equation} \label{symmetry}
  \alpha_{s1}(\mu_1) = \alpha_{s1}(\lambda_{12}\lambda_{21}\mu_{1})+ \left[
   \tilde{r}_{12}(\mu_1,\lambda_{21}\mu_{1}) + \tilde{r}_{21}(\lambda_{21}\mu_1, \lambda_{12} \lambda_{21}\mu_{1})\right] \alpha_{s1}^2(\lambda_{12} \lambda_{21}\mu_{1})+ \cdots.
 \end{equation}
  From the {\it reflexivity} property, if a scale setting is symmetric, i.e. satisfying Eq.(\ref{sysbas}), we will obtain
 \begin{equation} \label{sysrel}
 \tilde{r}_{12} (\mu_{1},\mu^*_{2}) + \tilde{r}_{21}(\mu_{2},\mu^*_{1})= 0 ,
 \end{equation}
  and vice versa. This shows that the {\it symmetry} property (\ref{sysbas}) and the relation (\ref{sysrel}) are mutually necessary and sufficient conditions.

\item {\it Transitivity}. Given three effective couplings $\alpha_{s1}(\mu_1)$, $\alpha_{s2}(\mu_2)$, and $\alpha_{s3}(\mu_3)$ under three renormalization schemes, we can expand any one of them in terms of the other; i.e.
  \begin{eqnarray}
  \quad\alpha_{s1}(\mu_1)&=&\alpha_{s2}(\mu_2)+ r_{12}(\mu_1,\mu_2)\alpha_{s2}^2(\mu_2)
       +\cdots , \nonumber\\
  \quad\alpha_{s2}(\mu_2)&=&\alpha_{s3}(\mu_3)+ r_{23}(\mu_2,\mu_3)\alpha_{s3}^2(\mu_3)
       +\cdots ,\nonumber\\
  \quad\alpha_{s3}(\mu_3)&=&\alpha_{s1}(\mu_1)+ r_{31}(\mu_3,\mu_1)\alpha_{s1}^2(\mu_1)
       + \cdots .\nonumber
  \end{eqnarray}
  After a scale setting, we obtain
 \begin{eqnarray}
\quad\alpha_{s1}(\mu_1)&=&\alpha_{s2}(\mu^{*}_2)+ \tilde{r}_{12}(\mu_1,\mu^{*}_2) \alpha_{s2}^2(\mu^{*}_2)+\cdots , \label{defr123} \\
\quad\alpha_{s2}(\mu_2)&=&\alpha_{s3}(\mu^{*}_3)+ \tilde{r}_{23}(\mu_2,\mu^{*}_3) \alpha_{s3}^2(\mu^{*}_3)+\cdots , \label{defr231} \\
\quad\alpha_{s3}(\mu_3)&=&\alpha_{s1}(\mu^{*}_1)+ \tilde{r}_{13}(\mu_3,\mu^{*}_1) \alpha_{s1}^2(\mu^{*}_1)+\cdots . \label{defr132}
 \end{eqnarray}
   Setting $\mu^{*}_2 =\lambda_{21}\mu_1$, $\mu^{*}_3 =\lambda_{32}\mu_2$ and $ \mu^{*}_1 = \lambda_{13}\mu_3$, if
  \begin{equation}\label{transbas}
  \lambda_{13}\lambda_{32}\lambda_{21} =1 \ .
  \end{equation}
  we say that the scale setting is {\it transitive}. \\

 {\it Explanation}: \\

 If $\mu^{*}_2 =\lambda_{21}\mu_1$, $\mu^{*}_3 =\lambda_{32}\mu_2$ and $ \mu^{*}_1 = \lambda_{13}\mu_3$, we obtain
 \begin{eqnarray}
  \alpha_{s1}(\mu_1)&=&\alpha_{s2}(\lambda_{21}\mu_{1})+
  \tilde{r}_{12}(\mu_1,\lambda_{21}\mu_{1})\alpha_{s2}^2(\lambda_{21}\mu_{1})+\cdots,\\
  \label{eq20}
  \alpha_{s2}(\mu_2)&=&\alpha_{s3}(\lambda_{32}\mu_{2})+
  \tilde{r}_{23}(\mu_2,\lambda_{32}\mu_{2})\alpha_{s3}^2(\lambda_{32}\mu_{2})+\cdots , \label{eq21}\\
  \alpha_{s3}(\mu_3)&=&\alpha_{s1}(\lambda_{13}\mu_{3})+
  \tilde{r}_{31}(\mu_3,\lambda_{13}\mu_{3})\alpha_{s1}^2(\lambda_{13}\mu_{3})+\cdots . \label{eq22}
 \end{eqnarray}
 As a combination of Eqs.(\ref{eq20},\ref{eq21},\ref{eq22}), we obtain
 \begin{eqnarray}
  \alpha_{s1}(\mu_1)&=&\alpha_{s1}(\lambda_{13}\lambda_{32}\lambda_{21}\mu_{1})+
  \alpha_{s1}^2(\lambda_{13}\lambda_{32}\lambda_{21}\mu_{1})\times \nonumber\\
  &&\left[ \tilde{r}_{31}(\lambda_{32}\lambda_{21}\mu_{1},
  \lambda_{13}\lambda_{32}\lambda_{21}\mu_1)+
  \tilde{r}_{23}(\lambda_{21}\mu_{1},\lambda_{32}\lambda_{21}\mu_1)+ \tilde{r}_{12}(\mu_{1},
  \lambda_{21}\mu_1)\right]+\cdots . \label{transiv}
 \end{eqnarray}

 If a scale setting is transitive, i.e. satisfying Eq.(\ref{transbas}), we obtain
 from the {\it reflexivity} property,
 \begin{equation} \label{transrel}
  \quad\quad\quad\tilde{r}_{12}(\mu_{1},\mu^*_{2}) + \tilde{r}_{23}(\mu^*_{2},\mu^*_{3}) + \tilde{r}_{31}(\mu^*_{3},\mu_{1})= 0 ,
 \end{equation}
  and vice versa. This shows that the {\it transitivity} property (\ref{transbas}) and the relation (\ref{transrel}) are mutually necessary and sufficient conditions. The {\it transitivity} property shows that under a proper scale setting method, we have $\lambda_{21}\equiv\lambda_{23} \lambda_{31}$, which means that the scale ratio $\lambda_{21}$ for any two couplings $\alpha_{s1}$ and $\alpha_{s2}$ is independent of the choice of a intermediate coupling $\alpha_{s3}$ under any renormalization scheme. Thus the relation between any two observables is independent of the choice of renormalization scheme. In fact, the {\it transitivity} property provides the theoretical foundation for the existence of commensurate scale relations among different physical observables~\cite{scale1}. The {\it transitivity} property is essential for self-consistent scale setting, and is a natural requirement from the RG invariance. It has already been pointed out that the {\it transitivity} property is the main reason why the renormalization group is called a ``group"~\cite{peter1,bogo,peter2}. The {\it transitivity} property (\ref{transbas}) can be extended to an arbitrary number of couplings; i.e. if we have $n$ couplings which are related with similar manner as above, then their transitivity relation is
 \begin{equation}
  \lambda_{1n}\lambda_{n(n-1)}\cdots\lambda_{32}\lambda_{21} =1 .
 \end{equation}

  One may observe that the {\it Symmetry} is a special case of {\it Transitivity}, since if setting $\alpha_{s3}(\mu_3)\equiv\alpha_{s1}(\mu_1)$, we have $\lambda_{11} \equiv 1$ and $\tilde{r}_{11}(\mu_{1},\mu_{1}) \equiv 0$ due to the {\it reflexivity}, which thus changes the transitive relation $\lambda_{13}\lambda_{32} \lambda_{21}=1$ into the symmetric relation $\lambda_{12}\lambda_{21}=1$.

\end{enumerate}

\begin{figure}[tb]
\begin{center}
\begin{minipage}[t]{8 cm}
\epsfig{file=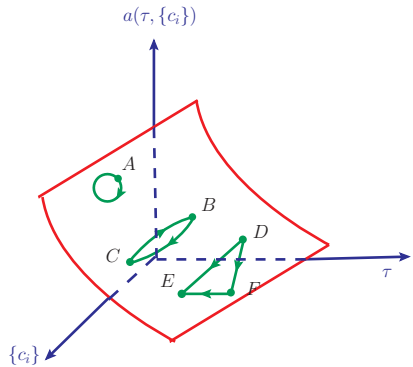,scale=0.5}
\end{minipage}
\begin{minipage}[t]{16.5 cm}
\caption{Pictorial representation of the self-consistency of the scale setting method through the universal coupling $a(\tau,\{c_i\})$. The point $A$ with a closed path represents the operation of {\it reflexivity}. The paths $\overline{BC}$ and $\overline{CB}$ represent the operation of {\it symmetry}, and the paths $\overline{DF}$,$\overline{FE}$ and $\overline{DE}$ represent the operation of {\it transitivity}. \label{fig3}}
\end{minipage}
\end{center}
\end{figure}

We present a more intuitive explanation of these requirements based on the universal coupling $a(\tau,\{ c_i \})$ and the extended renormalization group Eqs.(\ref{scale},\ref{scheme}). In the extended RG equations (\ref{scale},\ref{scheme}), there is no explicit reference to the QCD parameters, such as the number of colors or the number of active-flavors. Therefore, aside from its infinite dimensional character, $a(\tau,\{ c_i \} )$ is just a mathematical function like, say, Bessel functions or any other special functions~\cite{HJLu}. In practice, due to the unknown higher order scheme parameters $\{ c_i \}$, we need to truncate the beta function $\beta(a,\{ c_i \} )$ and solve the universal coupling $a(\tau,\{ c_i \} )$ in a finite-dimensional subspace; i.e. we need to evaluate $a(\tau,\{ c_i \})$ in a subspace where higher order $\{c_i\}$-terms are zero. In principle, this function can be computed to arbitrary degree of precision, limited only by the truncation of the fundamental $\beta$-function. In this formalism, any two effective couplings can be related by some evolution path on the hypersurface defined by $a(\tau,\{ c_i \} )$. In Fig.(\ref{fig3}) we illustrate the paths which represent the operations of {\it reflexivity}, {\it symmetry} and {\it transitivity}. We can pictorially visualize that the evolution paths satisfy all these self-consistency properties. A closed path starting and ending at the same point $A$ represents the operation of identity. Since the predicted value does not depend on the chosen path, if the effective coupling at $A$ is $a_A$, after completing the path we will also end up with an effective coupling $a_A$. Similarly, if we evolve $a_B$ at $B$ to a value $a_C$ at $C$, we are guaranteed that when we evolve $a_C$ at $C$ back to the point $B$, the result will be $a_B$. Hence, the evolution equations also satisfy {\it symmetry}. {\it transitivity} follows in a similar manner; i.e. going directly from $D$ to $E$ gives the same result as going from $D$ to $E$ through a third point $F$.

In summary, a scale setting method that satisfies {\it uniqueness} of the renormalization scale, {\it reflexivity}, {\it symmetry}, and {\it transitivity} effectively establishes equivalent relations among all the effective couplings, and thus, among all physical observables.

\section{Typical Scale-Setting Methods and Their Properties}
\label{secIV}

According to the RG invariance, physical quantities are renormalization scheme and scale independent. The exact renormalization scheme independence is respected only approximately for a perturbative calculation, which is the well-known renormalization scheme ambiguity. A resolution of renormalization scheme ambiguity is not simply to find a ``good expansion parameter for QCD". In fact, we should find a method that can provide the same estimate under any renormalization scheme for a fixed order calculation. There are some suggestions for such purpose, such as FAC~\cite{fac1,fac2,fac3,fac4}, PMS~\cite{pms1,pms2,pms3,pms4}, BLM~\cite{blm} and PMC~\cite{pmc3,pmc1,pmc2,pmc5,BMW}, which are also programmed to solve the renormalization scale ambiguity. A short review of FAC, BLM and PMS can be found in Ref.~\cite{book}. Even though all of them strive to eliminate the renormalization scheme ambiguity, they can lead to quite different results. For instance FAC and PMS are programmed to directly deal with the nature of the perturbative series, whose scale is determined by the total correction; BLM and PMC are programmed to improve the behavior of the coupling by absorbing only the part of the correction that is related to coupling constant renormalization (i.e. the $n_f$-terms or $\{\beta_i\}$-terms) into it, which then naturally improves the convergence of the perturbative series.

In this section, we make a detailed discussion on the scale setting methods FAC, PMS, BLM and PMC. We present their ideas and basic properties, and show how the self-consistency conditions, such as {\it reflexivity}, {\it symmetry}, and {\it transitivity}, are satisfied or broken by these methods.

\subsection{\it The Fastest Apparent Convergence: FAC Scale-Setting}

It is observed that the standard pQCD prediction for a physical quantity $\sigma=f(\mu_r/\Lambda_{QCD})$ usually gives its asymptotic expansion in powers of $1/\ln\left(\mu_r/\Lambda_{QCD}\right)$, which, inversely, means $\left(\mu_r/\Lambda_{QCD}\right)=f^{-1}(\sigma)$. Based on this fact, FAC
uses this fact to select the renormalization scale~\cite{fac1,fac2}. The advantage of dealing with the inverse function $f^{-1}(\sigma)$ other than the perturbative function $f(\mu_r/\Lambda_{QCD})$ lies in that it allows one to get rid of any ambiguity related to the definition of $\Lambda_{QCD}$ and $\mu_r$, since any redefinition of one of these two scales results only in a trivial overall rescaling of the inverse function $f^{-1}(\sigma)$. The inverse function $f^{-1}(\sigma)$ depends solely on the physical quantity considered, and is therefore a renormalization scheme independent object \footnote{Practically, the inverse function $f^{-1}(\sigma)$ is only an approximation due to a fixed-order calculation, there is residual scheme-dependence from the omitted higher order terms. }.

In practice, the FAC scale is determined by requiring all the higher order coefficients in Eq.(\ref{phyvalue}) to be zero; i.e. ${\cal C}_{i(\geq1)}(\mu^{FAC}_r)\equiv0$. It is for this reason, Stevenson called it ``Fastest Apparent Convergence" (FAC)~\cite{pms2}. It has been argued by Grunberg~\cite{fac3,fac4} and Krasnikov~\cite{FAC6} that it is really a renormalization group improved effective charge or effective coupling scheme, all the known-type of higher order corrections can be absorbed into an effective coupling through the RG equation in order to provide a reliable estimate, and this method is also applicable when there are large higher order corrections. Here, for simplicity, we follow Stevenson's naming for the method.

\subsubsection{\it Basic Arguments of FAC}

The expansion of $f^{-1}(\sigma)$ is obtained by introducing an effective coupling $\bar{\alpha}_s(\mu_r)$ of the particular renormalization scheme where all higher order corrections to $\sigma$ vanish~\cite{fac1}. If a physical observable in an arbitrary renormalization scheme can be written as
\begin{equation}
\sigma = A + B\left[\alpha_s(\mu_r)\right]^{d}\left[1+\sigma_1(\mu_r) \alpha_s(\mu_r) +{\cal O}(\alpha_s^2) \right] ,
\end{equation}
the effective coupling $\bar{\alpha}_s(\mu_r)$ is defined by the identity
\begin{equation}\label{eq:fac1}
\sigma=A+B\left[\bar\alpha_s(\mu_r)\right]^{d} ,
\end{equation}
where $A$ and $B$ are general perturbative or non-perturbative quantities predicted in principle by QCD, $d$ is the $\alpha_s$-order at the Born level and $\sigma_1(\mu_r)$ is the NLO coefficient. Consequently, $\bar\alpha_s(\mu_r)$ is the object effectively extracted from a LO analysis of the experimental data on $\sigma$. Next, we require such effective coupling also to satisfy the conventional RG equation; i.e. putting $\bar\rho\equiv\bar\alpha_s/(4\pi)$, we have
\begin{equation}
\mu_r^2\frac{\partial \bar\rho}{\partial\mu_r^2} =\bar\beta(\bar\rho) = -\beta_0\bar\rho^2-\beta_1\bar\rho^3 +{\cal O}(\bar\rho^4) .
\end{equation}
Its solution up to two-loop level is~\cite{fac1,fac2}
\begin{equation}\label{eq:fac2}
\beta_0\ln\left(\frac{\mu_r^2}{\Lambda_{QCD}^2}\right) = \frac{1}{\bar{\rho}}+ \frac{\beta_1}{\beta_0} \ln(\beta_0\bar\rho)+c_1 +\int_0^{\bar\rho} {\rm d}x \left[\frac{1}{x^2}-\frac{\beta_1}{\beta_0 x}+\frac{\beta_0}{\bar\beta(x)}\right] .
\end{equation}
For example, we can take $\Lambda_{QCD}$ as $\Lambda^{\overline{MS}}_{QCD}$ which is compensated by the $\overline{MS}$ value of $c_1^{\overline{MS}}$ to give a scheme independent estimate; i.e.
\begin{equation}\label{eq:fac3}
\bar\rho(\mu_r)=\rho_{\overline{MS}}(Q)\left[1+\left(c^{\overline{MS}}_{1}- \beta_0\ln\frac{\mu_r^2}{Q^2}\right)\rho_{\overline{MS}}(Q) +\cdots \right] .
\end{equation}
The result is independent of the choice of ${\overline{MS}}$. Comparison of Eq.(\ref{eq:fac1}) with Eq.(\ref{eq:fac2}) yields the inverse function $f^{-1}(\sigma)$. A simple two-loop approximation is obtained by dropping the integral in Eq.(\ref{eq:fac1}), giving
\begin{equation}\label{eq:fac4}
\beta_0\ln\left(\frac{\mu_r^2}{(\Lambda^{\overline{MS}}_{QCD})^2}\right) = \frac{1}{\bar{\rho}}+ \frac{\beta_1}{\beta_0} \ln(\beta_0\bar\rho)+c^{\overline{MS}}_1 +{\cal O}(\bar\rho) ,
\end{equation}
which equals to the two-loop expression in solution (\ref{scale2}) under suitable parameter transformations. Later on, a more complicated RG equation improved analysis was done by including the three-loop $\beta_2$-term~\cite{fac3,fac4}.

In this way a systematic expansion of $(\mu_r/\Lambda_{QCD})$ as a function of $\bar\rho$ (or equivalently, $\sigma$) is achieved. Since dimensional transmutation is implemented in most direct manner, the only free parameter in Eq.(\ref{eq:fac1}) is $\Lambda_{QCD}$ (the value of $c_1$ is just to compensate the choice of $\Lambda_{QCD}$ in order to provide a scheme-independent estimate at the considered perturbative order). Some more points regarding the FAC scale setting method are~\cite{fac2}:
\begin{itemize}

\item When the NLO correction in Eq.(\ref{eq:fac3}) is large, the use of Eq.(\ref{eq:fac4}) amounts to a resummation of the most important higher order corrections into Eq.(\ref{eq:fac3}): a RG improved perturbation theory is achieved. This is the main point of FAC.

\item From the RG equation, assuming the asymptotic expansion of $\bar\beta(\bar{\rho})$ is well-behaved, an unambiguous criterium for the validity of perturbative theory for each process is given by the condition that $(\beta_1/\beta_0)\bar\rho\ll 1$. With the help of Eq.(\ref{eq:fac4}), this alternatively means
 \begin{eqnarray}
  Q &\gg& \Lambda_{QCD}\exp\left(\frac{c_1}{2\beta_0}\right) \exp\left\{\frac{1}{2\beta_0} \left[\frac{1}{\bar{\rho}_{max}} + \frac{\beta_1}{\beta_0} \ln(\beta_0\bar\rho_{max}) \right]\right\}
 \end{eqnarray}
 with $\bar\rho_{max}\equiv\beta_0/\beta_1$.

\item The FAC method depends sensitively to which quantity it is applied. For instance, the prediction for the ratio $R=\sigma_1/\sigma_2$ of two cross sections $\sigma_1$ and $\sigma_2$ depends on whether the RG improvement is applied separately to $\sigma_1$ and $\sigma_2$, or directly to $R$. The first method is more reliable, since $\sigma_1$ and $\sigma_2$ are related directly to Feynman diagrams, whereas $R$ is a more artificial construct; this point of view also has the advantage of exploiting more completely the information contained in the expansion of $\sigma_1$ and $\sigma_2$.

\end{itemize}

\subsubsection{\it Properties of FAC}

It is straightforward to verify that FAC satisfies all the mentioned self-consistency requirements.

\begin{enumerate}

\item The {\it existence} and {\it uniqueness} of the renormalization scale $\mu_r$ are guaranteed, since the scale-setting conditions for FAC are often linear equations in $\ln \mu_r^2$, especially for lower order calculations.

  As a simple explanation, if the NLO coefficient ${\cal C}_1(\mu_r)$ for a physical observable, as defined in Eq.(\ref{phyvalue}), has the form
 \begin{equation}
  {\cal C}_1(\mu_r) = A + B \ln\mu_r^2 .
 \end{equation}
 The FAC scale is obtained by requiring, ${\cal C}_1(\mu^{\rm FAC}_r)=0$, which leads to
 \begin{equation}
  \mu^{\rm FAC}_r = \exp\left(- \frac{A}{2 B}\right) .
 \end{equation}

\item The FAC requires all $\ln(\mu_r^2/\mu'^2_r)$-terms in Eq.(\ref{asexp}) to vanish, thus we obtain $\mu'_r=\mu_r$. Then, the {\it reflexivity} is satisfied by FAC.

\item {\it Symmetry} is trivial. After FAC scale setting, two coefficients $\tilde{r}_{12}$ and $\tilde{r}_{21}$ which are defined in Eqs.(\ref{defr12},\ref{defr21}) satisfy
 \begin{displaymath}
 \tilde{r}_{12}(\mu_{1},\mu^*_{2})+\tilde{r}_{21}(\mu_{2},\mu^*_{1})=0 \,,
 \end{displaymath}
 where $\mu^*_{1}=\lambda_{12}\mu_2$ and $\mu^*_{2}=\lambda_{21}\mu_1$. It shows that these two NLO coefficients $\tilde{r}_{12}$ and $\tilde{r}_{21}$ only differ by a sign. Thus, requiring one of them to vanish is equivalent to requiring the other one to vanish. Furthermore, due to the reflexivity property, one can easily obtain $\lambda_{12}\lambda_{21}=1$.

\item {\it Transitivity} is also satisfied by FAC. In FAC the scales are so chosen such that the NLO term vanishes; i.e. after FAC scale setting, Eqs.(\ref{defr123},\ref{defr231}) change to
 \begin{eqnarray}
 \alpha_{s1}(\mu_1) &=& \alpha_{s2}(\mu^*_2) + {\cal O}(\alpha_{s2}^2) , \\
 \alpha_{s2}(\mu_2) &=& \alpha_{s3}(\mu^*_3) + {\cal O}(\alpha_{s3}^2)
 \end{eqnarray}
 As a combination, we obtain
 \begin{eqnarray}
 \alpha_{s1}(\mu_1) = \alpha_{s3}(\mu^*_3) + {\cal O}(\alpha_{s3}^2) \,,
 \end{eqnarray}
 where $\mu^*_{1}=\lambda_{13}\mu_3$, $\mu^*_{2}=\lambda_{21}\mu_1$ and $\mu^*_{3}=\lambda_{32}\mu_2$. Notice that this last equation does not contain the NLO term. Thus, the relationship between $\mu_1$ and $\mu_3$ is still given by the FAC condition (i.e., no NLO term), even when we have employed an intermediate scheme. These arguments ensure the transitive relation, $\lambda_{31}=\lambda_{32}\lambda_{21}$, be satisfied.

\end{enumerate}

\subsection{\it The Principle of Minimum Sensitivity: PMS Scale-Setting}

An ``unphysical" parameter, such as the renormalization scale or the renormalization scheme, means its value will not affect the true result of a physical observable. For an all-order calculation, it is true due to the RG invariance. However, for a fixed-order calculation, there is a remaining dependence on the ``unphysical" parameters underlying the conventional scale setting, which depends on the perturbative convergence of the process.

The PMS scale setting is designed to eliminate the renormalization scheme dependence. Given the result in some arbitrary initial renormalization scheme, the outcome of PMS is suggested to be a unique and optimum result, which is scheme independent~\cite{pms1,pms2,pms3,pms4}. It is based on the argument that if an estimate has to depend on some ``unphysical" parameters, then their values should be chosen in order to minimize the sensitivity of the estimate to small variations of these parameters; i.e. the scheme and scale must be chosen so as to minimize the sensitivity of the estimation to their small variations. It has later been argued, cf. Ref.~\cite{pmsrs}, that the perturbative convergence can also be improved by PMS. However, in practice this is not fulfilled.

More explicitly, the PMS requires the truncated series, i.e. the approximant of a physical observable, e.g. $\rho_n$ which is defined in Eq.(\ref{phyvalue}), to satisfy the following RG invariance,
\begin{eqnarray}
\frac{\partial \rho_n}{\partial \tau} = \left(\left. \frac{\partial}{\partial\tau}\right|_{\alpha_s} +\beta(\alpha_s)\frac{\partial}{\partial \alpha_s}\right) \rho_n &\equiv& 0 \label{pmsbasic} \\
\frac{\partial \rho_n}{\partial \beta_j}=\left(\left.{\partial \over \partial \beta_j}\right|_{\alpha_s} - \beta(\alpha_{\rm s}) \int _0^{\alpha_{\rm s}} {\rm d} \alpha^\prime {\alpha^{\prime {j+2}}\over [\beta(\alpha^\prime)]^2} {\partial \over \partial \alpha_{\rm s}} \right ) \rho_n &\equiv& 0 ,
\end{eqnarray}
where $\tau =\ln(\mu_r^2/\Lambda^2_{QCD})$ and $j\geq2$. Here, we have used the following equation, which is a transformation of Eq.(\ref{scheme}):
\begin{eqnarray}
{\partial \alpha_{\rm s} \over \partial \beta_j} &=& -\beta(\alpha_{\rm s}) \int_0^{\alpha_{\rm s}} {\rm d} \alpha^\prime {\alpha^{\prime j+2} \over [\beta(\alpha^\prime)]^2} = \frac{\alpha_{\rm s}^{j+1}}{\beta_0}\left(\frac{1}{j-1}-\frac{\beta_1}{\beta_0}
\frac{j-2}{j(j-1)}\alpha_{\rm s}+\dots\right) . \label{OPTR}
\end{eqnarray}
The functions $\beta(\alpha_s)$, $\alpha_s$, $\beta_j$ and etc. are scheme independent. Here for convenience, we have omitted the scheme labels in these equations.

\subsubsection{\it Basic Arguments of PMS}

Every renormalization scheme corresponds to a different $\beta(\alpha_s)$-series, and thus a different (effective) coupling. The PMS optimization \cite{pms1,pms2,pms3,pms4} for the perturbative series can be required in the variables that control such a scheme, e.g. the subtraction point $\mu_r$ and the scheme dependent coefficients $\beta_2$, $\beta_3$, $\dots$. For definiteness, following Ref.~\cite{pms1,pms2,prosperi,pmsrs}, we adopt the process $R\equiv\sigma(e^+e^-\to {\rm hadrons})/\sigma(e^+e^-\to\mu^+\mu^-)$ for an explanation of PMS. Detailed derivation of the process can be found in Refs~\cite{prosperi,pmsrs}. For self-consistency, we present their main results here, but will transform their notations to agree with our present conventions.

The quantity $R_{e^+e^-}(s)$ with an arbitrary choice of $\mu_r$ (for the moment different from the total energy $s$) and in an arbitrary renormalization scheme takes the form
\begin{equation}
R_{e^+e^-}(s)= \left(3\sum_f Q_f^2 \right) \left[1+\frac{\alpha_{s}(\mu_r)}{\pi}+r_2(s)
\left(\frac{\alpha_{s}(\mu_r)}{\pi}\right)^2 + r_3(s) \left(\frac{\alpha_{s}(\mu_r)}{\pi}\right)^3 + \dots \right]\,,
\end{equation}
where $Q_f$ stands for the electric charge of $f$ quark. According to PMS, the quantity $R_{e^+e^-}(s)$ should be renormalization scheme and renormalization scale independent even at the fixed order; i.e., it is stationary for the scale parameter $\tau$ and the scheme-dependent parameters $\beta_i$ ($i\geq2$). If we neglect the masses of active quarks, we obtain the scale-invariant and scheme-invariant equations:
\begin{eqnarray}
\left({\partial \over \partial \tau} + \beta(\alpha_{\rm s}){\partial
 \over \partial \alpha_{\rm s}} \right) R_{e^+e^-} &=& 0 \;, \label{OPTRG1}\\
\left ({\partial \over \partial \beta_j} - \beta(\alpha_{\rm s}) \int _0^{\alpha_{\rm s}}
 {\rm d}\alpha^\prime {\alpha^{\prime {j+2}}\over [\beta(\alpha^\prime)]^2}
{\partial \over \partial \alpha_{\rm s}} \right ) R_{e^+e^-} &=& 0\,, \label{OPTRG2}
\end{eqnarray}
where $j=2$, $3$, $\dots$, $\tau =\ln(\mu_r^2/\tilde\Lambda^2_{QCD})$. Similar to Eq.(\ref{relation}), $\tilde\Lambda_{QCD}$ is related to the conventional $\Lambda^{\overline{MS}}_{QCD}$ through the following relation
\begin{displaymath}
\tilde\Lambda_{QCD}=\left(\frac{\beta_1}{\beta_0^2}\right)^{-\beta_1/2\beta_0^2} \Lambda^{\overline{MS}}_{QCD} .
\end{displaymath}
Eqs.(\ref{OPTRG1},\ref{OPTRG2}) can be used, first to obtain $r_2$, $r_3$, $\dots$ in an arbitrary renormalization scheme, when we know this quantities in a specific renormalization scheme, and then to make the optimal choice for $\tau$, $\beta_2$, $\beta_3$, $\dots$.

From now on we will use the notation $a=\alpha_{\rm s}/\pi$. Replacing (\ref{OPTR}) in (\ref{OPTRG2}), asking that these equations for a given $\mu_r^2$ are satisfied for an arbitrary value of $a$, we obtain differential equations for $r_2$, $r_3$ and etc.. Restricting to $r_2$, $r_3$ and $j = 2$, we have
\begin{eqnarray}
&&{\partial r_2 \over \partial \tau}= {1 \over 4}\beta_0 \qquad \qquad \qquad
{\partial r_2 \over \partial \beta_2} = 0 \qquad\nonumber \\
&&{\partial r_3 \over \partial\tau}={1 \over 2}\beta_0 r_2
+ {1\over 16}\beta_1\qquad
{\partial r_3 \over \partial \beta_2} = - {1 \over 16}{1 \over \beta_0} \qquad
\end{eqnarray}
Integrating the above equations, we obtain
\begin{eqnarray}
r_2 &=& {1\over 4}\beta_0 \tau + \rho_2 \qquad \qquad\nonumber \\
r_3 &=& { 1 \over 16} \beta_0^2 \tau^2 + {1 \over 2}\beta_0 \rho_2
\tau + { 1 \over 16} \beta_1 \tau - { 1 \over 16}{ \beta_2 \over
\beta_0} + \rho_3^\prime = \left ( r_2 + {1 \over 8}{\beta_1 \over \beta_0} \right)^2 -
{1 \over 16 }{\beta_2 \over \beta_0} + \rho_3
\label{OPTcoeff} \,,
\end{eqnarray}
where $\rho_2$ and $\rho_3$ are integration constants independent of $\tau$, $\beta_2$, $\dots$ and are scheme independent. They can be calculated, e.g. equating $\beta_2$, $r_2$, $r_3$ to their expressions $\beta_2^{\rm \overline {MS}}$, $r_2^{\rm \overline {MS}}$, $r_3^{\rm \overline {MS}}$ in the $\rm \overline {MS}$ scheme~\cite{kataev5}; then, we have
\begin{eqnarray}
\rho_2 &=& r_2^{\rm \overline {MS}} - {1\over 4} \beta_0 \ln {s \over \tilde \Lambda^2_{QCD}}\,\,,\nonumber\\
\rho_3 &=& r_3^{\rm \overline {MS}} - \left ( r_2^{\rm \overline {MS}}+{1 \over 8}{\beta_1 \over \beta_0} \right )^2 + {1 \over 16}{ \beta_2^{\rm \overline {MS}}\over \beta_0}\,.
\label{OPTRGinv}
\end{eqnarray}
Note that $\rho_3$ turns out to be independent of $s$, and $r_2$ has the form
\begin{equation}
r_2 = -{1\over 4}\beta_0 \ln {s \over \tilde\Lambda^2_{QCD}}+ {1\over 4}
\beta_0 \tau + r_2^{\rm \overline {MS}}\,,
\label{OPTr2}
\end{equation}
while $r_3$ depends on $s$ and $\tau$ only through $r_2$.

Using the 3-loop expression for the $\{\beta_i\}$-functions, we have
\begin{equation}
\tau = {4\over \beta_0 a} + {\beta_1 \over \beta_0^2}\ln \left({\beta_1 a \over \beta_0}\right) - {\beta_1 \over 2\beta_0^2} \ln \left({16\beta_0+4\beta_1 a + \beta_2 a^2\over \beta_0}\right) + {2\beta_2 \beta_0 - \beta_1^2 \over 2\beta_0^2}\,f(a,\beta_2)
\label{running3loop}
\end{equation}
with
\begin{equation}
f(a,\beta_2)={1 \over \sqrt{D}} \ln {4\beta_0 + {1 \over 2} a (\beta_1+\sqrt D) \over 4\beta_0 + {1 \over 2}a (\beta_1-\sqrt D)}
\label{function}
\end{equation}
and $D=\beta_1^2-4\beta_2 \beta_0\,$. Note that the present complex equation (\ref{running3loop}) is the strict three-loop solution, as a simpler estimation, one can use its perturbative expansion (\ref{alphasold}) to do the following discussion.

Let us make the same replacement in Eqs.(\ref{OPTRG1},\ref{OPTRG2}) and truncate Eq.(\ref{OPTR}) at ${\cal O} (a^3)$. By requiring Eqs.(\ref{OPTRG1},\ref{OPTRG2}) be exactly satisfied, we obtain the following equations
\begin{eqnarray}
3 \beta_0 r_3 + {1 \over 2} \beta_1 r_2 + {1 \over 16} \beta_2
+\left(3\beta_1 r_3 + {1 \over 2}\beta_2 r_2\right){a\over4} + 3 \beta_2 r_3 {a^2\over16} &=& 0 \,, \label{optimization1} \\
\left[1 +\left({\beta_1\over4 \beta_0} +2r_2 \right)\, a \right] I(a,\beta_2) - a &=& 0 \,,
\label{optimization2}
\end{eqnarray}
where
\begin{displaymath}
I(a,\beta_2)={4\beta_0 \over D}\left[ {(4\beta_1^2 - 8\beta_2 \beta_0)a+\beta_2 \beta_1 a^2 \over 16\beta_0 + 4\beta_1 a + \beta_2 a^2} - 2\beta_2 \beta_0 \, f(a,\beta_2) \right]\,.
\end{displaymath}
Eq.(\ref{running3loop}) gives $a$ as a function of $\tau$, and then, Eqs.(\ref{optimization1},\ref{optimization2}) become equations in $\tau$ and $\beta_2$, which determine the optimal choice of $\bar{\tau}(s)$ and $\bar{\beta}_2(s)$ for every $s$. We obtain an optimized running coupling $\overline a(s)$ through this way, which together with the optimized values $\bar{r}_2(s)$ and $\bar{r}_3(s)$, can be used to evaluate the quantity $R_{e^+e^-}(s)$. Some more points regarding the PMS scale setting method are:
\begin{itemize}
\item Because of the scheme-equation (\ref{OPTRG2}), we can obtain the coefficients $r_2$, $r_3$, and etc. under any renormalization scheme, cf. Eq.(\ref{OPTRGinv}). This equation is helpful for determining the universal integration constants $\rho_2$, $\rho_3$, and etc. by using the results derived from the conventional $\overline{MS}$-scheme.

\item Eqs.(\ref{optimization1},\ref{optimization2}) are general under any renormalization, so the derived formula for the optimized scale $\bar{\tau}(s)$ is also general, different choice of renormalization scheme will lead to different optimized scale, but the final result for $R(e^{+}e^{-})$ will be the same. So a scheme independent estimate is obtained using the PMS. This is the key point of PMS. The optimized scale and hence the optimized running coupling can be evaluated numerically~\cite{pmsrs}.

\item Because of the scheme-independence of the effective PMS scale $\bar{\tau}(s)$, one can obtain a relation between the effective scales under different renormalization schemes, which could be a commensurate scale relation as suggested by Ref.~\cite{scale1}. However, according to the above derivation, the PMS scale is determined as an overall effective scale for all the considered perturbative contributions, so one can not obtain a scale relation as simple as that of BLM or PMC.

\item Following the same way, the PMS method can be extended to higher order approximant.

\item It has been argued that by using PMS, there is a strong correlation between renormalization scheme insensitivity and good apparent perturbative convergence~\cite{pmsrs}. As a naive argument, we rewrite Eq.(\ref{OPTr2}) as
 \begin{displaymath}
 \qquad r_2 = \left[r_2^{\rm \overline {MS}}-{1\over 4}\beta_0 \ln {s \over \mu^{2}_r}\right] .
 \end{displaymath}
  It shows that by using the PMS optimized scale, $r_2$ happens to subtract the $\beta_0$-term into the coupling, which is similar to the description of BLM and PMC, and then the pQCD convergence will be improved at this order in a similar way as that of BLM and PMC. In this sense, PMS is consistent with BLM or PMC. However, for even higher order calculations, e.g. for $r_3$, the question is much more involved and we have no such simple correspondence.

\end{itemize}

\subsubsection{\it The Properties of PMS}

Unlike the case of FAC, in general, there are no known theorems that guarantee the {\it existence} or the {\it uniqueness} of the PMS solution. In some processes there may not be a minimum or a maximum. Although for practical cases, PMS does provide solutions, and when there are more than one solution usually only one of them lies in the physically reasonable region~\cite{pms1,pms2,pms3,pms4}, these observations alone do not prove that PMS will be trouble-free for new processes.

To discuss PMS properties in a renormalization scheme-independent way, following the suggestion of Refs.~\cite{pmc6,self1,self2}, we adopt the 't Hooft scheme \cite{tH} to define the effective coupling. Under the 't Hooft scheme, the RG equation (\ref{scale}) simplifies to
\begin{equation}
\frac{{\rm d} a}{{\rm d}\tau}=-a^2(1+a) ,
\end{equation}
whose solution can be written as
\begin{equation}
\tau =\frac{1}{a} + \ln \left( \frac{a}{1+a} \right).
\end{equation}
In the above solution, for convenience, we have redefined $\tau$ as $\frac{\beta_0^2}{\beta_1} \ln\left(\frac{\mu^2_r}{\Lambda^{'tH 2}_{QCD}}\right)$, where $\Lambda^{'tH}_{QCD}$ is the asymptotic scale under the 't Hooft scheme.

Given two effective couplings $a_1$ and $a_2$ under the 't Hooft scheme, they are related by the perturbative series
\begin{equation}
a_1(\tau_1) = a_2(\tau_2) + (\tau_2-\tau_1) a_2^2(\tau_2) + \cdots .
\end{equation}
PMS proposes the choice of $\mu_2$ (or equivalently, $\tau_2$) at the stationary point, {\it i.e.}:
\begin{equation}
\frac{{\rm d} a_1}{{\rm d} \tau_2} = 0 = \frac{{\rm d}}{{\rm d} \tau_2} \left[ a_2(\tau_2) + (\tau_2-\tau_1)
  a_2^2(\tau_2) \right] .
\end{equation}
With the help of the above RG equation, we obtain
\begin{equation}
1+a_2 = \frac{1}{2(\tau_1-\tau_2)} .
\end{equation}
In order to express $\tau_2$ in terms of $\tau_1$, one must solve the last equation in conjunction with
\begin{equation}
\frac{1}{a_2} + \log \left( \frac{a_2}{1+a_2} \right) =\tau_2 .
\end{equation}

\begin{figure}[tb]
\begin{center}
\begin{minipage}[t]{8 cm}
\epsfig{file=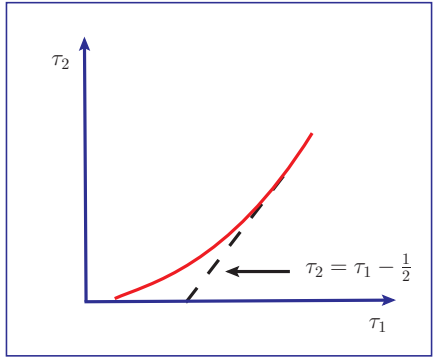,scale=0.6}
\end{minipage}
\begin{minipage}[t]{16.5 cm}
\caption{The dependence of the PMS scale parameter $\tau_2$ as a function the external scale parameter $\tau_1$. \label{fig2}}
\end{minipage}
\end{center}
\end{figure}

In Fig.\ref{fig2} we present the graphical solution of the PMS scale-parameter $\tau_2$ as a function of the external scale-parameter $\tau_1$. One may observe two points:
\begin{itemize}
\item $\tau_2 \geq \tau_1 -\frac{1}{2}$. Since $\tau_2 \neq \tau_1$ in any cases, so PMS explicitly violates {\it reflexivity}. For a fixed-order estimation, when one uses an effective coupling to predict itself, the application of PMS would lead to an inaccurate result.

\item In the large momentum region $(\tau_1 \gg 1)$, we obtain $a_2(\tau_2)\to 0$, and
 \begin{equation}\label{pmsrel}
  \tau_2 \simeq \tau_1 - \frac{1}{2} .
 \end{equation}
  Under the same renormalization scheme ${\cal R}$, we have the same asymptotic parameter $\Lambda^{'tH-{\cal R}}_{QCD}$ for both $a_1$ and $a_2$. Here $\Lambda^{'tH-{\cal R}}_{QCD}$ is the 't Hooft scale associated with the ${\cal R}$-scheme, where the word ``associated" means we are choosing the particular 't Hooft scheme that shares the same 't Hooft scale with the ${\cal R}$-scheme. Then the relation (\ref{pmsrel}) in terms of $\mu_1$ and $\mu_2$ becomes
 \begin{equation}
  {\mu_2}\simeq {\mu_1} \exp \left( -\frac{\beta_1}{4 \beta_0^2}\right ) .
 \end{equation}
\end{itemize}

More generally, it is found that after PMS scale setting, the scale displacement between any two scales $\mu_i$ and $\mu_j$ in the large momentum region is
\begin{equation}
\lambda_{ij} = \frac{\mu_i}{\mu_j} \simeq \exp\left(-\frac{\beta_1}{4\beta_0^2}\right) .
\end{equation}
This would mean that
\begin{eqnarray}
\lambda_{12} \lambda_{21} &\simeq& \exp\left(-\frac{\beta_1}{2 \beta_0^2}\right) \neq 1 , \\
\lambda_{13} \lambda_{32} \lambda_{21} &\simeq& \exp\left(-\frac{3\beta_1}{4 \beta_0^2}\right) \neq 1 . \label{pmsneq}
\end{eqnarray}
This shows that the PMS does not satisfy the {\it symmetry} and {\it transitivity} requirements. Let us point out that adding the scheme-parameter optimization in PMS does not change any of the above conclusions. It only makes the solution much more complicated~\cite{pmsrs}. The inability of PMS to meet these self-consistent requirements resides in that the derivative operations in general do not commute with the operations of {\it reflexivity}, {\it symmetry} and {\it transitivity}. This shows the necessity of further careful studies of theoretical principles underlying PMS~\cite{pmc6}.

As argued in Sec.\ref{secIII}, any truncated perturbative series will explicitly break RG-invariance (\ref{inv-scale}); i.e. Eq.(\ref{inv-scale}) can only be approximately satisfied for any fixed-order estimation. The precision depends on to which perturbative order we have calculated, the convergence of the perturbative series, and how we set the renormalization scale. As shown by Eq.(\ref{pmsbasic}), the PMS requires the truncated series, i.e. the approximant of a physical observable, to satisfy the RG-invariance near $\mu=\mu_{\rm PMS}$. This provides the underlying reason for why PMS does not satisfy the {\it reflexivity}, {\it symmetry} and {\it transitivity} properties. Phenomenological problems of PMS will be discussed in section \ref{comparison}.

\subsection{\it BLM Scale-Setting}

BLM is designed to improve the pQCD estimate by absorbing the $\{\beta_i\}$-terms into the running coupling using the $n_f$-terms as a guide. The PMC procedure, which we discuss in detail below, provides a rigorous setting for the BLM procedure. Since the invention of BLM by Brodsky-Lepage-Mackenzie~\cite{blm} in 1983, the BLM scale setting method has been widely accepted in the literature for dealing with high energy processes, such as the $e^+e^-\to $ hadrons, deep inelastic scattering, heavy meson or baryon productions or decays, the exclusive processes such as pion-photon transition form factors, and etc.. In 1992, Lepage and Mackenzie showed that the apparent failure of QCD lattice perturbative theory to account for Monte Carlo measurements of perturbative quantities is a result of choosing the bare lattice coupling constant as the expansion parameter~\cite{lepage2}. As a solution, they suggested an alternative procedure of BLM for determining the effective scale in lattice perturbation theory, which greatly enhances the predictive power of lattice perturbation theory. Later on, the reliability/importance of BLM has been emphasized in Ref.~\cite{pomeron}, where an interesting feature for the NLO BFKL Pomeron intercept function $\omega(Q^2,0)$ has been found; i.e. after using BLM scale setting, the intercept function $\omega(Q^2,0)$ has a very weak dependence on the gluon virtuality $Q^2$ in comparison with those derived from the conventional scale setting under the MOM scheme and $\overline{MS}$ scheme.

In addition, many of the favorable features of BLM have been observed in the literature, which will be listed in the following subsections. BLM presents a way to resolve the renormalization scheme-scale ambiguity, which results in a new criterion for the convergence of perturbative expansions in QED/QCD by unambiguously fixing the perturbative coefficients. In addition to eliminating the renormalization-scheme dependence, a better convergence has also been observed because of the absence of renormalons. More importantly, the renormalization scale can be determined without computing all higher-order corrections. Thus, the lower-order or even the LO analysis can be meaningfully compared with experiments.

BLM scale setting is inspired by QED. As has been discussed in the Introduction, the physical quantity within the QED framework can be expanded in perturbative series as
\begin{equation}\label{eq:blmstart}
\rho_n = {\cal C}_0 \alpha^p_{em}(\mu_r) + \sum_{i=1}^{n}{\cal C}_i(\mu_r) \alpha^{p+i}_{em}(\mu_r), \;\; (p\geq 0)
\end{equation}
where ${\cal C}_0$ is the tree-level term, ${\cal C}_i$ stands for the perturbative correction, and $p$ is the power of the coupling associated with the tree level. For Abelian theory as QED, since the variation of the effective coupling is due to vacuum polarization alone, the BLM method reduces to the standard criterion that only vacuum-polarization insertions contribute to the effective coupling. That is, after BLM scale setting, we have
\begin{eqnarray}
\rho_n &=& {\cal C}_0 \alpha_{em}^p(\mu^*_r) + \tilde{\cal C}_1(\mu^{**}_r) \alpha_{em}^{p+1}(\mu^{**}_r) + \tilde{\cal C}_2(\mu^{***}_r) \alpha_{em}^{p+2}(\mu^{***}_r)+\cdots,
\end{eqnarray}
where all photon self-energy corrections are absorbed into the couplings by an appropriate (unique) choice of effective scales $\mu_r^{*}$, $\mu_r^{**}$, $\cdots$. In fact, as will be shown by its underlying principle, PMC, the BLM scales at different orders are determined by dealing with different $\{\beta_i\}$-functions that will emerge in higher order calculation~\cite{pmc3,pmc1,pmc2,pmc5}. Since all dependence upon the number of the light-fermion flavors ($n_f$) usually enters through the photon self-energy in low orders, both the effective scales $\mu_r^{*}$, $\cdots$, and the low-order coefficients $\tilde{\cal C}_i$ are independent of $n_f$. The light-fermion loop corrections serve mainly to renormalize the coupling constant $\alpha_{em}$, as expected. Note that different from the previously introduced FAC and PMS methods, each perturbative order will usually have its own scale within BLM; there is no reason for running coupling at all orders to have the same scale. In fact, by taking the same BLM scales for all orders, serious problems occur, cf. Ref.~\cite{blma1}.

\subsubsection{\it Basic Arguments of BLM}

The BLM scale can be determined order by order in perturbation theory. We take the LO QCD scale setting as an explanation; i.e., to the first order, the physical observable can be re-expanded as~\cite{blm}
\begin{eqnarray}
\rho &=& C_{0}\alpha^{p}_{s,\overline{MS}}(\mu_r) \left[1+ \left(A n_{f}+B\right) \frac{\alpha_{s,\overline{MS}}(\mu_r)}{\pi} \right] \\
&=& C_{0}\alpha^{p}_{s,\overline{MS}}(\mu_r) \left[1+ \left(-\frac{3}{2}A \beta_{0} +\frac{33}{2}A +B \right)\frac{\alpha_{s,\overline{MS}}(\mu_r)}{\pi} \right],
\end{eqnarray}
where $\mu_r=\mu^{\rm init}_r$ stands for an initial renormalization scale, which practically can be taken as the typical momentum transfer of the process. The $n_f$ term is due to the quark vacuum polarization and we adopt the familiar $\overline{MS}$-scheme as an illustration. As will be shown later, by taking any other renormalization scheme, one can obtain the same estimate for the physical observable through proper scale displacement~\cite{scale1}. It shows that even though the expansion coefficients could be different under different renormalization schemes, after BLM scale setting, one can find a relation between the effective renormalization scales which ensures that the total result remain the same under different renormalization schemes.

At the NLO level, all $n_f$ terms should be resummed into the coupling. Using the well-known NLO $\alpha_s$-running formulae
\begin{equation}
\alpha_{s,\overline{MS}}(\mu^*_r)=\frac{\alpha_{s,\overline{MS}}(\mu_r)} {1+\frac{\beta_0}{4\pi}\alpha_{s,\overline{MS}}(\mu_r)\ln\frac{\mu^{*2}_r}{\mu^2_r}},
\end{equation}
we obtain
\begin{equation}
\rho=C_{0}\alpha^{p}_{s,\overline{MS}}(\mu^*_r) \left[1+ C^*_{1} \frac{\alpha_{s,\overline{MS}}(\mu^*_r)}{\pi} \right],
\end{equation}
where
\begin{equation}
\mu^*_r=\mu_r\exp\left(\frac{3A}{p}\right) \;\;{\rm and}\;\; C_1^*=\frac{33}{2}A +B \;.
\end{equation}
Both the effective scale $\mu_r^*$ and the coefficient $C_1^*$ are $n_f$ independent. The term $33A/2$ in $C_1^*$ serves to remove those contributions which renormalize the coupling constant into the effective coupling. Some more points of BLM scale setting are listed in the following :
\begin{itemize}
\item Using BLM scale setting, one eliminates all $n_f$-terms, so two renormalization schemes that differ only by an $n_f$-independent rescaling give identical perturbative expansions in $\alpha_{s}(\mu^*_r)$. Strictly speaking, one absorbs those $n_f$-terms that are related to the renormalization into the coupling by using the RG equation (\ref{basic-RG}). Thus, the differences between $MS$ and $\overline{MS}$, for example, are irrelevant in this approach. Note that for higher-order calculation, there are $n_f$-terms that are insensitive to the ultraviolet cutoff and thus have no relation to the $\beta$-function of the coupling, such as terms associated with Feynman diagrams with light-by-light quark loops. They should be identified and kept separate from the BLM scale setting.

\item The LO BLM scale is determined solely by the parameter $A$, which exactly comes from quark vacuum-polarization insertions. After BLM scale setting, perturbation theory can work well under high energy processes, such as $e^{+}e^{-}\to\;hadrons$, {\it deep-inelastic scattering}, $\eta_c$-{\it decays}, {\it heavy} $(Q\bar{Q})$-{\it potential} and etc., it has been found that the LO terms in $\alpha_{s,\overline{MS}}(\mu^{*}_r)$ for these processes are by themselves quite accurate~\cite{blm}.

\item Using BLM scale setting, the perturbative expansion will be unchanged in low orders as the important momenta vary across a quark threshold, since all vacuum-polarization effects due to a new quark are automatically absorbed into the effective coupling. This means, we can use a naive LO/NLO $\alpha_s$-running with the fixed active flavor number $n_f$ to do the calculation. In fact, after BLM scale setting, the value of $n_f$ can be correctly determined \cite{sjbnf}.

\item Reactions with gluon-gluon coupling are more difficult to analyze because of the quark loops appear in the higher-order corrections to the gluon-gluon vertex as well as in propagator insertions; i.e. it is not easy to separate the divergent part of the vertex from the finite process-dependent part in a unique and general fashion. For example, the BLM scale which appears in the three-gluon vertex is a function of the virtuality of the three external gluons $q^2_1$, $q^2_2,$ and $q^2_3$. It has been computed in detail in Ref.~\cite{binger}, where, by taking the subprocess $g g \to g \to Q \bar{Q}$ as an example, the authors show that when the virtualities of the gluons are very different, the energy scale for the process should be
 \begin{equation}
  \mu_r^{2} \propto {q^2_{\rm min} q^2_{\rm med} \over q^2_{\rm max}}
 \end{equation}
  where $|q^2_{\rm min}| < |q^2_{\rm med}| < |q^2_{\rm max}|$, $q^2_{\rm max}$ stands for the maximal virtuality and etc.. Such scale also correctly sets the effective number of quarks $(n_{f})$ which appear in the $\beta$-function controlling the three-gluon vertex renormalization. This example shows that it is critical to set the renormalization scale properly; a prediction based on the guessing scale such as $\mu^2_r \sim q^2_{max}$ will give misleading results.

 \begin{figure}[tb]
\begin{center}
\begin{minipage}[t]{8 cm}
\epsfig{file=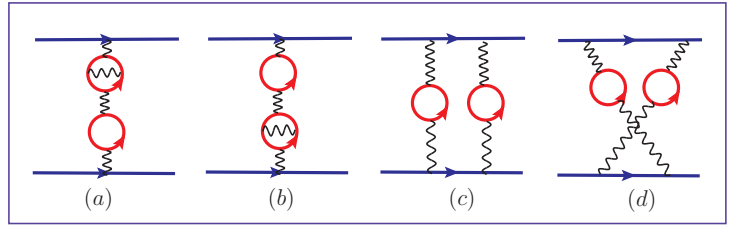,scale=0.6}
\end{minipage}
\begin{minipage}[t]{16.5 cm}
\caption{Typical $n_f^2$-terms for the electron-muon elastic scattering process at $\alpha^4_{em}$-order, where the solid circles stand for the light-lepton loops. Diagrams (a) and (b) are vacuum polarization contributions to the dressed photon propagator which will be absorbed into $\alpha_{em}(t)$ as shown by Eq.(\ref{qed}). Diagrams (c) and (d) introduce new type of $\{\beta_i\}$-terms and new PMC scales must be introduced. \label{newbeta}}
\end{minipage}
\end{center}
\end{figure}

\item As has been mentioned in the Introduction, there can be residual initial renormalization scale dependence due to the unknown higher-order $\{\beta_i\}$-terms. For example, for the simpler QED process of the electron-muon elastic scattering through the one-photon exchange only, there is one type of $\{\beta_i\}$-terms, which can be conveniently summed up to all orders and its renormalization scale can be unambiguously set as the virtuality of the exchanged photon as shown by Eq.(\ref{qed}). When two or more skeleton diagrams are involved, more than one types of $\{\beta_i\}$-terms will emerge; i.e. Fig.(\ref{newbeta}c,\ref{newbeta}d) shows the diagrams with two-photon exchange, and there are two types of $\{\beta_i\}$-terms which must be absorbed into two different PMC scales. Because of the unknown higher-order corrections for these two types of $\{\beta_i\}$-terms, there is still residual initial scale dependence.

\end{itemize}

\subsubsection{\it The Properties of BLM}

It is straightforward to verify that BLM satisfies all the self-consistent requirements outlined in Sec.\ref{secIII}.

\begin{enumerate}

\item The {\it existence} and {\it uniqueness} of the renormalization scale $\mu_r$ are guaranteed, since the scale setting conditions for BLM are often linear equations in $\ln \mu^2_r$. As a simple explanation, if the NLO coefficient ${\cal C}_1(\mu_r)$ in Eq.(\ref{eq:blmstart}) has the form
 \begin{equation}
  {\cal C}_1(\mu_r) = ( a + b\; n_f) + ( c + d\; n_f) \ln\mu^2_r ,
 \end{equation}
 with $a, b, c$ and $d$ are constants free of $n_f$, the LO BLM scale can be set as
 \begin{equation}
  \ln{\mu^{\rm BLM}_{r}} = - \frac{b}{2d} + {\cal O}(\alpha_s) ,
 \end{equation}
  where the omitted higher-order $\alpha_s$-terms will be determined by $n_f$-terms at the NLO-level or even higher levels.

\item {\it Reflexivity} is satisfied. The BLM requires all $\ln(\mu^2_r/\mu'^2_r)$-terms in Eq.(\ref{asexp}) to vanish, which are proportional to $n_f$-terms, thus we obtain
 \begin{displaymath}
  \mu'_r=\mu_r \,.
 \end{displaymath}

\item {\bf Symmetry} is trivial, because after BLM scale setting, we always have
 \begin{displaymath}
 \tilde{r}_{12}(\mu_{1},\mu^*_{2}) =- \tilde{r}_{21}(\mu_{2},\mu^*_{1})\,.
 \end{displaymath}
 That is, those two NLO coefficients only differ by a sign. Thus, requiring one of them to be $n_f$-independent is equivalent to requiring the other one also to be $n_f$-independent. This argument ensures the symmetric relation, $\lambda_{12}\lambda_{21}=1$, be satisfied after BLM scale setting.

\item {\bf Transitivity} is also satisfied by BLM. After BLM scale setting, the two coefficients $\tilde{r}_{12}(\mu_1,\mu^*_2)$ and $\tilde{r}_{23}(\mu^{*}_2,\mu^*_3)$ in the following two series
 \begin{eqnarray}
 \alpha_{s1}(\mu_1)&=& \alpha_{s2}(\mu^*_2) + \tilde{r}_{12}(\mu_1,\mu^*_2) \alpha_{s2}^2(\mu^*_2)+ {\cal O}(\alpha_{s2}^3) \label{Eq12}
 \end{eqnarray}
 and
 \begin{eqnarray}
 \alpha_{s2}(\mu^*_2) &=& \alpha_{s3}(\mu^*_3) + \tilde{r}_{23}(\mu^*_2,\mu^*_3) \alpha_{s3}^2(\mu^*_3)+ {\cal O}(\alpha_{s3}^3) \,, \label{Eq23}
 \end{eqnarray}
 should be independent of $n_f$. After substituting Eq.(\ref{Eq23}) into Eq.(\ref{Eq12}), we obtain
 \begin{equation}
 \alpha_{s1}(\mu_1) = \alpha_{s3}(\mu^*_3) + \left[\tilde{r}_{12}(\mu_1,\mu^*_2) + \tilde{r}_{23}(\mu^*_2,\mu^*_3)\right] \alpha_{s3}^2(\mu_3) + {\cal O}(\alpha_{s3}^3) \,.
 \end{equation}
 We see that the new NLO coefficient $[\tilde{r}_{12}(\mu_1,\mu^*_2) + \tilde{r}_{23}(\mu^*_2,\mu^*_3)]$ will also be $n_f$-independent, since it is the sum of two $n_f$-independent quantities. These arguments ensure the transitive relation, $\lambda_{31}=\lambda_{32}\lambda_{21}$, to be satisfied after BLM scale setting.

\end{enumerate}

\subsubsection{\it Commensurate Scale Relation in QCD}
\label{csrQCD}

The BLM prediction is renormalization-scheme independent, which is ensured by the commensurate scale relation (CSR)~\cite{scale1}, i.e. the specific value of the renormalization scale is rescaled according to the choice of the scheme so that the final result is scheme independent. All perturbatively calculable observables in QCD, such as the annihilation ratio $R_{e^+e^-}(Q^2)$, the heavy quark potential, the radiative corrections to the Bjorken sum rules and etc., can be related to each other at fixed relative scales. The CSR for the observables $A$ and $B$ in terms of their effective couplings ($\alpha_A$ and $\alpha_B$) takes the following form~\cite{scale1}
\begin{equation}
\alpha_A(Q_A) = \alpha_B(Q_B)\ \left[1 + r_{A/B} {\alpha_B(Q_B)\over \pi} +{\cal O}(\alpha^2_B)\right] .\label{BLMEqAB}
\end{equation}
The ratio of the renormalization scales $\lambda_{A/B}=Q_A/Q_B$ is so chosen that the coefficient $r_{A/B}$ is independent of the number of flavors $n_f$. This guarantees that the effective couplings for the observables $A$ and $B$ pass through new quark threshold at the same physical scale. The value of $\lambda_{A/B}$ is unique at LO, and due to the {\it transitivity} of BLM, the relative scales must satisfy the relation
\begin{equation} \label{blmrel}
\lambda_{A/B} = \lambda_{A/C}~\lambda_{C/B}\ .
\end{equation}
This ensures that predictions in pQCD are independent of the choice of an intermediate renormalization scheme $C$. In particular, the scale-fixed predictions can be made without reference to theoretically constructed renormalization schemes such as $\overline{MS}$.

As a simple explanation of CSR, let us now consider expanding any observable or effective coupling $\alpha_A$ in terms of $\alpha_{\cal V}$ (corresponding to an arbitrary intermediate renormalization scheme ${\cal V}$) up to NLO:
\begin{equation}
\alpha_A(Q_A) = \alpha_{\cal V}(Q_A) \left[1 + (C_{{\cal V}A} +D_{{\cal V}A}\, n_f)\, {\alpha_{\cal V}(Q_A) \over \pi} + {\cal O}(\alpha^2_{\cal V})\right] .
\end{equation}
Note $Q_{A}$ is a formal renormalization scale defined by the physical observable through the effective coupling $\alpha_{A}$. According to BLM scale setting, we must shift the scale $Q_{A}$ in the argument of $\alpha_{\cal V}$ to the scale $Q_{\cal V} = e^{3 D_{{\cal V}A}}Q_{A}$~\cite{blm}, and
\begin{equation}
\alpha_A(Q_A) = \alpha_{\cal V}({Q_{\cal V}}) \left[1 + r_{A/{\cal V}} {\alpha_{\cal V}(Q_{\cal V})\over \pi} + {\cal O}(\alpha^2_{\cal V}) \right]\ , \label{BLMEq}
\end{equation}
where $r_{A/{\cal V}} = C_{{\cal V}A} + (33/2) D_{{\cal V}P}$ is the NLO coefficient in the expansion of the observable $A$ in scheme ${\cal V}$. Thus, the ratio for the two relative scales between the observables $A$ and ${\cal V}$, $\lambda_{A/{\cal V}}=Q_A/Q_{\cal V}$, is fixed by the requirement that the coefficient $r_{A/{\cal V}}$ in the expansion of $\alpha_{\cal V}$ is independent of vacuum polarization corrections. Similarly, we can compute another observable or effective coupling $\alpha_B$ as an expansion in terms of $\alpha_{\cal V}$ :
\begin{equation}
\alpha_B(Q_B) = \alpha_{\cal V}(Q_{\cal V}) \left[1 + r_{B/{\cal V}} \, {\alpha_{\cal V}(Q_{\cal V}) \over \pi} + {\cal O}(\alpha^2_{\cal V}) \right]\ ,
\end{equation}
where $Q_{\cal V}=Q_B/\lambda_{B/{\cal V}}$, and again $r_{B/{\cal V}}$ must be independent of vacuum polarization contributions.

We can now substitute and eliminate $\alpha_{\cal V}(Q_{\cal V})$, which results in the required LO CSR (\ref{BLMEqAB}) with $Q_A/Q_B = \lambda_{A/B} = {\lambda_{A/{\cal V}} / \lambda_{B/{\cal V}}}$ and $r_{A/B}= r_{A/{\cal V}}-r_{B/{\cal V}}$. Note also the BLM {\it symmetry} property $\lambda_{A/B} \lambda_{B/A} = 1$. Alternatively, we can directly compute the commensurate scale $Q_A=\lambda_{A/B} Q_B$ by requiring $r_{A/B}$ to be $n_f$-independent, which is in agreement with the BLM {\it transitivity}.

The CSR given in Eq.(\ref{BLMEqAB}) provides a practical way to test QCD: One can compare two observables by checking that their effective couplings agree both in normalization and in their scale dependence. The ratio of commensurate scales $\lambda_{A/B}$ is fixed uniquely: it ensures that both observables $A$ and $B$ pass through heavy quark thresholds at precisely the same physical point. Calculations are often performed most advantageously in $\overline{\rm MS}$ scheme, but all references to such theoretically-constructed schemes may be eliminated when comparisons are made between observables. This also avoids the problem that one need not expand observables in terms of couplings which have singular or ill-defined functional dependence.

\begin{figure}[tb]
\begin{center}
\begin{minipage}[t]{8 cm}
\epsfig{file=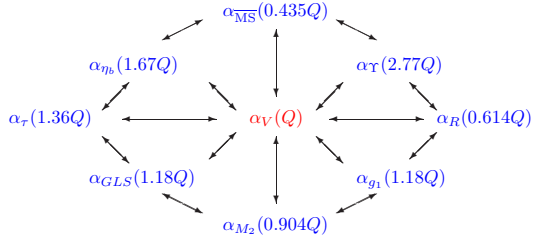,scale=0.7}
\end{minipage}
\begin{minipage}[t]{16.5 cm}
\caption{Leading order CSRs derived by Ref.~\cite{scale1} for various renormalization schemes that are defined through corresponding physical observables. Here $\alpha_V(Q)$ stands for the effective coupling defined from the heavy quark potential $V(Q^2)=-4\pi\alpha_{V}(Q)/Q^2$. \label{commensurate}}
\end{minipage}
\end{center}
\end{figure}

The intermediate renormalization scheme ${\cal V}$, which defines an effective coupling $\alpha_{\cal V}$, can be taken arbitrarily; i.e., in addition to our familiar $\overline{MS}$ scheme, any perturbatively calculable physical quantity can be used to define an effective coupling~\cite{fac1,fac2,fac3,fac4,gruta1,gruta2,eekataev}. In choosing any one of those schemes to predict another observable, the argument of the effective coupling is displaced from its (formal) physical value according to CSR, which ensures the final prediction to be independent of the renormalization scheme. The relative scale for a number of observables, which are summarized in Ref.~\cite{scale1}, is indicated in Fig.(\ref{commensurate}). Here for clarity, we have set $\alpha_{\cal V}(Q)$ to be the effective coupling $\alpha_V(Q)$ defined from the heavy quark potential $V(Q^2)=-4\pi\alpha_{V}(Q)/Q^2$. An essential feature of this scheme is that there is no renormalization scale ambiguity, since $Q^2=-t$, the photon/gluon virtuality, is the optimized scale in the {\rm GM-L} scheme.

Because of CSR, there is no difference of which renormalization scheme one chooses to do the calculation. A tricky point is that one sometimes can find a proper scheme which makes the expression much more simplified and more convergent. In particular, it has been found that up to light-by-light type corrections, all terms involving $\zeta_3, \zeta_5$ and $\pi^2$ in the relation between the annihilation ratio $R_{e^+e^-}$ and the Bjorken sum rule for polarized electroproduction are automatically absorbed into the renormalization scales~\cite{scale1}. Then, the final perturbative series becomes quite simple:
\begin{equation}
\widehat\alpha_{g_1}(Q) =\widehat\alpha_R(Q^*)- \widehat\alpha_R^2(Q^{**}) +\widehat\alpha_R^3(Q^{***}),
\end{equation}
where $\widehat\alpha = ({3 C_F / 4 \pi}) \alpha$, $Q^*$, $Q^{**}$ and $Q^{***}$ are LO, NLO and NNLO BLM scales accordingly, and the two effective couplings $\alpha_R$ and $\alpha_{g1}$ are defined in Eqs.(\ref{alphaRQ},\ref{alphag1Q}). This equation is called the generalized Crewther relation. The coefficients in CSR can be identified with those obtained in conformally invariant gauge theories as proven by Crewther~\cite{crewther1,crewther2,crewther3,crewther4,Mikhailov2}.

The CSR between observables can be tested at quite low momentum transfers, even at where pQCD expansion would be expected to break down~\cite{scale1}. It is likely that some of the higher twist contributions common to the two observables are also correctly represented by CSR. In contrast, expansions of any observable in $\alpha_{\overline{\rm MS}}(Q)$ must break down at low momentum transfer since $\alpha_{\overline{\rm MS}}(Q)$ becomes singular at $Q=\Lambda_{\overline{\rm MS}}$. For example, in the 't Hooft scheme~\cite{tH} where the higher order $\beta_n=0$ for $n=2,3,...$, $\alpha_{\overline{\rm MS}}(Q)$ has a simple pole at $Q=\Lambda^{'tH-\overline{MS}}_{QCD}$. The CSR allows one to test QCD without explicit reference to renormalization schemes such as $\overline{\rm MS}$. It is thus reasonable to expect that the perturbative series will be more convergent when one relates finite observables to each other.

As a summary, the key point of the CSR lies in that the scale displacement between different renormalization schemes is unique and does not depend on any intermediate scheme.

The above discussion on CSR is performed at LO level. In general, such scale relation (\ref{blmrel}) can be extended to any perturbative order. That is, even though the scale values maybe changed from their LO values because of the higher-order corrections, the relative relation among the scales must be remained unchanged due to the {\it transitivity} property of the BLM scale setting. As a demonstration, one needs to clarify the following two points :
\begin{itemize}
\item The LO BLM scale itself is a perturbative series with higher perturbative terms coming from a higher-order calculation of the physical observable, then we need to show that the scale relation (\ref{blmrel}) is always right for LO BLM scale after including those higher-order terms.
\item We should have similar scale relations for other higher order BLM scales, such as NLO, NNLO BLM scales.
\end{itemize}
After we demonstrate the first point, then the second point can be recursively demonstrated, since according to BLM procedure, if the LO terms have been settled down, we can separate them; while the remaining higher-order terms can be regarded as NLO correction to those known terms and then the previous LO procedures apply; and so on so forth, we can extend the LO demonstration procedure to any perturbative order. In the following, we present a demonstration of how the CSR for the LO BLM scales is satisfied even by including perturbative contributions up to NNLO.

In practice, most physical observable in pQCD are computed in $\overline{\rm MS}$ scheme, with the running coupling fixed at a physical scale of the process. For convenience, we adopt the $\overline{MS}$-scheme as the intermediate scheme. Specifically, up to NNLO, the perturbative series for two effective couplings $\alpha_1(Q)/\pi$ and $\alpha_2(Q)/\pi$ which correspond to the two physical observables $A$ and $B$ can be written as
\begin{equation}
\frac{\alpha_1(Q)}{\pi} = \frac{\alpha_{\overline{\rm MS}}(Q)}{\pi}
+ (A_1+B_1 n_f) \left( \frac{\alpha_{\overline{\rm MS}}(Q)}{\pi}\right)^2
+ (C_1+D_1 n_f+ E_1 n_f^2) \left( \frac{\alpha_{\overline{\rm MS}}(Q)} {\pi}
\right)^3 + \cdots
\label{Alpha1MS}
\end{equation}
and
\begin{equation}
\frac{\alpha_2(Q)}{\pi} = \frac{\alpha_{\overline{\rm MS}}(Q)}{\pi}
+(A_2+B_2 n_f)\left(\frac{\alpha_{\overline{\rm MS}}(Q)}{\pi}\right)^2
+(C_2+D_2 n_f+ E_2 n_f^2)\left(\frac{\alpha_{\overline{\rm MS}}(Q)}{\pi}
\right)^3 + \cdots .
\label{Alpha2MS}
\end{equation}
On the other hand, the effective coupling $\alpha_1(Q)/\pi$ can be directly expressed by $\alpha_2(Q)/\pi$; i.e.,
\begin{eqnarray}
\frac{\alpha_1(Q)}{\pi}&=&\frac{\alpha_2(Q)}{\pi}+(A_{12}+B_{12} n_f)
\left(\frac{\alpha_2(Q)}{\pi}\right)^2+(C_{12}+D_{12} n_f+ E_{12} n_f^2)
\left(\frac{\alpha_2(Q)}{\pi}\right)^3 + \cdots . \label{Alpha12}
\end{eqnarray}
As a combination of Eqs.(\ref{Alpha1MS},\ref{Alpha2MS},\ref{Alpha12}), the coefficients $A_{12}, B_{12}, C_{12}, D_{12}$ and $E_{12}$ read
\begin{eqnarray}
A_{12} &=& A_1-A_2 ,\label{relata12}\\
B_{12} &=& B_1-B_2 ,\label{relatb12}\\
C_{12} &=& C_1-C_2-2(A_1-A_2)A_2 ,\label{relatc12}\\
D_{12} &=& D_1-D_2-2(A_1 B_2+A_2 B_1)+4 A_2 B_2 ,\label{relatd12}\\
E_{12} &=& E_1-E_2 -2(B_1-B_2) B_2 , \label{relate12}
\end{eqnarray}

When taking the NNLO perturbative contributions into account to the physical observable, the LO BLM scale will have an NLO term. Following the standard BLM procedure, which will be shown in the following subsection, we obtain three LO BLM scales $Q^{LO}_1$, $Q^{LO}_2$ and $Q^{LO}_{12}$ from Eqs.(\ref{Alpha1MS},\ref{Alpha2MS},\ref{Alpha12}) :
\begin{eqnarray}
{Q^{LO}_1} &=& Q \, \exp \left[\frac{3 B_{1}}{2 T} + \frac{9\beta_0}{8 T^2}\left(B_{1}^2-E_{1} \right) \frac{\alpha_{\overline{MS}}(Q)}{\pi}+{\cal O}\left(\frac{\alpha^2_{\overline{MS}}}{\pi^2}\right)\right] \\
{Q^{LO}_2} &=& Q \, \exp \left[\frac{3 B_{2}}{2 T} + \frac{9\beta_0}{8 T^2} \left(B_{2}^2-E_{2} \right) \frac{\alpha_{\overline{MS}}(Q)} {\pi}+{\cal O}\left(\frac{\alpha^2_{\overline{MS}}}{\pi^2}\right) \right] \\
{Q^{LO}_{12}} &=&Q \, \exp \left[\frac{3 B_{12}}{2 T} + \frac{9\beta_0}{8 T^2} \left(B_{12}^2-E_{12} \right) \frac{\alpha_{\overline{MS}}(Q)} {\pi} +{\cal O}\left(\frac{\alpha^2_{\overline{MS}}}{\pi^2}\right) \right] ,
\end{eqnarray}
where in QCD, $C_A=3$, $T=1/2$, $C_F=4/3$. Setting $\lambda_{A/V}=Q^{LO}_1 /Q$, $\lambda_{B/V}=Q^{LO}_2 /Q$ and $\lambda_{A/B}=Q^{LO}_{12} /Q$, we obtain the required scale relation :
\begin{displaymath}
\lambda_{A/B}=\lambda_{A/V}/\lambda_{B/V}=\lambda_{A/V}\lambda_{V/B},
\end{displaymath}
where we have used the above relations (\ref{relata12}-\ref{relate12}) for $B_{12}$ and $E_{12}$. Here, the second step is due to the symmetric relation $\lambda_{B/V}\lambda_{V/B}=1$. This finishes our demonstration for the first point.

\subsubsection{\it An Analytic Extension of $\overline{MS}$-Scheme}

As has been discussed in Sec.\ref{secII}, in conventional $\overline{MS}$-scheme, the $\{\beta_i\}$-functions depend on the active number of ``massless'' quarks $(n_{f})$, which is usually a step function of the renormalization scale $\mu_r$; i.e., the quark masses do not enter into the $\{\beta_i\}$-functions since the running coupling is mass independent due to the decoupling theorem~\cite{decouple}.

An important property of BLM scale setting, is that the active number of flavors $n_f$ which goes into the $\beta$-function can be correctly determined by including the quark mass effect. As has been argued by Refs.~\cite{apt1,apt2,apt3,sjbnf,nf2,nf3,nf4}, there are a number of reasons to construct an analytic extension of the coupling under $\overline{MS}$-scheme, such as :
\begin{itemize}
\item The comparison of $\alpha_s$ determined from different experiments and at different momentum scales is an essential test of QCD. One source of error is the neglect of quark masses and in the subsequent running of $\alpha_s$ from the conventional reference scale, the $Z$-boson mass.

\item Lattice calculation for the heavy quarkonium spectra provides a most precise determination of $\alpha_s$ at low momentum scales, cf.Refs.~\cite{latas1,latas2,latas3,latas4,latas5}. It is important to know how finite quark mass effects enter into the $\alpha_s$ running to lower and higher energy scales with as small error as possible.

\item Finite mass threshold effects in supersymmetric grand unified theories are important when analyzing the running and unification of couplings over very large ranges, which has been discussed in Refs.~\cite{supas1,supas2,supas3}.
\end{itemize}

An analytic extension of the coupling under the $\overline{MS}$-scheme which incorporates the finite-mass quark threshold effects has been suggested in Ref.~\cite{sjbnf}, which is called as $\widetilde{\overline{MS}}$-scheme. Such an extension is obtained by connecting the coupling under the conventional $\overline{MS}$-scheme to the analytic and physically-defined $V$-scheme through the CSR derived from BLM; i.e.
\begin{equation}
\widetilde{\alpha}_{\overline{MS}}(Q) =\alpha_V(Q^*) + \frac{2N_C}{3} {\alpha_V^2(Q^{**})\over\pi} + \cdots , \label{alpmsbar2}
\end{equation}
where the commensurate scales $Q^*$ and $Q^{**}$ are given by~\cite{sjbnf}
\begin{eqnarray}
Q^* & = & Q\exp\left[\frac{5}{6}+{\cal O}\left(\frac{\alpha_V} {\pi}\right)\right] , \\
Q^{**} & = & Q\exp\left[\left(\frac{105}{128}-\frac{9}{8}\zeta_3\right)\frac{C_F}{N_C}
+\left(\frac{103}{192}+\frac{21}{16}\zeta_3\right) +{\cal O}\left(\frac{\alpha_V} {\pi}\right)\right] .
\end{eqnarray}
Here for the $V$-scheme, we mean that its effective coupling $\alpha_V(Q)$ is defined from the heavy quark potential $V(Q^2)=-4\pi\alpha_{V}(Q)/Q^2$. There is no renormalization scale ambiguity in $\alpha_V$-scheme, since $Q^2=-t$, the photon/gluon virtuality, is the optimized scale in the {\rm GM-L} scheme which automatically sums up all vacuum polarization contribution into the coupling (at high order $\alpha_V$ is infrared sensitive, so it is difficult to adopt as a standard QCD running coupling). Through a proper way, the $V$-scheme automatically includes the effects of finite quark masses in the same manner that lepton mass appear in Abelian QED~\cite{sjbnf}. So the extended $\widetilde{\overline{MS}}$-scheme can also include the quark masses by relating the $\overline{MS}$-scheme to $V$-scheme through CSR. Note in deriving the LO scale $Q^*$, the $\alpha_V$ correction to $Q^*$ is of less importance for our present analysis, so we do not write it out.

\begin{figure}[tb]
\begin{center}
\begin{minipage}[t]{8 cm}
\epsfig{file=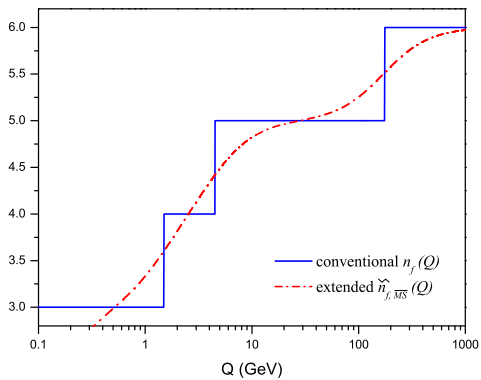,scale=0.7}
\end{minipage}
\begin{minipage}[t]{16.5 cm}
\caption{The continuous $\widetilde {n}_{f,\overline{MS}}(Q)$ in the analytic extension of the $\overline{MS}$-scheme as a function of the physical scale $Q$~\cite{sjbnf}. For reference the continuous $n_f$ is also compared with the conventional procedure of taking $n_f$ to be a step-function at the quark-mass thresholds. \label{fig:sjbnf}}
\end{minipage}
\end{center}
\end{figure}

Taking the logarithmic derivative of CSR given by Eq.~(\ref{alpmsbar2}) with respect to $\ln Q$, we can obtain the $\{\beta^{\overline{MS}}\}$-like function, which at LO level gives the following relation between the continuous $\widetilde{n}_{f,\overline{MS}}$ and $n_{f,V}$ for various quarks :
\begin{equation}
\widetilde{n}_{f,\overline{MS}}(Q) =n_{f,V}(Q^*) \simeq\left(1+\frac{5} {\rho_i\exp\left(\frac{5}{3}\right)} \right)^{-1},
\end{equation}
where $\rho_i=Q^2/m_i^2$ [for reference, the quark masses (in GeV) we used are $m_u=0.004$, $m_d=0.008$, $m_s=0.200$, $m_c=1.5$, $m_b=4.5$ and $m_t=175$ ]. Adding all flavors together gives the total $\widetilde {n}_{f,\overline{MS}}(Q)$ which is shown in Fig.(\ref{fig:sjbnf}). For reference, the continuous $n_f$ is also compared with the conventional procedure of taking $n_f$ to be a step-function at the quark-mass thresholds. The figure shows clearly that there are hardly any plateaus at all for the continuous $\widetilde {n}_{f,\overline{MS}}(Q)$ in between the quark masses. Thus, there is really no scale below 1 TeV where $\widetilde{n}_{f,\overline{MS}}(Q)$ can be approximated by a constant. We also note that if one would use any other scale than the BLM-scale for $\widetilde{n}_{f,\overline{MS}}(Q)$, the result would be to increase the difference between the analytic $n_f$ and the standard procedure of using the step-function at the quark-mass thresholds.

Numerically, it is found that taking finite quark mass effects analytically into account in the running, rather than using a fixed $n_f$ between thresholds, leads to effects of the order of one percent for the one-loop running coupling, with the largest differences occurring near thresholds~\cite{sjbnf}. These small differences are somewhat important for observables that are calculated by neglecting quark masses and could in principle turn out to be significant in comparing low and high energy measurements of the strong coupling. Moreover, the advantage of the modified scheme $\widetilde{\alpha}_{\overline{MS}}$-scheme is that it provides an analytic interpolation of conventional dimensional regularization expressions by utilizing the mass dependence of the physical $\alpha_V$-scheme. In effect, quark thresholds are treated analytically to all orders in $m^2_i/Q^2$; i.e., the evolution of our analytically extended coupling in the intermediate regions reflects the actual mass dependence of a physical effective charge and the analytic properties of particle production in a physical process. Just as in Abelian QED, the mass dependence of the effective potential and the analytically-extended $\widetilde{\overline{MS}}$-scheme reflects the analyticity of the physical thresholds for particle production in the crossed channel. Furthermore, the definiteness of the dependence in the quark masses automatically constrains the renormalization scale. Alternatively, one could connect $\widetilde{\overline{MS}}$-scheme to another physical coupling such as $\alpha_R$ defined from $e^+e^-$ annihilation.

By utilizing the BLM scale setting, based on the massless $n_f$ contribution, the analytic extension of the $\overline{MS}$-scheme correctly absorbs both massless and mass dependent quark contributions from QCD diagrams, such as the double bubble diagram, into the running coupling. This gives the opportunity to convert a calculation made in the $\overline{MS}$-scheme with massless quarks into an expression which includes quark mass corrections from QCD diagrams. In addition one can use this procedure to analytically discriminate the dependence of the coupling on time-like and space-like arguments.

\subsubsection{\it BLM Scale-Setting up to Four-Loop Level} \label{sec:IV:blm}

Based on the main idea of BLM, the method can be extended to higher orders in a systematic way, only one should be careful of how to deal with the $n_f$-series at each perturbative order consistently. Practically, in doing the BLM extension, the following points must be respected; i.e.,
\begin{itemize}
\item All $n_f$-terms, which are associated with the $\beta$-function in the renormalization of the coupling constant, must be absorbed into the coupling, while those $n_f$-terms that have no relation to the $\beta$-function should be identified and kept separate. After BLM scale setting, the perturbative series for the physical observable becomes a conformal series, all non-conformal terms should be absorbed into the effective coupling in a consistent manner.

\item There are always new $n_f$-terms (corresponding to new $\{\beta_i\}$-terms) emerging at each perturbative order, so we should introduce new BLM scales at each perturbative order so as to absorb all $n_f$-terms into the coupling consistently. There is no strong reason to use a unified effective scale for all perturbative orders. In fact, as has already been pointed out in Ref.~\cite{blma1} that if taking only one effective BLM scale for the whole perturbative series, one can not fix all unknown parameters uniquely even at the NLO level, since one does not have enough constraints to achieve the goal.

\item The BLM scales themselves should be a RG-improved perturbative series~\cite{blm,pmc2,pmc3,pmc5,scale1,blma1}. The length of the perturbative series for each BLM scale depends on how many new $n_f$-terms (or $\{\beta_i\}$-terms) we have from the higher-order calculation and to what perturbative order we have performed.
\end{itemize}

As a combination of these points, it is interesting to find that the BLM scale setting leads to the correct expansion coefficients in the ``conformal limit". This inversely shows that the later developed PMC scale setting is more essential, since, as will be shown in the following sections, by dealing with the perturbative series according to $\{\beta_i\}$-terms, the results are optimized and unique, which also provide an unambiguous principle to set the BLM scale up to all perturbative orders.

In the literature, several ways for extending the BLM method beyond the NLO have been suggested, such as the dressed skeleton expansion, the large $\beta_0$-expansion, the BLM expansion with an overall renormalization scale, the sequential BLM (seBLM), an extension over the sequential BLM (xBLM) and etc.~\cite{blma2,blma3,blma4,Mikhailov1,blma1,Mikhailov2,dse1,dse2,blma5,blma6,blma7,blma8}. However it can be found that the purpose of most of these references is just to eliminate the $n_f$-terms (mostly by introducing an overall effective BLM scale for all perturbative orders), but they do not respect all the above listed points simultaneously, so even though they do make some improvements in understanding BLM, the criticism of BLM made in some of these references are incorrect due to improper understanding/use of BLM. Especially, they can not obtain the most important BLM feature that the BLM prediction should be independent of the choice of initial renormalization scale. If a method to extend the BLM setting up to all orders still depends on the choice of initial renormalization scale, then one will still have a (transferred) renormalization scale uncertainty, which is, in principle, not an essential improvement of the conventional scale setting method.

On the other hand, the correct and unambiguous way to the BLM scales up to the two-loop level has been suggested by Brodsky and Lu in 1995~\cite{scale1}, and recently such way has been improved up to four-loop level~\cite{pmc2}. These two references, especially the second reference, can be used as a useful guidance for setting BLM scales up to any perturbative order. To be a useful reference and to clarify some misunderstandings of BLM in the literature, we present the technical details in the following.

\begin{figure}[tb]
\begin{center}
\begin{minipage}[t]{9 cm}
\epsfig{file=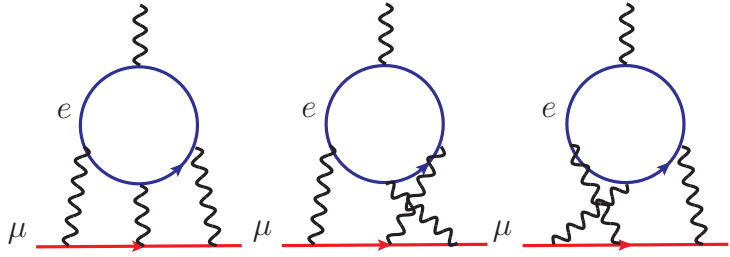,scale=0.6}
\end{minipage}
\begin{minipage}[t]{16.5 cm}
\caption{The finite electron-loop light-by-light diagrams contributing to the muon's anomalous magnetic moment~\cite{extraloop}. Three more are obtained by reversing the direction of the electron loop. \label{nonUVloop}}
\end{minipage}
\end{center}
\end{figure}

Generally, separating the $n_f$-terms explicitly at each perturbative order, the pQCD prediction for a physical observable $\rho$ can be rewritten as
\begin{eqnarray}
\rho &=& r_0 \left[ a^n_s(Q) +(A_{1}+A_{2} n_f) a^{n+1}_s(Q) + (B_{1} +B_{2}n_f + B_{3}n_f^2 ) a^{n+2}_s(Q) + \right. \nonumber\\
&& \quad\left. (C_{1} +C_{2}n_f + C_{3}n_f^2 + C_{4} n_f^3 ) a^{n+3}_s(Q)+ \cdots \right ] \label{eq_blm}
\end{eqnarray}
where $a_s(Q)={\alpha_s(Q)}/{\pi}$ and the overall tree-level parameter $r_0$ is scale-independent and is free of $a_s(Q)$. Here, $Q$ stands for the initial renormalization scale, $n (\geq1)$ stands for the initial $\alpha_s$-order at the tree level. After BLM scale setting, all $n_f$-terms in the perturbative expansion can be summed into the running coupling. Here, we shall concentrate on those processes in which all $n_f$-terms are associated with the $\{\beta_i\}$-terms. There are $n_f$-terms coming from the Feynman diagrams with the light-by-light quark loops which are irrelevant to the ultra-violet cutoff in higher-order processes. However, we should remind that there may still be large higher-order corrections not associated with renormalization; an important example in QED case is the electron-loop light-by-light contribution to the sixth-order muon anomalous moment, its Feynman diagrams are shown by Fig.(\ref{nonUVloop}), which is of order $(\alpha/\pi)^3\ln(m_{\mu}/m_{e})$ and is sizable~\cite{extraloop}.

The BLM scales for the pQCD prediction of $\rho$ can be determined in a general scheme-independent way. The generalization of the BLM procedure to higher order assigns a different renormalization scale for each order in the perturbative series, which can be fixed order-by-order. We can shift the initial renormalization scale $Q$ into effective ones until we fully absorb those higher-order terms with $n_f$-dependence into the running coupling. According to the following steps, one can set the BLM scales up to NNLO :
\begin{itemize}
\item The first step is to set the effective scale $Q^{*}$ at LO
\begin{equation}
\rho = r_0\Big[ a^n_s(Q^*) + \widetilde{A}_{1} a^{n+1}_s(Q^*) + (\widetilde{B}_{1} + \widetilde{B}_{2}n_f) a^{n+2}_s(Q^*) + (\widetilde{C}_{1} +\widetilde{C}_{2}n_f +\widetilde{C}_{3}n_f^2) a^{n+3}_s(Q^*)+\cdots \Big] . \label{first}
\end{equation}
\item The second step is to set the effective scale $Q^{**}$ at NLO
\begin{eqnarray}
\rho &=& r_0\Big[ a^n_s(Q^*) + \widetilde{A}_{1} a^{n+1}_s(Q^{**}) + \widetilde{\widetilde{B}}_{1} a^{n+2}_s(Q^{**}) + (\widetilde{\widetilde{C}}_{1} +\widetilde{\widetilde{C}}_{2}n_f) a^{n+3}_s(Q^{**})+\cdots \Big]\ , \label{second}
\end{eqnarray}
\item The final step is to set the effective scale $Q^{***}$ at NNLO
\begin{eqnarray}
\rho &^=& r_0\Big[ a^n_s(Q^*) + \widetilde{A}_{1} a^{n+1}_s(Q^{**}) + \widetilde{\widetilde{B}}_{1} a^{n+2}_s(Q^{***}) +\widetilde{\widetilde{\widetilde{C}}}_{1} a^{n+3}_s(Q^{***})+\cdots \Big]\ . \label{third}
\end{eqnarray}
\end{itemize}

When performing the scale shifts $Q\to Q^*$, $Q^* \to Q^{**}$ and $Q^{**}\to Q^{***}$, we eliminate the $n_f$-terms associated with the $\{\beta_i\}$-terms completely. At the same time, we also have to modify the coefficients, since the net changes to the coefficients are proportional to $\{\beta_i\}$-functions. Note that to set the effective scale for $a^{n+3}_s$, one needs even higher order information and here, as has been discussed previously, a sensible choice is $Q^{***}$, since this is the renormalization scale after shifting the scales up to NNLO. The effective renormalization scales up to NNLO can be written as
\begin{eqnarray}
\ln\frac{Q^{*2}}{Q^2} &=& \ln \frac{Q^{*2}_0}{Q^2} + \frac{x \beta_0}{4} \ln \frac{Q^{*2}_0}{Q^2} a_s(Q) +\frac{y}{16}\left(\beta^2_0 \ln^2 \frac{Q^{*2}_0}{Q^2} -\beta_1 \ln\frac{Q^{*2}_0}{Q^2}\right)a^2_s(Q) + {\cal O}(a^3_s)\\
\ln\frac{Q^{**2}}{Q^{*2}} &=& \ln\frac{Q^{**2}_0}{Q^{*2}} + \frac{z \beta_0}{4} \ln \frac{Q^{**2}_0}{Q^{*2}} a_s(Q)+ {\cal O}(a^2_s) \\
\ln\frac{Q^{***2}}{Q^{**2}} &=& \ln \frac{Q^{***2}_0}{Q^{**2}}+ {\cal O}(a_s)
\end{eqnarray}
where the effective scales $Q_0^{*,**,***}$ are determined so as to eliminate $A_2 n_f$, $\widetilde{B}_2 n_f$ and $\widetilde{\widetilde{C}}_2 n_f$-terms completely, the parameters $x$ and $z$ are used to eliminate the $B_3 n_f^2$ and the $\widetilde{C}_3 n_f^2$ terms respectively, and the parameter $y$ is used to eliminate the $C_4 n_f^3$-term. It is found that
\begin{eqnarray}
\ln \frac{Q^{*2}_0}{Q^2} &=& \frac{6A_2}{n}\;,\;\;
\ln \frac{Q_0^{**2}}{Q^{*2}} = \frac{6\widetilde{B}_2}{(n+1)\widetilde{A}_1} \;,\;\;
\ln \frac{Q_0^{***2}}{Q^{**2}} = \frac{6\widetilde{\widetilde{C}}_2}{(n+2)\widetilde{\widetilde{B}}_1}
\end{eqnarray}
and
\begin{eqnarray}
x &=& \frac{3(n+1)A_2^2 -6 n B_3}{n A_2} \\
y &=& \frac{(n+1)(2n+1)A_2^3 -6n(n+1)A_2 B_3 +6n^2 C_4}{n A^2_2} \\
z &=& \frac{3(n+2)\widetilde{B}_2^2 -6(n+1)\widetilde{A}_1 \widetilde{C}_3}{(n+1) \widetilde{A}_1 \widetilde{B_2}}
\end{eqnarray}
If $x=0$, or $y=0$, or $z=0$, this shows that there is no new $\{\beta_i\}$-terms that will change the value of the effective BLM scale. The exponential form shows that after BLM scale setting, it will not change the properties of the initial choice of scale; i.e. its space-like and time-like nature will not be changed.

The step-by-step coefficients are listed in the following
\begin{eqnarray}
\widetilde{A}_1 &=& A_1 +\frac{33}{2}A_2 \;,\; \widetilde{\widetilde{B}}_1 = \widetilde{B}_1 +\frac{33}{2}\widetilde{B}_2 \;,\; \widetilde{\widetilde{\widetilde{C}}}_1 = \widetilde{\widetilde{C}}_1 +\frac{33}{2}\widetilde{\widetilde{C}}_2\\
\widetilde{B}_1 &=& \frac{1}{4n}\Bigg[1089(n+1)A_2^2 +153n A_2 +66(n+1)A_1 A_2+(4B_1 -1089 B_3)n\Bigg] \\
\widetilde{B}_2 &=&\frac{-1}{4n}\Bigg[66(n+1)A_2^2 +19n A_2 +4(n+1)A_1 A_2 -4n(B_2 +33B_3)\Bigg] \\
\widetilde{C}_1 &=& \frac{1}{64{A_2}n^2} \Bigg[-40392{C_4}n^3 + 143748{{A_2}}^4(3+5n+2n^2) + 8{A_2}n^2 ( 8{C_1} + 35937{C_4} + \nonumber\\
&& 5049{B_3}n ) -13464{{A_2}}^3n( n^2-3n -7)+72{A_1}{A_2}(1+n)(34{A_2}n - 242{B_3}n +\nonumber\\
&& 121{{A_2}}^2( 3 + 2n )) + 3{{A_2}}^2n( 2857n + 352{B_1}( 2 + n ) - 95832{B_3}( 3 + 2n ) ) \Bigg]\\
\widetilde{C}_2 &=& \frac{1}{192 {A_2}n^2}\Bigg[22392{C_4}n^3 - 52272{{A_2}}^4(3+5n+2n^2 ) - 24{A_2}n^2 ( -8{C_2} + \nonumber\\
&& 6534{C_4} + 933{B_3}n ) - 48{A_1}{A_2}( 1 + n ) ( 19{A_2}n - 132{B_3}n + 66{{A_2}}^2( 3 + 2n ) ) + \nonumber\\
&& {A_2}^2n( -5033n - 192{B_1}( 2 + n ) + 3168{B_2}( 2 + n ) + 52272{B_3}( 8 + 5n ))+ \nonumber\\
&& 24{{A_2}}^3n( -1871 + n( -627 + 311n ) ) \Bigg]\\
\widetilde{C}_3 &=& \frac{1}{576{A_2}n^2}\Bigg[ -2736{C_4}n^3 + 4752{{A_2}}^4(3+5n+2n^2) + 144{A_2}n^2  ( 4{C_3} + 198{C_4} + \nonumber\\
&& 19{B_3}n ) - 912{{A_2}}^3(n^3 -4n)+288{A_1}{A_2}(1+n)(-2{B_3}n + {{A_2}}^2(3 + 2n )) \nonumber\\
&& -{{A_2}}^2n( -325n + 576{B_2}( 2 + n ) + 9504{B_3}( 5 + 3n ) )\Bigg] \\
\widetilde{\widetilde{C}}_1 &=& \frac{1}{4(n+1)\widetilde{A}_1}\Bigg[33(n+2)\widetilde{B_2}(2\widetilde{B_1} +33\widetilde{B_2}) +(n+1)(153\widetilde{B_2}+4\widetilde{C_1}-1089\widetilde{C_3})\widetilde{A_1} \Bigg]\\
\widetilde{\widetilde{C}}_2 &=&\frac{-1}{4(n+1)\widetilde{A}_1}\Bigg[2(n+2)\widetilde{B_2}(2\widetilde{B_1} +33\widetilde{B_2}) +(n+1)(19\widetilde{B_2}-4(\widetilde{C_2}+33\widetilde{C_3}))\widetilde{A_1} \Bigg]
\end{eqnarray}
One may observe that even though the perturbative coefficients at the final step $\widetilde{A}_1$, $\widetilde{\widetilde{B}}_1$ and $\widetilde{\widetilde{\widetilde{C}}}_1$ are different, the formulas in deriving them are similar, which shows clearly that the BLM scale setting can be done in a recursive way. In deriving the above NNLO formulae, the following equation is implicitly adopted, i.e. the value of $a_s$ at any scale $Q^*$ can be obtained from its value at the scale $Q$,
\begin{equation}\label{alphaexp}
a_{s}(Q^*) = a_{s}(Q)-\frac{\beta_{0}}{4} \ln\left(\frac{Q^{*2}}{Q^2}\right) a^2_{s}(Q) +\frac{1}{4^2} \left[\beta^2_{0}\ln^2\left(\frac{Q^{*2}}{Q^2}\right) - \beta_{1} \ln\left(\frac{Q^{*2}}{Q^2}\right) \right] a^3_{s}(Q) +{\cal O}(a^4_{s}).
\end{equation}
For even higher-order corrections, we should use the $\alpha_s$ running behavior derived from Eq.(\ref{alphasbeta}) to do the calculation, since it shows which $\{\beta_i\}$-terms should be kept in the $\alpha_s$-expansion series.

All perturbative coefficients $A_{i}$, $B_{i}$, $C_{i}$ and etc. are renormalization-scheme dependent, so different renormalization schemes lead to different BLM scales $Q^{*,**,***}$; however the final result for $\rho$ should be scheme independent due to CSRs among different observables. Calculating the observable $\rho$ by its corresponding effective coupling and changing $a_s$ to be another effective coupling, starting from Eq.(\ref{eq_blm}) and following the same procedures, one can naturally obtain the CSRs up to NNLO. Moreover, by using the relations between $Q^{*,**,***}$ and $Q$, one can find the needed scale displacement among the effective scales which are derived under different schemes or conventions so as to ensure the scheme-independence of the observables. For example, from the relation between $Q^{*}$ and $Q$, one can easily obtain the well-known one-loop relation for the coupling \cite{blm}, $\alpha_s^{\overline{MS}}(e^{-5/3}Q^2) =\alpha_s^{GM-L}(Q^2)$.

\subsubsection{\it Example of BLM scale setting for $R_{e^{+}e^{-}}(Q)$ at the Four Loop Level} \label{anni}

\begin{table}[htb]
\begin{center}
\begin{minipage}[t]{16.5 cm}
\caption{Coefficients for the perturbative expansion of $R_{e^{+}e^{-}}(Q)$ before and after BLM scale setting. }
\label{tab:ee}
\end{minipage}
\begin{tabular}{|c||c||c||c|}
\hline
 ~~ ~~ & ~~$n_f =3$~~ & ~~$n_f =4$~~ & ~~$n_f =5$~~ \\
\hline\hline
$A$ & 1.6401 & 1.5249 & 1.4097 \\
$B$ & -10.2840 & -11.6857 & -12.8047 \\
$C$ & -106.8960 & -92.9124 +$2\kappa/15$ & -80.0075 +$\kappa/33$ \\
\hline\hline
$\widetilde{A}$ & 0.0849 & 0.0849 & 0.0849 \\
$\widetilde{\widetilde{B}}$ & -23.2269 & -23.3923 & -23.2645 \\
$\widetilde{\widetilde{\widetilde{C}}}$ & 82.3440 & 82.3440 +$2\kappa/15$ & 82.3440+$\kappa/33$ \\
\hline
\end{tabular}
\end{center}
\end{table}

Measurements of the cross sections for electro-positron annihilation into hadrons provides one of the most precise determination of $\alpha_s$. The explicit expression for $R_{e^{+}e^{-}}(Q)$ up to $\alpha_s^4$-order under the $\overline{MS}$-scheme can be found in Refs.~\cite{Ralphas1,Ralphas2,Ralphas3}. One finds
\begin{equation}
R_{e^{+}e^{-}}(Q) = 3\sum_q e_q^2 \Bigg[ 1 + \left(a^{\overline{MS}}(Q)\right) + A \left(a^{\overline{MS}}(Q)\right)^2 + B \left(a^{\overline{MS}}(Q)\right)^3 + C \left(a^{\overline{MS}}(Q)\right)^4 \Bigg] ,
\end{equation}
where
\begin{eqnarray}
A &=& 1.9857 - 0.1152 n_f , \nonumber\\
B &=& -6.63694 - 1.20013 n_f - 0.00518 n_f^2 -1.240 \eta , \nonumber\\
C &=& -156.61 + 18.77 n_f - 0.7974 n_f^2 + 0.0215 n_f^3 +\kappa \eta . \nonumber
\end{eqnarray}
Here $\eta={\left(\sum_q e_q\right)^2}/{\left(3\sum_q e_q^2\right)}$, $e_q$ is the electric charge for the active flavors. The coefficient $\kappa$ is yet to be determined, whose contribution will be further suppressed by $\eta$, so we set its value to zero in the following numerical calculation \footnote{ Recently, $\kappa$ has been calculated by Refs.~\cite{Ralphas4,Ralphas5}, which will slightly affect our estimations and its UV-finite $n_f$-dependent terms will not affect our BLM treatment.}. The values of $A$, $B$ and $C$ for $n_f=3$, $4$ and $5$ are presented in Table~\ref{tab:ee}. At the present perturbative order, all $n_f$-terms in the above equation should be absorbed into $\alpha_s$. After applying BLM up to NNLO, we obtain
\begin{equation}
R_{e^{+}e^{-}}(Q) = 3\sum_q e_q^2 \Bigg[ 1 + \left(a^{\overline{MS}}_s(Q^*)\right) + \widetilde{A} \left(a^{\overline{MS}}_s(Q^{**})\right)^2 + \widetilde{\widetilde{B}} \left(a^{\overline{MS}}_s(Q^{***})\right)^3 +\widetilde{\widetilde{\widetilde{C}}} \left(a^{\overline{MS}}_s(Q^{***})\right)^4 \Bigg] ,
\end{equation}
where all the coefficients and effective scales can be calculated with the help of the formulae listed in the Sec.\ref{sec:IV:blm}. The coefficients are presented in Table \ref{tab:ee}, slight differences for $\widetilde{\widetilde{B}}$ and $\widetilde{\widetilde{\widetilde{C}}}$ with varying $n_f$ are caused by the charge-dependent parameter $\eta$.

From the experimental value, $r_{e^+ e^-}(31.6GeV)=\frac{3}{11} R_{e^+ e^-}(31.6GeV)=1.0527\pm0.0050$ \cite{rexp}, we obtain
\begin{equation}
\Lambda^{'tH-\overline{MS}}_{QCD} = 412^{+206}_{-161} {\rm MeV} \;,\;\;
\Lambda^{\overline{MS}}_{QCD}  = 359^{+181}_{-140} {\rm MeV} .
\end{equation}
With the help of the four-loop coupling (\ref{alphas}), we obtain $\alpha^{\overline{MS}}_s(M_Z)=0.129^{+0.009}_{-0.010}$. It is consistent with the values obtained from $e^+ e^-$ collider; i.e., $\alpha^{\overline{MS}}_s(M_Z)=0.13\pm 0.005\pm0.03$ by the CLEO Collaboration \cite{cleo} and $\alpha^{\overline{MS}}_s(M_Z)=0.1224\pm 0.0039$ from the jet shape analysis \cite{jet}. One may observe that a smaller central value of the world average for $\alpha^{\overline{MS}}_s(M_Z)$ also results from the measurements of $\tau$-decays, $\Upsilon$-decays, the jet production in the deep-inelastic-scattering processes, and from heavy quarkonia based on the unquenched QCD lattice calculations \cite{aveas}. A larger $\Lambda_{\overline{MS}}$ leads to a larger $\alpha^{\overline{MS}}_s(M_Z)$, and vice versa. For example, if we set $\alpha^{\overline{MS}}_s(M_Z)$ to the present world average, we obtain $$\Lambda^{'tH}_{\overline{MS}}|_{n_f=5}= 245^{+9}_{-10}\; {\rm MeV} \;,\;\; \Lambda_{\overline{MS}}|_{n_f=5}= 213^{+19}_{-8}\; {\rm MeV} .$$ It is found that after BLM scale setting, the perturbative expansion of $R_{e^+ e^-}(Q=31.6\;{\rm GeV})$ becomes more convergent. In particular, we find $Q^*=\left(0.757\pm0.008\right)Q$ which leads to $a^{\overline{MS}}_s(Q^*) /a^{\overline{MS}}_s(Q) =1.060\pm0.004$. This conclusion is consistent with that of Ref.\cite{blma1}.

\subsection{\it The Principle of Maximum Conformality: PMC Scale-Setting}

Since its invention in 1983, the BLM scale setting has achieved much success in dealing with high energy processes. As an extension of BLM scale setting, a program to deal with higher order $n_f$-terms associated with renormalization up to NNLO level has been proposed in Ref.~\cite{blma1}, which suggests that one can expand the effective scale itself as a perturbative series. Later on, an enhanced discussion of this suggestion up to NNLO level has been presented in Ref.~\cite{scale1}, where the $n_f^2$-term at the NNLO is first identified with $\beta^2_0$-term and then is absorbed into the running coupling \footnote{Strictly speaking, it has been observed that such $n_f^2$-term together with the $n_f$-term at the same order should be rearranged into a proper linear combination of $\beta_1$-term and $\beta_0^2$-term. The $\beta_0^2$-term is then absorbed into the LO BLM scale and the $\beta_1$-term is absorbed into the NLO BLM scale accordingly~\cite{pmc2}.}. However, BLM in its previous form is difficult to be applied to even higher order calculations; i.e. it is not clear how to deal with the $n_f$-term, the $n_f^2$-term, etc. in those higher order corrections in its original version~\cite{blm}.

The two most important question for extending the BLM scale setting consistently and unambiguously up to any perturbative order are: 1) how to deal with the $n_f$-series in the perturbative coefficients at each order in an unambiguous way, and what is the underlying principle? 2) how to set the perturbative series in the BLM scales themselves in a consistent order-by-order manner? It has been found that the Principle of Maximum Conformality (PMC) provides the foundations underlying BLM scale setting~\cite{pmc3,pmc1,pmc2,pmc5,pmc6,BMW,pmc7,pmc4}. It inherits all the favorable features of BLM, and it provides the answer for solving the above two points. Ref.~\cite{pmc1} gives one suggestion, where one can use a single global PMC scale at LO by proper weighting the separate scales for each skeleton graph, such as $t$-channel or $s$-channel, to deal with the pQCD cross-section. Later on, the PMC scales were shown to be related to the BLM scales in the NNLO analysis through a particular matching of the $n_f$-terms to $\{\beta_i\}$-terms, called the PMC-BLM correspondence principle \cite{pmc2}. Recently, taking PMC as a first principle a new scale setting method was presented, which provides a systematic all-orders method to set the effective scales and in a way, which can be readily automatized \cite{BMW}.

\subsubsection{\it Basic Arguments of PMC}

The purpose of the running coupling in any gauge theory is to sum all terms involving the $\beta^{\cal R}$-function. Here, we show the scheme dependence in the $\beta$-function explicitly, the superscript ${\cal R}$ stands for an arbitrary renormalization scheme. In fact, when the renormalization scales at each perturbative order are set properly within PMC, all non-conformal $\{\beta^{\cal R}_{i}\}\ne 0$ terms in a perturbative expansion arising from renormalization are summed into the running coupling. The remaining terms in perturbative series are then identical to that of a conformal theory; i.e., the theory with $\{\beta^{\cal R}_i\}\equiv0$. Through this treatment, the divergent ``renormalon" series of order $(\alpha_s^n (\beta^{\cal R}_i)^n n!)$ does not appear in the conformal series. Thus as in QED, the renormalization scales are determined unambiguously by PMC.

\begin{figure}[tb]
\begin{center}
\begin{minipage}[t]{8 cm}
\epsfig{file=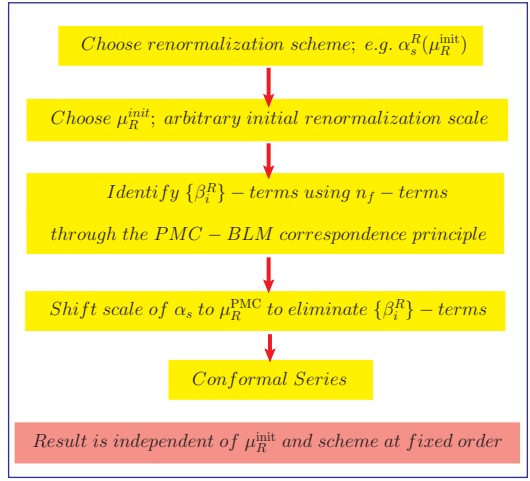,scale=0.5}
\end{minipage}
\begin{minipage}[t]{16.5 cm}
\caption{A ``flow chart" which illustrates the PMC procedure. \label{fig1}}
\end{minipage}
\end{center}
\end{figure}

A ``flow chart" which illustrates the PMC procedure is presented in Fig.(\ref{fig1}). The PMC provides an unambiguous and systematic way to set the optimized renormalization scale up to all orders. We first arrange all the coefficients, which are usually in $n_f$-power-series at each perturbative order, into $\{\beta^{\cal R}_i\}$-terms or non-$\{\beta^{\cal R}_i\}$-terms. Then, we absorb all $\{\beta^{\cal R}_i\}$-terms into the running coupling. Note that in practice we can directly deal with $n_f$-terms of the coefficients without changing them into $\{\beta^{\cal R}_i\}$-terms, and eliminate the $n_f$-terms from the highest power to none also in an order-by-order manner; the results are the same due to the PMC-BLM correspondence principle~\cite{pmc2}. Different types of $\{\beta^{\cal R}_i\}$-term will be absorbed into different PMC scales, and the PMC scales themselves will be a perturbative expansion series in $\alpha_s$. After these procedures, all non-conformal $\{\beta^{\cal R}_i\}$-terms in the perturbative expansion are summed into the running coupling so that the remaining terms in the perturbative series are identical to that of a conformal theory; i.e., the corresponding theory with $\{\beta^{\cal R}_i\} \equiv \{0\}$.

As a simple explanation of PMC, for the coefficient ${\cal C}_1(\mu_r)$ of the pQCD expansion (\ref{eq:blmstart}) at the NLO level, we have
\begin{eqnarray}
{\cal C}_1(\mu_r) &=& {\cal C}_{10}(\mu_r) + {\cal C}_{11}(\mu_r) n_f = \tilde{{\cal C}}_{10}(\mu_r) + \tilde{{\cal C}}_{11}(\mu_r) \beta_0
\end{eqnarray}
where $\mu_r$ stands for an arbitrary initial renormalization scale, the coefficients ${\cal C}_{10}(\mu_r)$ and ${\cal C}_{11}(\mu_r)$ are $n_f$-independent, $\tilde{{\cal C}}_{10}={\cal C}_{10}+\frac{33}{2}{\cal C}_{11}$, and $\tilde{{\cal C}}_{11}=-\frac{3}{2}{\cal C}_{11}$. The LO PMC scale $\mu^{\rm PMC,LO}_r$ is then set by the condition
\begin{equation}\label{pmcbasic}
\tilde{{\cal C}}_{11}(\mu^{\rm PMC,LO}_r) = 0.
\end{equation}
This prescription ensures that, as in QED, vacuum polarization contributions due to the light-fermion pairs are absorbed into the running coupling.

\subsubsection{\it PMC - BLM Correspondence Principle}

A procedure for setting PMC scale at LO has been suggested in Ref.~\cite{pmc1}, which is adaptable to any NLO calculations. Its idea is to distinguish the nonconformal terms from the conformal terms by the variation of the cross section with respect to $\ln(\mu^{\rm init}_r)^2$ ($\mu^{\rm init}_r$ stands for some initial renormalization scale). Given the analytic form of the hard process amplitude or cross section as a series in $\alpha_s(\mu^{\rm init}_r)$, one can identify the LO PMC scale through the following way:
\begin{enumerate}
\item There is only one type of $\beta^{\cal R}$-function, i.e. $\beta_0$, emerges at the NLO level. The variation of the cross section with respect to $\ln(\mu^{\rm init}_r)^2$ can be used to distinguish the conformal terms versus the nonconformal terms.
\item The identified nonconformal terms always have the form $\beta_0 \ln\left(p^2_{ij}/(\mu^{\rm init}_r)^2\right)$ where $p^2_{ij} = p_i \cdot p_j$ are scalar product invariants ($i \ne j$) which enter the hard subprocess. In practice, these terms can be identified as coefficients of $n_f$; i.e., the flavor dependence arising from the light-quark loops associated with coupling constant renormalization.
\item The scale is then shifted from $\mu^{\rm init}_r$ to $\mu^{\rm PMC}_r$ in order to eliminate the non-conformal terms in the new perturbative series. Thus, when the scale is correctly set, the coefficients of $\alpha_s(\mu^{\rm PMC}_r)$ become independent of $\beta_0$ and $\ln(\mu^{\rm PMC}_r)^2$. The series is then identical to that of the conformal theory where $\{\beta^{\cal R}_i\}=0$.
\end{enumerate}

At LO, there is only the $\beta_0$-term, and the nonconformal terms always have the form of $\beta_0 \ln(\mu^{\rm init}_r)^2$, so one can determine the nonconformal terms exactly following the above procedures. However, at higher orders, the $\ln(\mu^{\rm init}_r)^2$-terms usually appear in a power series as $\beta_0 \ln(\mu^{\rm init}_r)^2$, $\beta_1 \ln(\mu^{\rm init}_r)^2$, $\beta_0^2 \ln^2(\mu^{\rm init}_r)^2$, etc.. So this method is no longer adaptable to deal with the higher-order corrections, because the derivative with respect to a single $\ln(\mu^{\rm init}_r)^2$ cannot distinguish all the emerged $\{\beta^{\cal R}_i\}$-terms.

As noted above, the purpose of the running coupling in any gauge theory is to sum up all terms involving the $\beta^{\cal R}$-function, conversely, one can find all the needed $\{\beta^{\cal R}_i\}$-terms at any relevant order from the general expansion (\ref{alphaexp}). This fact should be respected in constructing the perturbative $\{\beta^{\cal R}_i\}$-series in both the PMC scales and the perturbative coefficients of a physical observable. Furthermore, using this fact and also the known relation between $\{\beta^{\cal R}_i\}$ and $n_f$, one can obtain the PMC scales from the BLM scale setting method. This is the PMC and BLM correspondence principle~\cite{pmc2}. Since $\{\beta^{\cal R}_i\}$ ($i\geq2$) are scheme-dependent, the PMC and BLM correspondence depends on the renormalization scheme beyond the two-loop level \footnote{It is noted that another suggestion to extend the BLM to all perturbative orders have been given in Refs.~\cite{Mikhailov1,Mikhailov2,Mikhailov3}, which is close but different than PMC. At each perturbative order additional $\{\beta^{\cal R}_i\}$-terms are considered. One then introduces new free parameters to make the correspondence between $n_f$-terms and $\{\beta^{\cal R}_i\}$-terms.}.

More explicitly, up to NNLO, the physical observable $\rho$ defined in Eq.(\ref{eq_blm}) can be re-expanded in $\{\beta^{\cal R}_i\}$-series as,
\begin{eqnarray}
\rho &=& r_0 \Big[a^n_s(Q) +(A^0_{1}+ A^0_{2} \beta_0) a^{n+1}_s(Q) +(B^0_{1}+ B^0_{2} \beta_1 + B^0_{3} \beta_0^2) a^{n+2}_s(Q) \nonumber\\
&& \qquad +(C^0_{1}+ C^0_{2} \beta^{\cal R}_2 + C^0_{3} \beta_0 \beta_1 + C^0_{4}\beta_0^3) a^{n+3}_s(Q)\Big] . \label{eq_pmc}
\end{eqnarray}
The results for PMC can be naturally obtained from the BLM scale setting through the following unique parameter correspondence; i.e.,
\begin{eqnarray}
A_{1} &=& A^0_{1}+11 A^0_{2} \label{BLMPMC}\\
A_{2} &=& -\frac{2}{3}A^0_{2} \\
B_{1} &=& B^0_{1}+102 B^0_{2}+121 B^0_{3} \\
B_{2} &=& -\frac{2}{3}(19B^0_{2}+22B^0_{3}) \\
B_{3} &=& \frac{4}{9}B^0_{3} \\
C_{1} &=& C^0_{1}+\frac{2857}{2}C^0_{2}+1122C^0_{3}+1331C^0_{4} \\
C_{2} &=& -\frac{1}{18}(5033C^0_{2}-3732C^0_{3}-4356C^0_{4}) \\
C_{3} &=& \frac{1}{54}(325C^0_{2}+456C^0_{3}+792C^0_{4}) \\
C_{4} &=& -\frac{8}{27}C^0_{4}
\end{eqnarray}
which are obtained by comparing the Eqs.(\ref{eq_blm},\ref{eq_pmc}) and the four-loop $\{\beta^{\cal R}_i\}$-terms under the $\overline{MS}$ scheme (${\cal R}=\overline{MS}$), whose expressions have been given in Eqs.(\ref{beta00},\ref{beta01},\ref{beta02},\ref{beta03}).

\subsubsection{\it A Systematic All-Orders Method for PMC Scale-Setting}

Recently another way, i.e. a systematic all-orders method, to set the PMC scales has been suggested in Ref.~\cite{BMW}. In comparison to the method of using PMC - BLM correspondence principle, this new method has the advantage that the explicit $\alpha_s$-expansion of the effective scales is avoided and the scheme independence can be derived without introducing the commensurate scale relation. Moreover, the new scale setting processes is easier to automatize.

The starting point for this method is to introduce a generalization of the conventional schemes used in dimensional regularization, where logarithmically divergent integrals are regularized by the following transformation of the integration measure:
\begin{equation}
\label{integral}
\int {\rm d}^4 p \to \mu^{2\epsilon} \int {\rm d}^{4-2\epsilon}p \ ,
\end{equation}
where $\mu$ is an \emph{arbitrary} mass scale. Divergences are then separated as $1/\epsilon$ poles, which can be absorbed into redefinitions of the couplings. The choice of subtraction defines the \emph{renormalization scheme} and can be chosen at the theorist's convenience. The arbitrary mass scale becomes the \emph{initial renormalization scale} $\mu^{\rm init}_r$ of the running couplings constants. In the minimal subtraction ($MS$) scheme one absorbs the $1 / \epsilon$ poles appearing in loop integrals which come in powers of
\begin{displaymath}
\ln \frac{\left(\mu^{\rm init}_{r}\right)^2}{Q^2} + \frac{1}{\epsilon} + c \ ,
\end{displaymath}
where $Q$ is some typical scale and $c$ is the finite part of the integral. The widely used $\overline{ MS}$-scheme differs from the $MS$-scheme by an additional subtraction of the term $\ln(4 \pi) - \gamma_E$ together with the $1/\epsilon$ pole.

One can generalize this by subtracting a constant $-\delta$ in addition to the standard subtraction $(\ln 4 \pi - \gamma_E)$ of the $\overline{MS}$-scheme. This amounts to redefining the renormalization scale by an exponential factor; i.e. $\left(\mu^{\rm init}_r\right)^2 = \left(\mu^{\rm init}_{r}\right)^2\exp(\ln 4\pi - \gamma_E-\delta)$. In particular, the MS-scheme is recovered for $\delta = \ln 4 \pi - \gamma_E$. Another particular $\cal R_\delta$-scheme suggested in the literature is the $G$-scheme~\cite{kataev-beta,gscheme}, which is obtained for $\delta = - 2$. The $\delta$-subtraction defines an infinite set of renormalization schemes called \mbox{$\delta$-$\cal R$enormalization (${\cal R}_\delta$)} schemes; i.e.
\begin{eqnarray}
{\cal R}_0 = \overline{\rm MS} \;,\; {\cal R}_{\ln 4\pi-\gamma_E} = {\rm MS} \;,\; {\cal R}_{-2} = {\rm G} \ .
\end{eqnarray}
Moreover, since all ${\cal R}_\delta$ schemes are connected by scale-displacements, e.g. $\mu_{\delta_2}^2 = \mu_{\delta_1}^2 \exp({\delta_2 - \delta_1})$ for any two $R_\delta$-schemes $R_{\delta_1}$ and $R_{\delta_2}$~\cite{BMW2}, the $\beta$-function of the QCD coupling $\alpha_s$ in any ${\cal R}_\delta$-scheme is the same. In this subsection we use for brevity the notations $a = \alpha_s/4\pi$ and $\beta$ as the coupling $\beta$-function in any ${\cal R}_\delta$-scheme.

It is found that the $\delta$-terms in the perturbative series will always accompany $\{\beta_i\}$-terms, and thus the elimination of $\delta$-terms is equivalent to the elimination of $\{\beta_i\}$-terms. Therefore the PMC estimate can be achieved directly through a proper treatment of $\delta$-terms. This leads to a systematic prescription of setting the scales to all-orders, and opens the opportunity to start a program for automatically setting the PMC scales.

More explicitly, using this generalization, it was shown that a physical observable in any ${\cal R}_\delta$-scheme reads:
\begin{eqnarray}
\label{rhodeltas}
\rho_\delta(Q^2) &=& {\cal C}_0 + {\cal C}_1 a_1(\mu_1) + \left({\cal C}_2 + \beta_0 {\cal C}_1 \delta_1 \right) a^2_2(\mu_2) + \left[{\cal C}_3 +\beta _1 {\cal C}_1 \delta_1+ 2 \beta _0 {\cal C}_2\delta_2+ \beta _0^2 {\cal C}_1 \delta_1^2 \right] a^3_3(\mu_3) \\
& & + \left[{\cal C}_4 +\beta _2 {\cal C}_1\delta_1 +2 \beta _1 {\cal C}_2\delta_2 +3 \beta _0 {\cal C}_3\delta_3 +3 \beta _0^2 {\cal C}_2\delta_2^2 +\beta _0^3 {\cal C}_1 \delta_1 ^3 +\frac{5}{2} \beta _1 \beta _0 {\cal C}_1\delta_1^2 \right] a^4_4(\mu_4) + {\cal O}(a^5) \ . \nonumber
\end{eqnarray}
where $\mu_i \equiv Q e^{\delta_i/2}$, the initial scale is for simplicity set to $\mu_{r}^{\rm init} = Q$ and we defined ${\cal C}_i(Q) = {\cal C}_i$. An artificial index is introduced on each $a$ and correspondingly on each $\delta$ to keep track of which coupling each $\delta$ term is associated with. The more general expansion with higher tree-level power in $a$ can be readily derived~\cite{BMW2} and does not change the conclusions and results.

The expression in Eq.(\ref{rhodeltas}) exposes the pattern of $\{\beta_i\}$-terms in the coefficients at each order. Since there is nothing special about a particular value of $\delta$, one concludes that some of the coefficients of the $\{\beta_i\}$-terms are degenerate; e.g. the coefficient of $\beta_0 a(Q)^2$ and $\beta_1 a(Q)^3$ can be set equal. Thus, the ${\cal R}_\delta$-scheme not only illuminates the $\{\beta_i\}$-pattern, but also exposes a \emph{special degeneracy} of coefficients at different orders. Therefore, for any scheme, the expression for $\rho$ can be put to the form:
\begin{eqnarray}
\rho(Q^2) &= &r_{0,0} + r_{1,0} a(Q) + \left[r_{2,0} + \beta_0 r_{2,1} \right] a^2(Q)
+ \left[r_{3,0} + \beta_1 r_{2,1} + 2 \beta_0 r_{3,1} + \beta _0^2 r_{3,2} \right] a^3(Q) \nonumber\\
&& +\left[r_{4,0} + \beta_2 r_{2,1} + 2\beta_1 r_{3,1} + \frac{5}{2} \beta_1 \beta_0 r_{3,2} +3\beta_0 r_{4,1}+ 3 \beta_0^2 r_{4,2} + \beta_0^3 r_{4,3} \right] a^4(Q) + { \cal O }(a^5) \label{betapattern}
\end{eqnarray}
where $r_{i,0}$ are the conformal parts of the perturbative coefficients; i.e.
$r_{i} = r_{i,0} + {\cal O}(\{\beta_i\})$. In particular, it follows that $r_{0,0} = {\cal C}_0$ and $r_{1,0} = {\cal C}_1$, while the higher order coefficients ${\cal C}_{i\geq 2}$ are identified with the full brackets. The artificial indices on $a_i$ and $\delta_i$ in Eq.(\ref{rhodeltas}) reveals how the $\{\beta_i\}$-terms must be absorbed into the running coupling. The different $\delta_k$'s keep track of the power of the $1/\epsilon$ divergence of the associated diagram at each loop order in the following way: the $\delta_k^p a^n$-term indicates the term associated to a diagram with $1/\epsilon^{n-k}$ divergence for any $p$. Grouping the different $\delta_k$-terms one, recovers in the $N_C \to 0$ Abelian limit the dressed skeleton expansion \cite{qed1}. Resumming the series according to this expansion thus correctly reproduces the QED limit of the observable and matches the conformal series with running couplings evaluated at effective scales at each order.

Using this information from the $\delta_k$-expansion, it can be shown that the order $a^k(Q)$ coupling must be resummed into the effective coupling $a^k(Q_k)$, given by:
\begin{eqnarray}
r_{1,0} a(Q_1) &=& r_{1,0} a(Q) - \beta(a) r_{2,1} + \frac{1}{2} \beta(a) \frac{\partial \beta}{\partial a} r_{3,2} + \cdots + \frac{(-1)^n}{n!} \frac{{\rm d}^{n-1}\beta}{({\rm d} \ln\mu^2)^{n-1}} r_{n+1,n} \ , \nonumber \\
r_{2,0} a^2(Q_2) &=& r_{2,0} a^2(Q) - 2 a(Q) \beta(a) r_{3,1} + \left [a(Q) \frac{{\rm d}\beta}{{\rm d} \ln\mu^2}+ \beta(a)^2 \right ] r_{4,2} + \cdots + \Delta_2^{(n-1)}(a) r_{n+2,n} \ , \nonumber \\
& \hspace{2mm} \vdots &\nonumber \\
r_{k,0} a^{k}(Q_k) &=& r_{k,0} a^k(Q) + r_{k,0}\ k \ a^{k-1}(Q) \beta(a) \left \{
R_{k,1} +\Delta_k^{(1)}(a) R_{k,2} + \cdots + \Delta_k^{(n-1)}(a) R_{k,n} \right \} \ ,
\label{korder}
\end{eqnarray}
which defines the PMC scales $Q_k$, and where we introduced
\begin{eqnarray}
R_{k,j} &=& (-1)^{j}\frac{r_{k+j, j}}{r_{k,0}} \ , \\
\Delta_k^{(1)}(a) &=& \frac{1}{2} \left [ \frac{\partial \beta}{\partial a} + (k-1) \frac{\beta}{a}\right] \ , \\
\Delta_k^{(2)}(a) & =& \frac{1}{3!}\left [ \beta \frac{\partial^2 \beta}{\partial a^2} +Ê \left( \frac{\partial\beta}{\partial a} \right )^2 + 3(k-1) \frac{\beta}{a}\frac{\partial\beta}{\partial a} + (k-1)(k-2) \frac{\beta^2}{a^2}\right ] \ . \\
& \hspace{2mm} \vdots \nonumber
\end{eqnarray}
Eq.(\ref{korder}) is systematically derived by replacing the $\ln^j Q_1^2/Q^2$ by $R_{k,j}$ in the logarithmic expansion of $a^k(Q_k)$ up to the highest known $R_{k,n}$-coefficient in pQCD. The resummation can be performed iteratively using the RG equation for $a$ and leads to the effective scales for an NNNLO prediction:
\begin{eqnarray}
\label{exactscales}
\ln \frac{Q_{k}^2}{Q^2} &=& \frac{R_{k,1} + \Delta_k^{(1)}(a) R_{k,2}+\Delta_k^{(2)}(a) R_{k,3}}{1+ \Delta_k^{(1)}(a) R_{k,1} + \left({\Delta_k^{(1)}(a)}\right)^2 (R_{k,2} -R_{k,1}^2) + \Delta_k^{(2)}(a)R_{k,1}^2 } \ .
\end{eqnarray}
The final pQCD prediction for $\rho$ after setting the PMC scales $Q_i$ then reads
\begin{equation}
\rho(Q^2) = r_{0,0} + r_{1,0} a(Q_1) + r_{2,0} a^2(Q_2) +r_{3,0} a^3(Q_3) + r_{4,0}a^4(Q_4)+{ \cal O }(a^5) \ ,
\end{equation}
Here $Q_4$ remains unknown, since it requires the knowledge of $r_{5,1}$ in the coefficient of $a^5$. It is noted that in contrast to the PMC-BLM correspondence principle, in this method, all effective PMC scales are resummed at once, instead of a step-wise process. Moreover, the effective scales naturally become functions of the coupling through the $\beta$-function, in principle, to all orders.

This method systematically sums up all known non-conformal terms, in principle to all-orders, but is in practice truncated due to the limited knowledge of the $\beta$-function. It is easy to see that the LO values of the effective scales are independent of the initial renormalization scale. This follows since taking $\mu_{r}^{\rm init} \neq Q$, we must replace $R_{k,1} \to R_{k,1} + \ln Q^2/(\mu_{r}^{\rm init})^2$ and thus the LO effective scales read, $\ln Q_{k, \rm LO}^2/(\mu_{r}^{\rm init})^2 = R_{k,1} + \ln Q^2/(\mu_{r}^{\rm init})^2$, where $\mu_{r}^{\rm init}$ cancels and Eq.(\ref{exactscales}) at LO is recovered. More generally the effective scales do not depend on the initial renormalization scale at any order if the $\beta$-function is known. In practice, since the $\beta$-function is not known to all orders, there is residual renormalization scale dependence, which however is highly suppressed. The effective scales contain all the information of the non-conformal parts of the initial pQCD expression for $\rho$ in Eq.(\ref{betapattern}), which is exactly the purpose of the running coupling.

In a conformal theory, where $\{\beta_i\}=\{0\}$, the $\delta$-dependence vanishes in Eq.(\ref{rhodeltas}). Therefore, by absorbing all $\{\beta_i\}$ dependence into the running coupling, we obtain a final result independent of the initial choice of scale and scheme. It is found that the use of ${\cal R}_\delta$ scheme allows us to put this on rigorous grounds. From the explicit expression in Eq.(\ref{rhodeltas}) it is easy to confirm that
\begin{eqnarray}
\frac{\partial \rho_\delta}{\partial \delta} = -\beta(a) \frac{\partial \rho_\delta}{\partial a} \ .
\end{eqnarray}
The scheme-invariance of the physical prediction requires that $\partial \rho_\delta /\partial \delta = 0$. Therefore the scales in the running coupling must be shifted and set such that the conformal terms associated with the $\beta$-function are removed; the remaining conformal terms are by definition renormalization scheme independent. The numerical value for the prediction at finite order is then scheme independent as required by the renormalization group. The scheme-invariance criterion is a theoretical requirement of the renormalization group; it must be satisfied at any truncated order of the pertubative series, and is different from the formal statement that the all-orders expression for a physical observable is renormalization scale and scheme invariant; i.e. ${\rm d} \rho /{\rm d} \mu^{\rm init}_r = 0$. The final series obtained corresponds to the theory for which $\beta(a) = 0$; i.e. the conformal series. This demonstrates the concept of PMC to any order, which states that all non-conformal terms in the perturbative series must be resummed into the running coupling.

\subsubsection{\it Systematic All-Orders PMC Scale Setting for $R_{e^+e^-}(Q)$}

We take $R_{e^+e^-}(Q)$ as an example to show how to do the PMC scale setting by using the systematic all-orders method. The perturbative series matches the generic form of Eq.(\ref{betapattern}). It can be derived by analytically continuing the Adler function, $D$, into the time-like region \cite{Ralphas4,Ralphas5}:
\begin{equation}
R_{e^+e^-}(s) = \frac{1}{2\pi i} \int_{-s-i\epsilon}^{-s+i\epsilon} \frac{D( Q^2)}{Q^2} {\rm d} Q^2 \ ,
\end{equation}
with $D$ given by: $D(Q^2) = \gamma(a) - \beta(a) \frac{d}{da} \Pi(Q^2,a)$, where $\gamma$ is the anomalous dimension of the vector field and $\Pi$ is the vacuum polarization function. They can be written as perturbative expansions:
\begin{equation}
\gamma (a) = \sum_{n=0}^\infty \gamma_n a^n \ , \quad
\Pi(Q^2, a) = \sum_{n=0}^\infty \Pi_n (Q^2) a^n \ .
\end{equation}
It is then easy to show that to order $a^4$ the perturbative expression for $R_{e^+e^-}(Q)$ in terms of $\gamma_n$ and $\Pi_n$ reads (we suppress the $Q$ dependence of $\Pi_i$ in the following to simplify the notation):
\begin{eqnarray}
R_{e^+e^-}(Q) &=& \gamma_0 + \gamma_1 a(Q) + \left[\gamma_2 - \beta_0 \Pi_1\right] a^2(Q) + \left[\gamma_3 + \beta_1 \Pi_1 + 2 \beta_0 \Pi_2- \beta_0^2 \frac{\pi^2\gamma_1}{3} \right ] a^3(Q) \\
& & + \left[\gamma_4 + \beta_2 \Pi_1 + 2 \beta_1 \Pi_2 + 3\beta_0 \Pi_3 - \frac{5}{2} \beta_0\beta_1 \frac{\pi^2\gamma_1}{3} - 3 \beta_0^2 \frac{\pi^2 \gamma_2}{3} - \beta_0^3 \pi^2 \Pi_1 \right] a^4(Q) + {\cal O}(a^5) \ . \nonumber
\end{eqnarray}
This expression has exactly the form of Eq.(\ref{betapattern}), with the coefficients:
\begin{equation}
r_{i,0} = \gamma_i \ (i\geq1) ,\; r_{i,1} = \Pi_{i-1} \ (i\geq2) , \;
r_{i,2} = - \frac{\pi^2}{3} \gamma_{i-2} \ (i\geq3) ,\;
r_{i,3} = -\pi^2 \Pi_{i-3} \ (i\geq4) .
\end{equation}
The $\gamma_i$-coefficients contain $N_f$-terms in the dimensional regularization schemes, but since they are independent of $\delta$ to any order, they are kept fixed in the scale-setting procedure. The expression for the coefficient $\gamma_i$ and $\Pi_i$ can be found in Refs.~\cite{Ralphas4,Ralphas5}, while the four-loops $\beta$-function is given in Ref.~\cite{beta2}. Now we can set the effective scales $Q_1$, $Q_2$ and $Q_3$ up to the NNNLO. By convention, the argument of $a$ is space-like; however, the $\pi^2$-terms appearing in $R_{e^+e^-}$ can be avoided by using a coupling with time-like argument, leading to a more convergent series~\cite{pennington}. The last unknown scale can in this case be estimated \cite{BMW2}. It turns out that $Q_4 \sim Q $ which is the value we have used. The final result with five active flavors reads:
\begin{equation}
\label{Rnum}
\frac{3}{11} R_{e^+e^-}(Q) = 1 + \frac{\alpha_s(Q_1)}{\pi} + 1.84 \left(\frac{\alpha_s(Q_2)}{\pi}\right)^2
 - 1.00 \left(\frac{\alpha_s(Q_3)}{\pi}\right)^3 - 11.31 \left(\frac{\alpha_s(Q_4)}{\pi}\right)^4 \ ,
\end{equation}
which is explicitly free of any ${\cal R}_\delta$ scheme and scale ambiguities up to strongly suppressed residual ones.

\begin{figure}[htb]
\begin{center}
\begin{minipage}[bt]{6 cm}
\epsfig{file=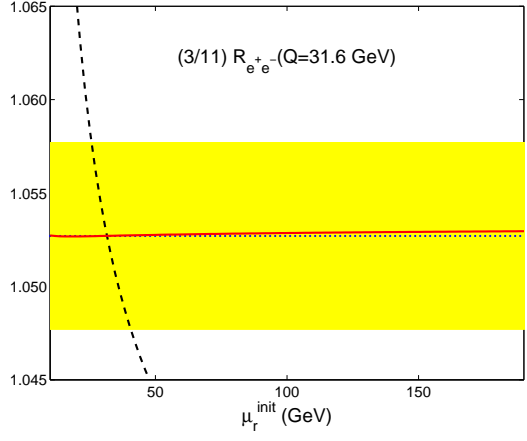,scale=0.4}
\end{minipage}
\begin{minipage}[t]{16.5 cm}
\caption{The PMC result for $R_{e^+e^-}(Q)$ as a function of the initial renormalization scale $\mu_r^{\rm init}$ (solid line), demonstrating the initial scale-invariance of the final prediction, up to strongly suppressed residual dependence. The shaded region is the experimental bounds~\cite{rexp} with the central value given by the dotted line. We also show the result before scale setting (dashed line), which is very sensitive to the choice of the initial renormalization scale. \label{Rsmu}}
\end{minipage}
\end{center}
\end{figure}

To find numerical values for the effective scales, we must determine the asymptotic scale, $\Lambda_{QCD}$. From the experimental value, $r_{e^+ e^-}(31.6~GeV) =\frac{3}{11} R_{e^+ e^-}(31.6~GeV)=1.0527\pm0.0050$~\cite{rexp}, we obtain
\begin{equation}
\Lambda^{'tH-\overline{MS}}_{QCD} = 481^{+255}_{-193} ~{\rm MeV} \;,\;\;
\Lambda^{\overline{MS}}_{QCD}  = 419^{+222}_{-168} ~{\rm MeV} .
\end{equation}
The asymptotic scale of ${\cal R}_\delta$ can be taken to be the same for any $\delta$, since they share the same $\beta$-function. The effective scales are found to be: $Q_1 = 1.3 ~Q$, $Q_2 = 1.2 ~Q$ and $Q_3 \approx 5.3 ~Q$. These values are independent of the initial renormalization scale up to some residual dependence coming from the truncated $\beta$-function. The final PMC result for $R_{e^+e^-}(Q)$ as a function of the initial renormalization scale $\mu^{\rm init}_r$ are shown in Fig.(\ref{Rsmu}), which demonstrates the initial scale-invariance of the final prediction up to strongly suppressed residual dependence. In Fig.(\ref{Rsmu}), the shaded region is the experimental bounds with the central value given by the thin dashed line. As a comparison, we also show the result before PMC scale setting, which however is very sensitive to the choice of the initial renormalization scale.

For completeness, we use our final result to predict the QCD coupling at the scale of the Z-boson mass, $M_Z$: $\alpha_s(M_Z) = 0.132^{+0.010}_{-0.011}$. This value is consistent with that of subsection \ref{anni}.

We have checked against the QED case, where $R_{e^+e^-}$ can be seen as the imaginary part of the QED four loop 1PI vacuum polarization diagram by the optical theorem, and find in this case nearly complete renormalization scale independence of all three scales to NNNLO due to the small value of the coupling. Numerically, we obtain for three (lepton) flavors:
\begin{equation}
\label{RnumQED}
\frac{1}{3} R^{e^+e^- \to \ell}_{\rm QED}(Q) = 1 + 0.24\alpha_e(Q_1) -0.08 \alpha_e(Q_2)^2
- 0.13 \alpha_e(Q_3)^3 + 0.05 \alpha_e(Q_4)^4 \ ,
\end{equation}
where $\alpha_e = e^2/4\pi$ and $\{\frac{Q_1}{Q},\frac{Q_2}{Q},\frac{Q_3}{Q}\} = \{1.1, 0.6, 0.5\}$. It is straightforward to apply our present analysis to the $\tau$-decay into hadrons ratio~\cite{Lam:1977cu},$
R_{\tau} = \sigma_{\tau \to \nu_{\tau }+{\rm hadrons}}/\sigma_{\tau \to \nu _{\tau }+\bar{\nu}_e+e^-} $, a similar observation has been found~\cite{BMW2}.

\subsubsection{\it Discussion on the Factorization and Renormalization Scale Dependence}
\label{sec:fullscale}

When one applies the PMC scale setting to renormalizable hard subprocesses, the initial and final quark and gluon lines are taken to be on-shell so that the amplitude of the hard subprocess is gauge invariant. Thus, the application of PMC to hard subprocesses does not involve the factorization scale, and no single or double logarithms which involve the factorization scale enter.

\begin{figure}[htb]
\begin{center}
\begin{minipage}[bt]{8 cm}
\epsfig{file=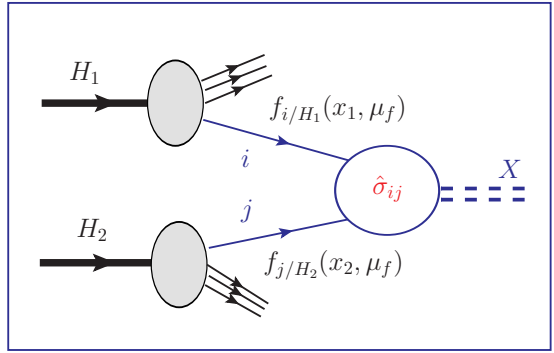,scale=0.7}
\end{minipage}
\begin{minipage}[t]{16.5 cm}
\caption{Diagram for calculating the total cross-section of a general hadroproduction process, $H_1 H_2 \to {X}$, where $X$ stands for any final states. It is obtained from the convolution of the partonic subprocess cross-section $\hat \sigma_{ij}$ with the parton distribution functions. \label{hadronicproduction}}
\end{minipage}
\end{center}
\end{figure}

However, for a general hadronic production process such as $H_1 H_2 \to {X}$ ($X$ stands for any final state), which is graphically shown by Fig.(\ref{hadronicproduction}), we need to be careful of how to set both the renormalization scale and the factorization scale consistently. It is important for any correct scale setting method to derive the full renormalization and factorization scale dependent terms. The two scales are independent quantities. The factorization scale is even necessary even in the theory where the coupling does not run. The factorization scale is introduced to identify the domain where
the initial and final quarks and gluons can be treated as on-shell partons.
The total cross-section for a hadronic production process can be generally written as
\begin{equation}
\sigma_{H_1 H_2 \to X} = \sum_{i,j} \int\int \frac{{\rm d}\hat{s}{\rm d}s}{S\hat{s}} f_{i/H_1}\left(x_1,\mu_f\right) f_{j/H_2}\left(x_2,\mu_f\right) \hat \sigma_{ij}(s,M,R) , \label{topbasic}
\end{equation}
where $x_1= {\hat{s} / S}$ and $x_2= {s / \hat{s}}$. The subprocess cross section $\hat \sigma_{ij}$ depends on the renormalization scale $\mu_r$ and the factorization scale $\mu_f$, with the definitions $M=\mu_f^2/Q^2$ ($Q$ stands for some typical energy scale~\footnote{For examples, $Q=m_H$ for Higgs production, $Q=m_t$ for top-quark pair production and etc., so as to eliminate the large logs.}) and $R=\mu_r^2/\mu_f^2$. Here $S$ denotes the hadronic center-of-mass energy squared and $s=x_1 x_2 S$ is the subprocess center-of-mass energy squared. The functions $f_{i/H_{1,2}}(x_\alpha,\mu_f)$ ($\alpha=1,2$) are the parton distribution functions (PDFs) describing the probability to find a parton of type $i$ with a momentum fraction between $x_\alpha$ and $x_{\alpha} +dx_{\alpha}$ in the hadron $H_{1,2}$.

The factorization scale is the scale entering PDF and the fragmentation functions. It is common and simple practice to identify the factorization scale with the renormalization scale, i.e. $\mu_{f}\equiv\mu_{r}$, and then to deal with the process following the same way as that of renormalization scale. The factorization scale should be chosen to match the nonpertubative bound-state dynamics with perturbative DGLAP evolution~\cite{dglap1,dglap2,dglap3}. This could be done explicitly using nonperturbative models such as AdS/QCD and light-front holography where the light-front wavefunctions of the hadrons are known, a recent report of which can be found in Ref.~\cite{adsQCD}. To fix one's attention on the elimination of renormalization dependence, one can fix the factorization scale $\mu_f$ to be the value that can eliminate large logs, such as setting $\mu_f \equiv Q$.

It is important to derive the full renormalization and factorization scale dependence, especially those terms from $\mu_f\neq\mu_r$ (or $R\neq1$), in order to achieve the renormalization scale independence in a consistent way.

As a first step, one can use the fact that the total hadronic cross-section is independent of the factorization scale, to derive the first derivative of the subprocess cross-section over the factorization scale, $\mu_f^2 \frac{{\rm d}}{{\rm d} \mu_f^2} \hat \sigma_{ij}(s,M,1)$; i.e. to retrieve the factorization scale dependence of the subprocess from $\hat\sigma_{ij}(s,1,1)$ by fixing $\mu_r=\mu_f$ (or $R=1$). For convenience, we rewrite Eq.(\ref{topbasic}) to a simpler notation with the direct-product symbol $\otimes$:
\begin{equation}
\sigma_{H_1 + H_2 \to X} = \sum \limits_{ij} f_{i/H_1}(\mu_f) \otimes \hat \sigma_{ij}(s,M,1) \otimes f_{j/H_2}(\mu_f) .
\end{equation}
The physical hadronic total cross-section will not depend on $\mu_f$,
\begin{equation}
\mu_f^2 \frac{\partial}{\partial \mu_f^2} \sigma_{H_1 + H_2 \to H + X} \equiv 0 ,
\end{equation}
which leads to the equation
\begin{equation}\label{dglapfac}
0\equiv \sum \limits_{ijk} f_{i/H_1}(\mu_f) \otimes \left [ a_s(\mu_f) P_{ik} \otimes \hat \sigma_{kj}(s,M,1) + \mu_f^2 \frac{\partial}{\partial \mu_f^2} \hat \sigma_{ij}(s,M,1) + a_s(\mu_f) \hat \sigma_{ik}(s,M,1)\otimes P_{kj} \right ] \otimes f_{j/H_2}(\mu_f) ,
\end{equation}
where $a_s=\alpha_s/4\pi$ and we have implicitly used the DGLAP evolution equation~\cite{dglap1,dglap2,dglap3},
\begin{equation}
\mu_f^2 \frac{\partial}{\partial \mu_f^2} f_{i/H_1} (\mu_f) = a_s(\mu_f)\sum_{j}\left [ P_{ij} \otimes f_{j/H_1}(\mu_f) \right ] .
\end{equation}
The splitting functions up to three loops can be found in Refs.~\cite{dglap1,dglap2,dglap3,split2,split3}. Equation (\ref{dglapfac}) should hold for an arbitrary $\mu_f$, therefore, the expression in the square brackets should be identically zero for any choice of $i$ and $j$ partons, which inversely yields the following ``evolution equation'' for the subprocess cross-section :
\begin{equation}
\mu_f^2 \frac{\partial}{\partial \mu_f^2} \hat \sigma_{ij}(s,M,1) = \frac{\partial}{\partial \ln\mu_f^2} \hat \sigma_{ij}(s,M,1) = - a_s(\mu_f) \sum \limits_{k}\Bigg[ P_{ik} \otimes \hat \sigma_{kj}(s,M,1) + \hat \sigma_{ik}(s,M,1) \otimes P_{kj}\Bigg]. \label{eqmuf}
\end{equation}
We can solve Eq.(\ref{eqmuf}) through an order-by-order manner using the partonic cross-sections $\sigma_{ij}(s,1,1)$ as the boundary condition. Initially we can set $\mu_r =\mu_f=Q$ ($M=R=1$). The splitting function can be expanded as
\begin{equation}\label{splitmuf}
P_{ij}(Q)=\frac{a_s(Q)}{4}P^{(0)}_{ij}(Q) + \frac{a^2_s(Q)}{4^2}P^{(1)}_{ij}(Q)+ \frac{a^3_s(Q)}{4^3}P^{(2)}_{ij}(Q) + \cdots .
\end{equation}
Then we obtain
\begin{equation}
\sigma_{ij}(s,M,1)=\sigma_{ij}(s,1,1)+\left(\frac{\partial}{\partial \ln\mu_f^2} \hat \sigma_{ij}(s,M,1)\right)_{M=1} L_{M} + \frac{1}{2!}\left(\frac{\partial^2} {\left(\partial \ln\mu_f^2\right)^2} \hat \sigma_{ij}(s,M,1)\right)_{M=1} L^2_{M} + \cdots
\end{equation}
where $L_{M}=\ln \mu_f^2/Q^2$.

As a second step, one can adopt the RG equation to retrieve $\hat\sigma_{ij}(s,M,R)$ from $\hat\sigma_{ij}(s,M,1)$. This step can be done by directly replacing the coupling at the renormalization scale $\mu_f$ to $\mu_r$; i.e. using the following formula,
\begin{equation} \label{alphasfull}
a_{s}(\mu_f^2)=a_{s}(\mu_r^2)+\frac{\beta_0}{4}\ln\frac{\mu_r^2}{\mu_f^2} a^2_{s}(\mu_r^2) + \frac{1}{16}\left\{\beta_0^2\ln^2 \frac{\mu_r^2}{\mu_f^2}+ \beta_1\ln\frac{\mu_r^2}{\mu_f^2}\right\} a^3_{s}(\mu_r^2) +{\cal O}(a_s^4) ,
\end{equation}
where $a_s=\alpha_s/\pi$. The full renormalization and factorization scale dependence for the splitting function can be retrieved by using this equation; i.e.~\cite{split3},
\begin{eqnarray}
P_{ij}(\mu_r,\mu_f) &=& \frac{a_s(\mu_r)}{4}P^{(0)}_{ij}(\mu_f) + \frac{a^2_s(\mu_r)}{4^2} \left[P^{(1)}_{ij}(\mu_f)-\beta_0 \ln\frac{\mu_f^2}{\mu_r^2} P^{(0)}_{ij}(\mu_f)\right] + \nonumber\\
&& \frac{a^3_s(\mu_r)}{4^3} \left[P^{(2)}_{ij}(\mu_f)+\beta^2_0 \ln\frac{\mu_f^2}{\mu_r^2} P^{(0)}_{ij}(\mu_f) -\beta_1 \ln\frac{\mu_f^2}{\mu_r^2} P^{(0)}_{ij}(\mu_f) -2\beta_0 \ln\frac{\mu_f^2}{\mu_r^2} P^{(1)}_{ij}(\mu_f)\right]+\cdots .
\end{eqnarray}
One can then apply e.g. PMC to set the renormalization scale for the splitting function.

The above two procedures can be extended up to any perturbative order. After doing these two steps, one can obtain the required $\hat\sigma_{ij}(s,M,R)$ with full renormalization and factorization scale dependence.

\subsection{\it A Comparison of FAC, PMS and BLM/PMC}
\label{comparison}

\begin{figure}[tb]
\begin{center}
\begin{minipage}[t]{8 cm}
\epsfig{file=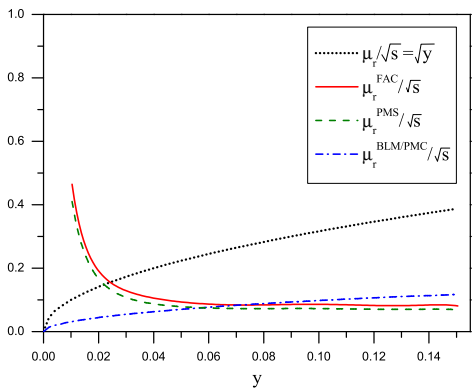,scale=0.7}
\end{minipage}
\begin{minipage}[t]{16.5 cm}
\caption{The scale $\mu_r/\sqrt{s}$ according to the BLM/PMC, PMS, FAC and the usual $\sqrt{y}$-procedures for the three-jet rate in $e^+e^-$ annihilation, as computed by Kramer and Lampe~\cite{Kramer1,Kramer2}. Notice the strikingly different behavior of the BLM/PMC scale from the PMS and FAC scales at low $y$.
The PMS and FAC scales increase at low jet virtuality, which is
the incorrect physical behavior.
\label{FigKramer}}
\end{minipage}
\end{center}
\end{figure}

As shown above, because the scale setting methods, such as FAC, PMS and BLM/PMC, have quite different starting points, they can give strikingly different results in practical applications. For example, Kramer and Lampe have analyzed the application of the FAC, PMS, and BLM/PMC methods for the prediction of jet production fractions in $e^+e^-$ annihilation in pQCD~\cite{Kramer1,Kramer2}. Usually, jets are defined by clustering particles with invariant mass less than $\sqrt{y s}$, where $y$ stands for the resolution parameter and $\sqrt{s}$ is the total center-of-mass energy. Physically, one expects the renormalization scale $\mu_r$ to reflect the invariant mass of the jets, that is, $\mu_r$ should be of order $\sqrt {y s}$. For example, in the analogous problem in QED, the maximum virtuality of the photon jet which sets the argument of the running coupling $\alpha_s$ cannot be larger than $\sqrt{y s}$. Thus one expects $\mu_r$ to decrease as $y \to 0$. However, as shown by Fig.(\ref{FigKramer}), the scales chosen by the FAC and PMS methods do not reproduce this physical behavior: The predicted scales $\mu^{PMS}_r$ and $\mu^{FAC}_r$ rise without bound at small values for the jet fraction $y$. This shows that the FAC and PMS can not get the right physical behavior in this limit, since they have summed physics into the running coupling not associated with renormalization. On the other hand, the BLM/PMC scale has the correct physical behavior as $y \to 0$. Since the argument of the running coupling becomes small using the BLM/PMC method, standard pQCD theory in $\alpha_s$ will not be convergent in the low $y$ domain~\cite{BMee}. In contrast, the scales chosen by PMS and FAC give no sign that the perturbative results break down in the soft region.

The real power of FAC is the concept of the effective charge, which allows one to define a scheme defined from a physical observable. As we described, the commensurate scale relations can be used to relate one effective charge to another. Furthermore, we list the main differences for PMS and BLM/PMC in the following:

\begin{itemize}
\item The PMS chooses the renormalization scale such that the first derivative of the fixed-order calculation with respect to the scale vanishes, However, this criterion of minimal sensitivity gives predictions which are not the same as the conformal prediction. As shown in Ref.~\cite{pmsrs}, by using the PMS together with the scheme-equations (\ref{scheme}) and the scheme-independent equation (\ref{inv-sch}), the renormalization scheme dependence can be reduced to a certain degree through an order-by-order manner. But there are still residual scheme dependence due to unknown higher order corrections, and in principle, the PMS prediction depends on the choice of renormalization scheme~\footnote{It is hard to estimate the contributions from those unknown higher-order terms within the framework of PMS. While for BLM/PMC, such dependence can be analyzed by using the extended renormalization group~\cite{HJLu}. }, and it disagrees with QED scale setting in the Abelian limit. Most important, the PMS does not satisfy the RG-properties {\it symmetry}, {\it reflexivity}, and {\it transitivity}, so that relations between observables depend on the choice of the intermediate renormalization scheme. Hence, when we successively express one effective charge in terms of others, PMS would lead to inconsistent scale choices.

\item At the present the BLM has been widely accepted for dealing with high energy processes. The PMC, being the principle underlying BLM, sums all $\{\beta_i\}$-terms in the fixed-order prediction into the running coupling, leaving the conformal series. The PMC is equivalent to BLM through the PMC-BLM correspondence principle. It satisfies all the RG-properties, such as {\it reflexivity}, {\it symmetry}, and {\it transitivity}. The PMC prediction is thus scheme-independent, and it automatically assigns the correct displacement of the intrinsic scales between schemes. The variation of the prediction away from the PMC scale exposes the non-zero $\{\beta_i\}$-dependent terms. The PMC prediction does have small residual dependence on the initial choice of scale due to the truncated unknown higher order $\{\beta_i\}$-terms, which will be highly suppressed by proper choice of PMC scales.
\end{itemize}

\section{Applications of PMC}
\label{secV}

In this section, we present some recent examples for the PMC scale setting. Some more subtle points in using the PMC scale setting are presented, which are useful references for future applications.

\subsection{\it Top-Quark Pair Total Cross Section at the NNLO Level}
\label{topquarkCS}

The total top-pair production cross-section $\sigma_{t \bar{t}}$ has been measured at the Tevatron with a precision $\Delta \sigma_{t \bar{t}}/\sigma_{t \bar{t}}\sim\pm 7\%$~\cite{cdft,d0t} and the two LHC experiments have reached similar sensitivity~\cite{atlas,cms}. Theoretically, the total cross-section for the top-pair production has been calculated up to NLO within the $\overline{MS}$-scheme in Refs.~\cite{nason1,nason2,nason3,beenakker1,beenakker2,czakon1}. Large logarithmic corrections associated with the soft gluon emission have been investigated and resummed up to next-to-next-to-leading-logarithmic corrections~\cite{nason3,czakon1,czakon2,moch1,moch2,moch3,beneke1,beneke2,andrea,vogt,hathor,czakon4,nik}. These results provide the foundation for deriving a more precise estimation by using PMC.

\subsubsection{\it Basic Formulas}

According to Eq.(\ref{topbasic}), The total top-pair production cross-section can be written as:
\begin{equation}
\sigma_{H_1 H_2 \to {t\bar{t} + X}} = \sum_{i,j}\frac{1}{S}\int\limits_{4m^2_{t}}^{S} {\rm d}s \int\limits_s^S \frac{{\rm d}\hat{s}}{\hat{s}} f_{i/H_1}\left(x_1,\mu_f\right) f_{j/H_2}\left(x_2,\mu_f\right) \hat \sigma_{ij}(s,\alpha_s(\mu_r),\mu_f) \ ,
\end{equation}
where $x_1= {\hat{s} / S}$, $x_2= {s / \hat{s}}$. The top-quark mass $m_{t}$ is the mass renormalized in the on-shell (pole-mass) scheme. Setting $s=4m_t^2 (S/4m_t^2)^{y_1}$ and $\hat{s}=s(S/s)^{y_2}$, we can transform the two-dimensional integration over $s$ and $\hat{s}$ into an integration over two variables $y_{1,2}\in[0,1]$, which can be calculated by using the improved VEGAS program~\cite{vegas,bcvegpy,genxicc}. The partonic cross-section $\hat \sigma_{ij}$ can be decomposed in terms of dimensionless scaling functions $f_{ij}$. Up to NNLO, it takes the form
\begin{equation}
\hat\sigma_{ij} = {1 \over m^2_{t}} \Big\{f_{ij}^{0}(\rho,Q) a^2_s(Q) + f_{ij}^{1}(\rho,Q) a^3_s(Q) + f_{ij}^2(\rho,Q) a^4_s(Q) \Big\} , \label{subcs}
\end{equation}
where $\rho={4m_t^2}/{s}$, $a_s(Q)={\alpha_s(Q)}/{\pi}$ and $(ij) = \{(q{\bar q})$, $(gg)$, $(gq)$, $(g\bar{q})\}$ stands for the four production channels respectively. $Q$ stands for the typical energy scale of the process. When applying PMC to the renormalizable hard subprocesses which enter the pQCD leading-twist factorization procedure, the initial and final quark and gluon lines are taken to be on-shell so that the hard subprocess amplitude is gauge invariant. Thus, the application of PMC to hard subprocesses does not involve the factorization scale. It is convenient to fix the factorization scale $\mu_f\equiv Q$. In principle this uncertainty can be removed if one knows the bound state wave function. As for the initial renormalization scale $\mu^{\rm init}_r$, we also set its value to be $Q$.

The scale functions $f_{ij}^{0,1,2}(\rho,Q)$ can be directly read from the HATHOR program~\cite{hathor}. According to the PMC scale setting, we need to find the explicit terms that are $n_f$- or $n^2_f$- dependent, which should be absorbed into the $\alpha_s$- running. The QCD Coulomb-type correction may also provide sizable contributions in the threshold region~\cite{coul1,coul2}, so terms that are proportional to $\pi/v$ or $(\pi/v)^2$ ($v=\sqrt{1-\rho}$, the heavy quark velocity) should be treated separately~\cite{brodsky1}; i.e., we need to introduce new PMC scales for the Coulomb type terms. Then, the NLO and NNLO scaling functions can be rearranged as
\begin{eqnarray}
f_{ij}^{1}(\rho,Q) &=& \left[A_{1ij} + B_{1ij} n_f \right] + D_{1ij} \left(\frac{\pi}{v}\right) \\
f_{ij}^{2}(\rho,Q) &=& \left[A_{2ij} + B_{2ij} n_f + C_{2ij} n^2_f\right] + \left[D_{2ij} +E_{2ij} n_f \right] \left(\frac{\pi}{v}\right) + F_{2ij}\left(\frac{\pi}{v}\right)^2 .
\end{eqnarray}
Substituting them into Eq.(\ref{subcs}), the partonic cross-section $\hat \sigma_{ij}$ changes to
\begin{eqnarray}
m_t^2 \; \hat\sigma_{ij} &=& A_{0ij} a^2_s(Q) + \left\{\left[A_{1ij} + B_{1ij} n_f \right] + D_{1ij} \left(\frac{\pi}{v}\right)\right\} a^3_s(Q) + \nonumber\\
&& \left\{\left[A_{2ij} + B_{2ij} n_f + C_{2ij} n^2_f\right] + \left[D_{2ij} +E_{2ij} n_f \right] \left(\frac{\pi}{v}\right) + F_{2ij}\left(\frac{\pi}{v}\right)^2\right\} a^4_s(Q) , \label{eq:pmctopstart}
\end{eqnarray}
where $A_{0ij}=f^{0}_{ij}(\rho,Q)$.

The PMC scales can be determined in a general scheme-independent way as described in Sec.\ref{sec:IV:blm}. We shift the renormalization scale $Q$ to absorb the $\{\beta_i\}$-dependent terms, using the $n_f$-dependence as a guide:
\begin{eqnarray}
m_t^2 \; \hat\sigma_{ij} &=& A_{0ij} a^2_s(Q_1^*) + \left[\tilde{A}_{1ij}\right] a^3_s(Q_1^{**}) + \left[\tilde{\tilde{A}}_{2ij} \right] a^4_s(Q_1^{**}) + \left(\frac{\pi}{v}\right) D_{1ij} \left[\frac{2\kappa}{1-\exp(-2\kappa)}\right] a^3_s(Q_2^*) ,
\end{eqnarray}
where $\kappa=\frac{\tilde{D}_{2ij}}{D_{1ij}} a_s(Q_2^*) + \frac{F_{2ij}}{D_{1ij}} \left(\frac{\pi}{v}\right) a_s(Q_2^*)$. $Q^{*,**}_{1}$ are LO and NLO PMC scales for the usual part and $Q^{*}_2$ is the LO PMC scale for the Coulomb part. For the usual part, the two PMC scales $Q^{*,**}_{1}$ satisfy
\begin{equation}
\ln\frac{Q^{*2}_1}{Q^2} = \frac{3B_{1ij}}{A_{0ij}} +\frac{9B^2_{1ij}-12A_{0ij}C_{2ij}}{8A_{0ij} B_{1ij}} \beta_0 \left(\frac{3B_{1ij}}{A_{0ij}}\right) a_s(Q),\;\; \ln\frac{Q^{**2}_{1}}{Q_{1}^{*2}} = \frac{2\tilde{B}_{2ij}}{\tilde{A}_{1ij}} \;,
\end{equation}
where the coefficients are~\cite{pmc3}
\begin{eqnarray}
\tilde{A}_{1ij} &=& \frac{2A_{1ij}+33B_{1ij}}{2},\;\; \tilde{\tilde{A}}_{2ij} =\frac{2\tilde{A}_{2ij} +33\tilde{B}_{2ij}}{2} , \\
\tilde{A}_{2ij} &=& \frac{1}{8A_{0ij}}\left[ 99B_{1ij}(2A_{1ij}+33B_{1ij})+ A_{0ij}(8A_{2ij}+ 306B_{1ij}-2178C_{2ij})\right] , \\
\tilde{B}_{2ij} &=& \frac{1}{4A_{0ij}}\left[4 A_{0ij}(B_{2ij}+33C_{2ij})- B_{1ij}(19A_{0ij} +6A_{1ij}+99B_{1ij})\right] .
\end{eqnarray}
While for the Coulomb part, we have adopted the Sommerfeld-Gamow-Sakharov rescattering formula~\cite{sommerfeld1,sommerfeld2,sommerfeld3} to sum up the higher-order $\pi/v$ terms. The overall factor $({\pi}/{v})$ before $D_{1ij}$ shall be canceled by a $v^{1}$-factor from the phase space. Its LO PMC scale $Q^*_2$ satisfies
\begin{equation}
\ln\frac{Q^{*2}_{2}}{Q^2} = \frac{2E_{2ij}}{D_{1ij}}
\end{equation}
and the coefficient
\begin{equation}
\tilde{D}_{2ij} = (2D_{2ij}+33E_{2ij})/{2} .
\end{equation}

Since the channels $(ij) = \{(q{\bar q}), (gg), (gq), (g\bar{q})\}$ are distinct and non-interfering, their PMC scales should be set separately.

\begin{itemize}

\item For $(q\bar{q})$ channel, all the coefficients $A_{0q\bar{q}}$, $A_{1q\bar{q}}$ and etc. are non-zero. Following PMC scale setting, the $n_f$-terms that are associated with the $\{\beta^{\overline{MS}}_{i}\}$-terms are absorbed into the $\alpha_s$-coupling:
\begin{itemize}
\item When performing the scale shift $Q\to Q^{*}_1$, the first type of $\{\beta^{\cal R}_{i}\}$-terms $B_{1q\bar{q}}$ and $C_{2q\bar{q}}$ are eliminated. Part of $B_{2q\bar{q}}$ which contains the same type $\{\beta^{\cal R}_{i}\}$-term is also absorbed into $\alpha_s$ running. The remaining part of $B_{2q\bar{q}}$ is compensated by $A_{1q\bar{q}}$ and $B_{1q\bar{q}}$ to ensure that the first type of $\{\beta^{\cal R}_{i}\}$-terms are absorbed into $\alpha_s$ -coupling exactly, which results in a new variable $\tilde{B}_{2q\bar{q}}$. Because $(B_{1q\bar{q}}/A_{0q\bar{q}})$ shows a monotone increase with the increment of collision energy $\sqrt{s}$, $Q^{*}_1$ shall show the same trend versus $\sqrt{s}$.

\begin{figure}[tb]
\begin{center}
\begin{minipage}[t]{8 cm}
\epsfig{file=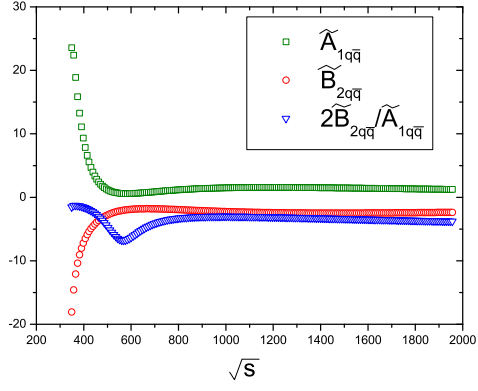,scale=0.7}
\end{minipage}
\begin{minipage}[t]{16.5 cm}
\caption{PMC coefficients of the $(q\bar{q})$ channel versus the subprocess collision energy $\sqrt{s}$. $m_t=172.9$ GeV. \label{qqcoe}}
\end{minipage}
\end{center}
\end{figure}

\item When performing the scale shift $Q^{*}_1\to Q^{**}_1$, the second type of $\{\beta_{i}\}$-terms, i.e. $\tilde{B}_{2q\bar{q}}$ are eliminated. As shown by Fig.(\ref{qqcoe}), the value of $\tilde{B}_{2q\bar{q}}$ is always negative and $\tilde{A}_{1q\bar{q}}$ has a minimum value in lower $\sqrt{s}$, then one can find that the NLO PMC scale $Q^{**}_1$ shall be suppressed to a certain degree in comparison to $Q$.
\item The Coulomb type correction provides a distinct contribution to the total cross-section in the threshold region, which should be treated separately from the usual part. Similarly, when performing the scale shift $Q\to Q^{*}_2$ for the Coulomb type contribution, $E_{2q\bar{q}}$ is eliminated.
\end{itemize}

\item For $(gg)$ channel, we have $C_{2gg}=0$, while other coefficients $A_{0gg}$, $A_{1gg}$ and etc. are non-zero. It can be treated in a similar way as the $(q\bar{q})$ channel. It is found that in distinction to the $q+\bar{q}$ case, both $2\tilde{B}_{2gg}/\tilde{A}_{1g\bar{g}}$ and $3B_{1gg}/A_{0g\bar{g}}$ are close to zero, and thus its LO and NLO PMC scales ($Q^{*}_1$ and $Q^{**}_1$) become close to $Q$
with appropriate modifications due to $\sqrt{s}$.

\item For $(gq)$ or $(g\bar{q})$ channel, we have $A^{0}_{ij}=0$, $D_{1gq}= 0$, $E_{2gq}=F_{2gq} \equiv 0$, which shows that the Coulomb type corrections start only at the NNLO order. We only need to set one LO PMC scale for these two channels. That is,
\begin{eqnarray}
m_t^2 \hat\sigma_{ij} &=& \left[A_{1ij}\right] a^3_s(Q) + \left\{\left[A_{2ij} + B_{2ij} n_f \right] + \left[D_{2ij} \right] \left(\frac{\pi}{v}\right) \right\} a^4_s(Q) \nonumber\\
&=& A_{1ij} a^3_s(Q_3^*) + \left[\tilde{A}_{2ij}\right] a^4_s(Q_3^*) + D_{2ij} \left(\frac{\pi}{v}\right) a^4_s(Q) , \nonumber
\end{eqnarray}
where $(ij)=(gq)$ or $(g\bar{q})$, $\tilde{A}_{2ij} = A_{2ij}+{33B_{2ij}}/{2}$ and $\ln {Q_3^{*2}}/{Q^2} = {2B_{2ij}}/{A_{1ij}}$.

\end{itemize}

\subsubsection{\it Numerical Analysis for the Total Cross Section}

To do the numerical calculation, we adopt $m_t = 172.9\pm1.1$ GeV~\cite{pdg} and the CTEQ PDFs of version 2010, i.e. CT10~\cite{cteq} \footnote{The CT10 is a global fit for general-purposes based on a partly NNLO fit to data. Only very recently, the CTEQ group released the CT10NNLO version, and a similar quality of agreement with the fitted experiment data sets in the NNLO fit as those of NLO had been observed~\cite{cteqn}. Since the change of CT10 to CT10NNLO only leads to very small numerical differences, we adopt our previous choice of CT10 to do the analysis. }. The combined PDF and $\alpha_s$ uncertainty are set by using different PDF sets determined by varying $\alpha_s(m_Z) \in [0.113, 0.230]$. At present, to keep our attention on the renormalization scale, we set as usual, $\mu_f \equiv m_t$. As initial choice, we set $\mu_r^{\rm init}=m_t$. After PMC scale setting, the PMC scales are usually different from $m_t$, so one must use the formulas listed in Sec.\ref{sec:fullscale} to get the full renormalization and factorization scale dependence before applying PMC. This point is very important for the later initial-scale-independent analysis. In the literature, the full renormalization and factorization scale dependence for the top-quark pair production up to NNLO can be found in Refs.~\cite{moch3,hathor}.

\begin{table}
\begin{center}
\begin{minipage}[tb]{16.5 cm}
\caption{Total cross-sections for the top-pair production before and after PMC scale setting for $\sqrt{s}=1.96$ TeV. $m_t=172.9$ GeV and the central CT10 as the PDF \cite{cteq}. }
\label{tab:tev}
\end{minipage}
\begin{tabular}{|c||c|c|c|c||c|c|c|c|}
\hline
& \multicolumn{4}{c||}{before PMC scale setting} & \multicolumn{4}{c|}{after PMC scale setting} \\
\hline
~ ~ & ~LO~ & ~NLO~ & ~NNLO~ & ~ total ~& ~LO~ & ~NLO~ & ~NNLO~ & ~ total ~ \\
\hline
$q+\bar{q}$ (pb) & 4.989 & 0.975 & 0.489 & 6.453 & 4.841 & 1.756 & -0.063 & 6.489 \\
$g+g$ (pb)  & 0.522 & 0.425 & 0.155 & 1.102 & 0.520 & 0.506 & 0.148 & 1.200 \\
$g+q$ (pb)  & 0.000 &-0.0366 & 0.0050& -0.0316 & 0.000 & -0.0367 & 0.0050 & -0.0315 \\
$g+\bar{q}$ (pb) & 0.000 &-0.0367 & 0.0050& -0.0315 & 0.000 & -0.0366 & 0.0050 & -0.0316 \\
sum (pb)   & 5.511 & 1.326 & 0.654 & 7.492 & 5.361 & 2.188 & 0.095 & 7.626 \\
\hline
\end{tabular}
\end{center}
\end{table}

\begin{table}
\begin{center}
\begin{minipage}[tb]{16.5 cm}
\caption{Total cross-sections for the top-pair production before and after PMC scale setting for $\sqrt{s}=14$ TeV. $m_t=172.9$ GeV and the central CT10 as the PDF \cite{cteq}. }
\label{tab:lhc}
\end{minipage}
\begin{tabular}{|c||c|c|c|c||c|c|c|c|}
\hline
& \multicolumn{4}{c||}{before PMC scale setting} & \multicolumn{4}{c|}{after PMC scale setting} \\
\hline
~ ~ & ~LO~ & ~NLO~ & ~NNLO~ & ~ total ~& ~LO~ & ~NLO~ & ~NNLO~ & ~ total ~\\
\hline
$q+\bar{q}$ (pb) & 73.45 & 9.00 & 5.16 & 87.58 & 69.33 & 21.08 & -2.98 & 86.96 \\
$g+g$ (pb)  & 487.52 & 262.96 & 49.87 & 800.68 & 485.51 & 303.69 & 37.35 & 835.41 \\
$g+q$ (pb)  & 0.00 & 9.30 & 7.69 & 17.00 & 0.00 & 9.37 & 7.69 & 16.97 \\
$g+\bar{q}$ (pb) & 0.00 & 0.02 & 1.92 & 1.95 & 0.00 & -0.03 & 1.92 & 17.06 \\
sum (pb)   & 560.96& 281.29& 64.63 & 907.20 & 554.84 & 334.11 & 43.98 & 941.26 \\
\hline
\end{tabular}
\end{center}
\end{table}

We first present the total cross-sections for the top-quark pair production using the PMC scale setting by fixing all the input parameters to their central values. The results are presented in Tables \ref{tab:tev} and \ref{tab:lhc}, where for comparison, the total cross-sections for the conventional scale setting method ($\mu_r\equiv m_t$) are also presented.

\begin{itemize}

\item Tables \ref{tab:tev} and \ref{tab:lhc} show that the pQCD convergence is improved after PMC scale setting. This is due to the fact that we have resummed the universal and gauge invariant higher-order corrections associated with the $\{\beta^{\overline{MS}}_i\}$-terms into the LO and NLO -terms by suitable choice of PMC scales. It is also the reason why after PMC scale setting, the total cross-section $\sigma_{t\bar{t}}$ is increased by $\sim 2\%$ at the Tevatron and $\sim 4\%$ at the LHC. This small change in the total cross-section after PMC scale setting means that the naive choice of $\mu_r \equiv m_t$ is a viable approximation for estimating the total cross-section.

\begin{figure}[tb]
\begin{center}
\begin{minipage}[t]{12 cm}
\epsfig{file=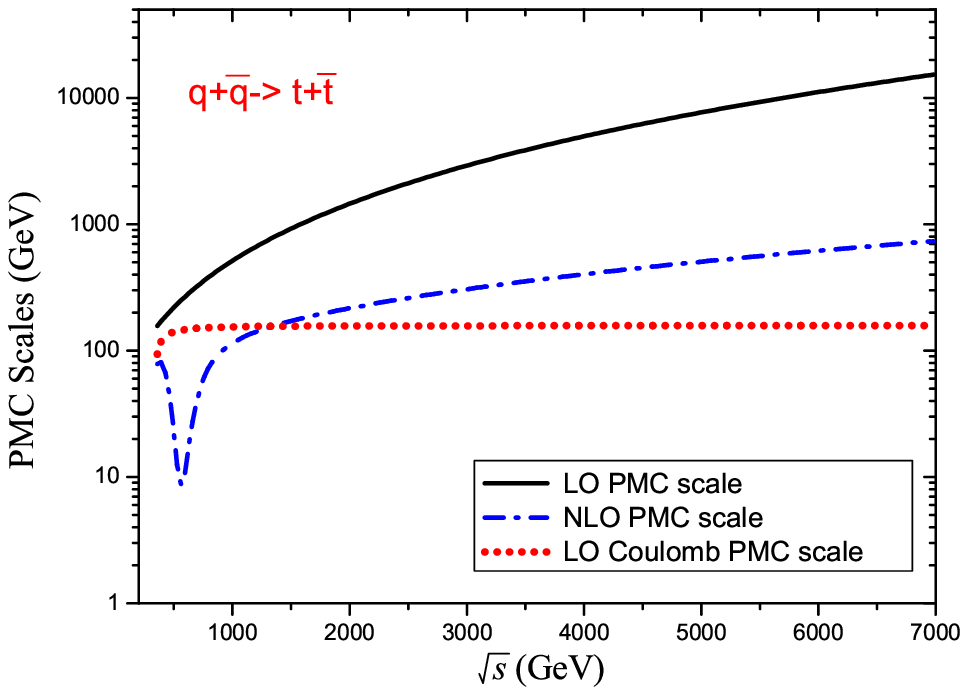,scale=0.6}
\hspace{1cm}
\epsfig{file=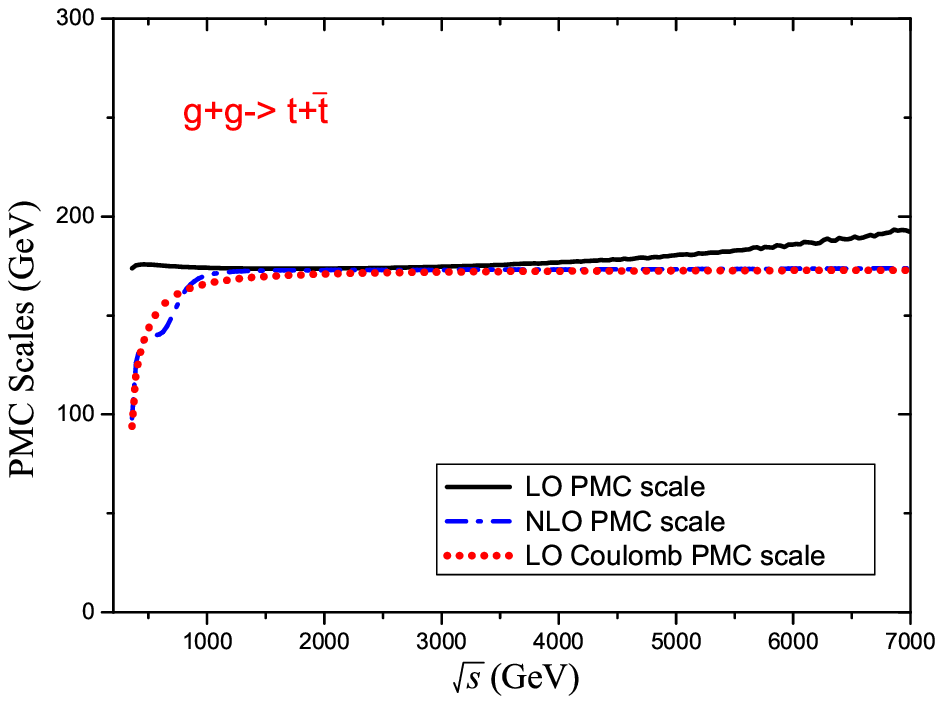,scale=0.6}
\end{minipage}
\begin{minipage}[t]{16.5 cm}
\caption{PMC scales versus the sub-process collision energy $\sqrt{s}$ for the top-quark pair production up to $\sqrt{s}=7$ TeV, where we have set the initial renormalization scale $\mu^{\rm init}_r=Q$. Here $Q =m_t=172.9$ GeV. \label{lhcscale}}
\end{minipage}
\end{center}
\end{figure}

\item Since different channels have quite different behaviors, it is necessary to use different PMC scales for each channel. The PMC scales are functions of $\sqrt{s}$, whose behaviors up to $\sqrt{s}=7$ TeV are presented in Fig.(\ref{lhcscale}). Because of the behaviors of the PMC coefficients, the LO PMC scale for the $(q\bar{q})$-channel increases with $\sqrt{s}$ and is much larger than $m_t$ for large $\sqrt{s}$. As a result, its LO cross-sections at the Tevatron and LHC are decreased by $3\%-5\%$ relative to the standard guess under the conventional scale setting. Because $|B_{1gg}/A_{0gg}|\ll1$, the LO PMC scale for the $(gg)$-channel is slightly different from $m_t$ and its LO cross-section remains almost unchanged. It is noted that there is a dip for the NLO scale of the $(q\bar{q})$-channel, which is caused by the fact that the NLO conformal term $\tilde{A}_{1q\bar{q}}$ reaches its smallest value and hence the factor $\tilde{B}_{2q\bar{q}}/\tilde{A}_{1q\bar{q}}$ reaches its highest negative value when $\sqrt{s} \simeq [\sqrt{2}\exp(5/6)]m_t \sim 563$ GeV; this results in an exponential suppression to the NLO scale. The NLO PMC scale for the $(q\bar{q})$-channel is smaller than $m_t$ by about one order of magnitude in low $x$-region. As a result, its NLO cross-section will be considerably increased; i.e. it is a factor of two times larger than its value derived from the conventional scale setting. As for the $(gg)$-channel, its NLO PMC scale slightly increases with $\sqrt{s}$, but it is smaller than $m_t$ for $\sqrt{s}\ll1$ TeV, so that its NLO cross-sections at the Tevatron and LHC are increased by $15\%- 20\%$.

\begin{figure}[tb]
\begin{center}
\begin{minipage}[t]{12 cm}
\epsfig{file=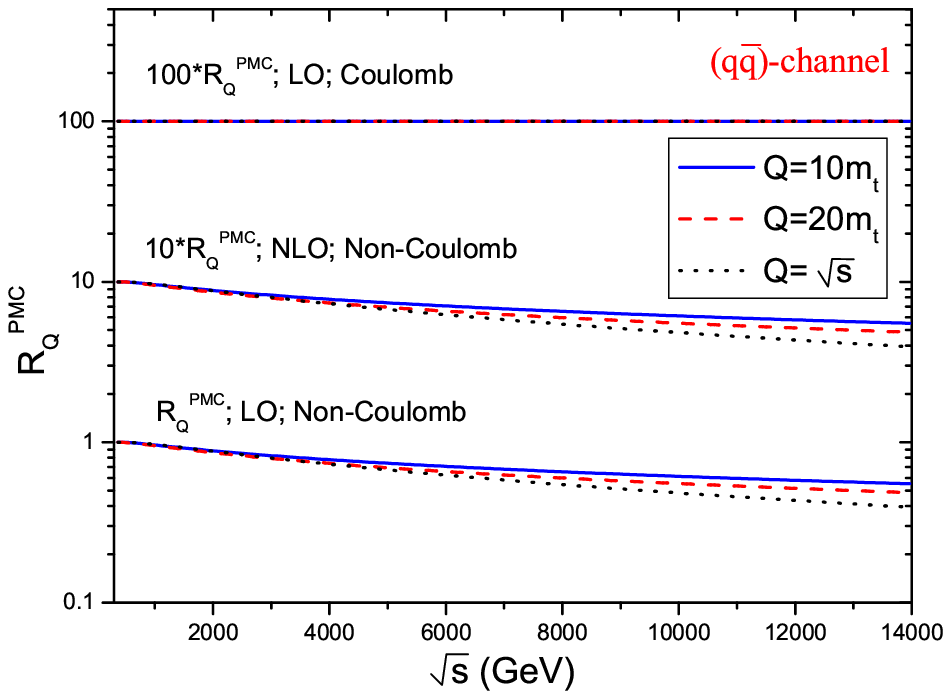,scale=0.6}
\epsfig{file=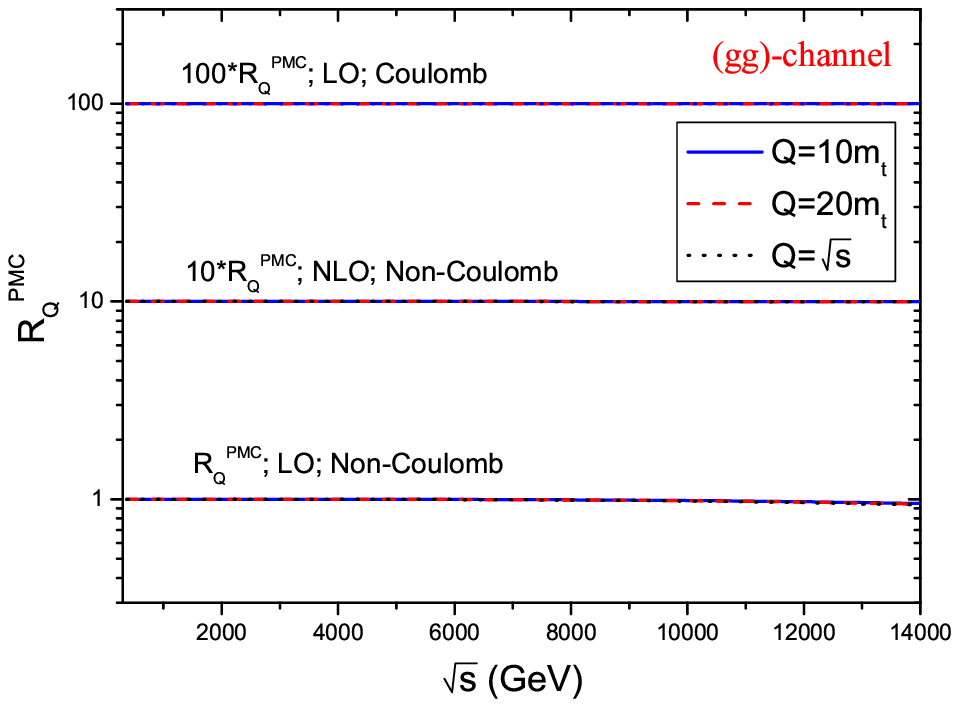,scale=0.6}
\end{minipage}
\begin{minipage}[t]{16.5 cm}
\caption{The ratio $R_Q^{\rm PMC}= {{\mu^{\rm PMC}_{r}|_{\mu^{\rm init}_{r}=Q}}\over{\mu^{\rm PMC}_{r}|_{\mu^{\rm init}_r = m_t} }} $ versus the sub-process collision energy $\sqrt{s}$ up to $14$ TeV, where $Q=10\,m_t$, $20\,m_t$ and $\sqrt{s}$ respectively. Here $m_t=172.9$ GeV. These results show that the renormalization scales for $t \bar t$ production determined by PMC scale setting at finite order is insensitive to the choice of very disparate initial scales. \label{pmcscalet}}
\end{minipage}
\end{center}
\end{figure}

\item There is residual initial renormalization scale dependence because of the unknown-higher-order $\{\beta^{\overline{MS}}_i\}$-terms. Because the PMC scales themselves must be a perturbative series of $\alpha_s$, the residual initial renormalization scale uncertainty can be greatly suppressed due to the fact that those higher-order $\{\beta^{\overline{MS}}_i\}$-terms are absorbed into the PMC scales' higher-order terms. We define a ratio $R^{\rm PMC}_{Q}$ to show how the change of initial renormalization scale affects the PMC scales; i.e.
 \begin{equation}
  R_Q^{\rm PMC}=\frac{\mu^{\rm PMC}_r|_{\mu^{\rm init}_{r}=Q}}{\mu^{\rm
  PMC}_r|_{\mu^{\rm init}_{r}=m_t}} ,
 \end{equation}
  where $\mu^{\rm PMC}_r|_{\mu^{\rm init}_{r}=Q}$ stands for the PMC scales determined under the condition of $\mu^{\rm init}_{r}=Q$, which is $Q^*_1$ (LO scale for the non-Coulomb part), $Q^{**}_1$ (NLO scale for the non-Coulomb part) or $Q^{*}_2$ (LO scale for the Coulomb part) respectively. In Fig.(\ref{pmcscalet}), the ratios for the dominant $q\bar{q}$- and $gg$- channels are presented. In order to amplify the differences, we take three disparate scales to draw the curves, i.e. $Q=10\,m_t$, $20\,m_t$ and $\sqrt{s}$ respectively.

  As shown in Fig.(\ref{pmcscalet}), the LO PMC scale $Q^{*}_2$ for the Coulomb-term in both channels are unchanged under different choice of $Q$. Among these choices, $Q=\sqrt{s}$ usually gives the largest deviation from the case of $Q=m_t$. The residual initial scale dependence for the $(gg)$-channel is small, $R_Q^{\rm PMC}\sim 1$, only for the LO non-Coulomb PMC scale $Q^*_1$, it has sizable effect. As an example, for the case of $Q=\sqrt{s}$, its $Q^{*}_1$ deviates from that of $Q=m_t$ by $\sim 1\%$ at $\sqrt{s}=7$ TeV, and it is raised only up to $\sim 7\%$ at $\sqrt{s}=14$ TeV. In the case of the $(q\bar{q})$-channel, the residual scale dependence of the LO/NLO PMC scale for the non-Coulomb part is somewhat larger; i.e. the deviation is about $12\%$ for the case of $Q=\sqrt{s}$ at $\sqrt{s}=2$ TeV, and the deviation reaches up to $\sim 60\%$ at $\sqrt{s}=14$ TeV. (We expect that this dependence on $Q$ will be greatly reduced at NNNLO.) However in such high collision region ($\sqrt{s}>2$ TeV), the total cross-sections are highly suppressed by the parton luminosities and their values are almost unchanged by using very different initial scales.

\begin{table}
\begin{center}
\begin{minipage}[tb]{16.5 cm}
\caption{Dependence on the initial scale of the total $t \bar t$ production cross-sections (in unit: pb) at the Tevatron and LHC. Here $m_t=172.9$ GeV and the central CT10 as the PDF~\cite{cteq}. The number in the parenthesis shows the Monte Carlo uncertainty in the last digit. }
\label{scaleun}
\end{minipage}
\begin{tabular}{|c||c|c|c||c|c|c|}
\hline
& \multicolumn{3}{c||}{PMC scale setting} & \multicolumn{3}{c|}{Conventional scale setting} \\
\hline
& $Q=m_t/4$ & $Q=m_t$ &$Q=4\,m_t$ & $\mu_r\equiv m_t/2$ & $\mu_r\equiv m_t$ & $\mu_r\equiv 2\,m_t$ \\
\hline
Tevatron (1.96 TeV) & 7.620(5) & 7.626(3) & 7.623(6) & 7.742(5) & 7.489(3) & 7.199(5) \\
LHC (7 TeV) & 171.6(1) & 171.8(1) & 171.7(1) & 168.8(1) & 164.6(1) & 157.5(1) \\
LHC (14 TeV)& 941.8(8) & 941.3(5) & 941.4(8) & 923.8(7) & 907.4(4) & 870.9(6) \\
\hline
\end{tabular}
\end{center}
\end{table}

  Total cross-sections with several typical initial renormalization scale $\mu^{\rm init}_r=Q$ are presented in Table \ref{scaleun}. At the NNLO level, it is found that the residual scale uncertainty to the total cross-section is less than $10^{-3}$ by setting $Q=4\,m_t$ or $Q =m_t/4$. In fact, even by setting $Q=20\,m_t$ and $\sqrt{s}$, such residual scale uncertainty is still less than $10^{-3}$~\cite{pmc5}. As a comparison, we also present the results for the conventional scale setting in Table \ref{scaleun}; by varying the renormalization scale within the region of $[m_t/2,2\,m_t]$, we obtain a large renormalization scale-uncertainty $\left({}^{+3\%}_{-4\%}\right)$ at the Tevatron and LHC, which agrees with the previous results derived in the literature, c.f. Refs.~\cite{moch2,moch3}. This shows that the renormalization scale uncertainty is greatly suppressed and essentially eliminated using PMC at the NNLO level.

\item We can analyze the combined PDF and $\alpha_s$ uncertainty by using different CTEQ PDF sets, i.e. CT10~\cite{cteq}, which are global fits of experimental data with varying $\alpha_s(m_Z) \in [0.113, 0.230]$. As for the total cross-section after PMC scale setting, we obtain
 \begin{eqnarray}
 \sigma_{\rm Tevatron,\;1.96\,TeV} &=& 7.626^{+0.705}_{-0.610} \;{\rm pb}\\
 \sigma_{\rm LHC,\;7\,TeV} &=& 171.8^{+19.5}_{-16.2} \;{\rm pb}\\
 \sigma_{\rm LHC,\;14\,TeV} &=& 941.3^{+83.3}_{-77.1} \;{\rm pb}
 \end{eqnarray}
 where the errors are caused by the PDF+$\alpha_s$ uncertainty. Here a larger PDF+$\alpha_s$ error than that of Refs.~\cite{moch3,beneke2} is due to the choice of PDFs with a wider range of $\alpha_s(m_Z)$. If taking the present world average $\alpha_s(m_Z)\simeq 0.118\pm0.001$~\cite{pdg}, we will obtain a much smaller PDF+$\alpha_s$ error; i.e.
\begin{eqnarray}
 \sigma_{\rm Tevatron,\;1.96\,TeV} &=& 7.626^{+0.143}_{-0.130} \;{\rm pb}\\
 \sigma_{\rm LHC,\;7\,TeV} &=& 171.8^{+3.8}_{-3.5} \;{\rm pb}\\
 \sigma_{\rm LHC,\;14\,TeV} &=& 941.3^{+14.6}_{-15.6} \;{\rm pb}
 \end{eqnarray}

\begin{figure}[tb]
\begin{center}
\begin{minipage}[t]{14 cm}
\epsfig{file=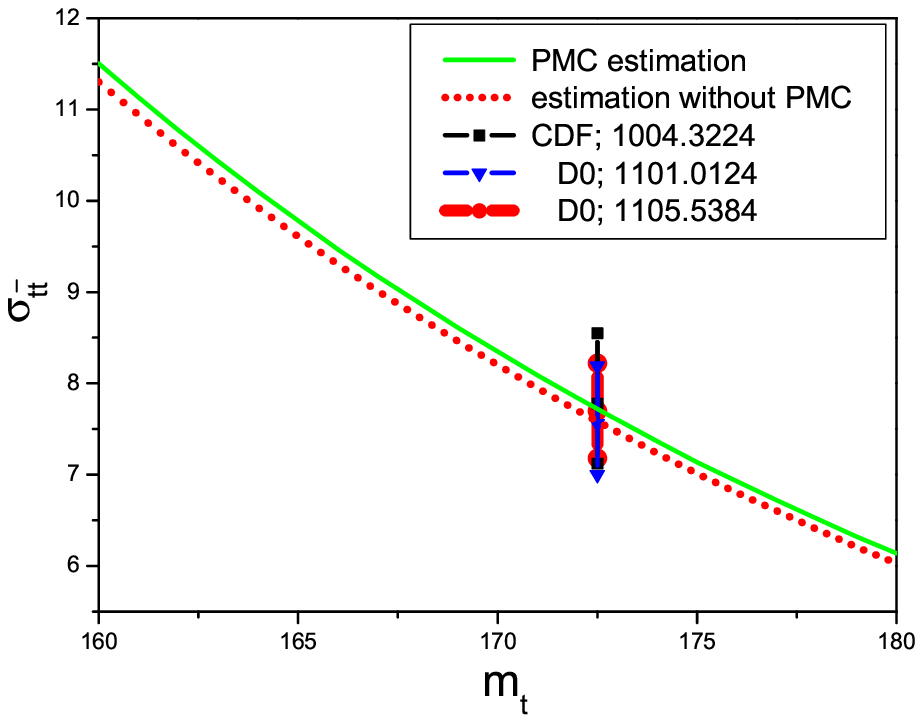,scale=0.7}
\epsfig{file=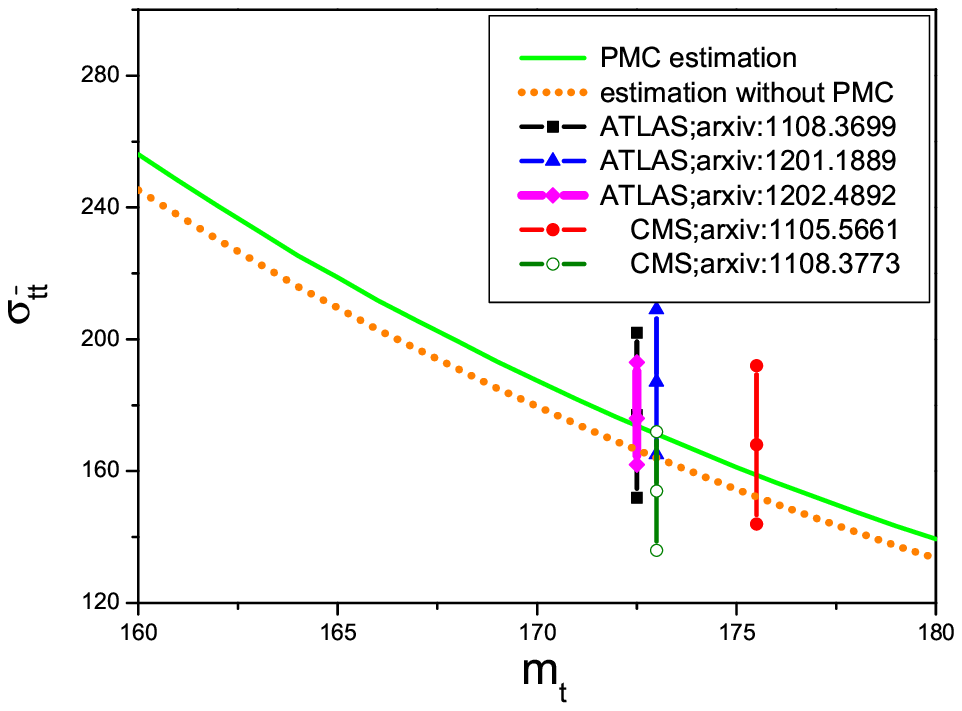,scale=0.7}
\end{minipage}
\begin{minipage}[t]{16.5 cm}
\caption{Total cross-section $\sigma_{t\bar{t}}$ for the top-pair production, with or without PMC scale setting, versus top-quark mass. The experimental data are adopted from Refs.~\cite{cdft,d0t,atlas,cms}. \label{mass}}
\end{minipage}
\end{center}
\end{figure}

\item The total cross-section $\sigma_{t\bar{t}}$ is sensitive to the top-quark mass, and it is found that the total cross-sections decrease with the increment of top-quark mass. After PMC scale setting, by varying $m_t=172.9\pm 1.1$ GeV~\cite{pdg}, we predict
 \begin{eqnarray}
 \sigma_{\rm Tevatron,\;1.96\,TeV} &=& 7.626^{+0.265}_{-0.257} \;{\rm pb}\\
 \sigma_{\rm LHC,\;7\,TeV} &=& 171.8^{+5.8}_{-5.6} \;{\rm pb}\\
 \sigma_{\rm LHC,\;14\,TeV} &=& 941.3^{+28.4}_{-26.5} \;{\rm pb}
 \end{eqnarray}
 where the errors are caused by the top-quark mass uncertainty. In Fig.(\ref{mass}) we present the total cross-section $\sigma_{t\bar{t}}$ as a function of $m_t$, where $\sigma_{t\bar{t}}$ with or without PMC scale setting are shown explicitly. After PMC scale setting, the value of $\sigma_{t\bar{t}}$ is more closer to the central value of the experimental data~\cite{cdft,d0t,atlas,cms}, which shows a better agreement with the experimental data.

\end{itemize}

As a summary, after PMC scale setting, we obtain the following points:
\begin{itemize}
\item A larger value for $\sigma_{t\bar{t}}$ is obtained, which agrees with the present Tevatron and LHC experimental data well. This is achieved because we have resummed the universal and gauge invariant higher-order corrections which are associated with the running of the coupling into the LO- and NLO- terms by using suitable PMC scales.

\item After PMC scale setting, a more convergent pQCD series expansion is obtained and the resulting LO- and NLO- terms are conformally invariant and do not depend on the choice of renormalization scheme. The slight change of PMC scales will lead to large effects due to the explicit breaking of the conformal invariance~\cite{pmc3}.

\item In principle, the PMC scale and the resulting renormalized amplitude is independent of the choice of the initial renormalization scale. The residual scale-uncertainty will be greatly suppressed when the PMC scales have been set suitably. Then, the usual renormalization scale uncertainty $\Delta\sigma_{t\bar{t}}/\sigma_{t\bar{t}} \sim\left({}^{+3\%}_{-4\%}\right)$ at the Tevatron and LHC is greatly suppressed or even eliminated by PMC scale setting.

\end{itemize}

\subsection{\it Top-Quark Pair Backward-Forward Asymmetry}

The top-quark pair forward-backward asymmetry which originates from charge asymmetry physics~\cite{cskim,Kuhn} has been studied at the Tevatron and LHC. Two options for the asymmetry have been used for experimental analysis; i.e. the $t\bar{t}$-rest frame asymmetry
\begin{equation} \label{Afbtt}
A_{FB}^{t\bar{t}}= \frac{\sigma(y^{t\bar{t}}_t > 0) - \sigma(y^{t\bar{t}}_t < 0)}{\sigma(y^{t\bar{t}}_t > 0) + \sigma(y^{t\bar{t}}_t < 0)}
\end{equation}
and the $p\bar{p}$-laboratory frame asymmetry
\begin{equation} \label{Afblab}
A_{FB}^{p\bar{p}}= \frac{\sigma(y^{p\bar{p}}_{t} > 0) - \sigma(y^{p\bar{p}}_{t} < 0)}{\sigma(y^{p\bar{p}}_{t} > 0) + \sigma(y^{p\bar{p}}_{t} < 0)} ,
\end{equation}
where $y^{t\bar{t}}_t$ ($y^{p\bar{p}}_{t}$) is the top quark rapidity in the $t\bar{t}$-rest frame ($p\bar{p}$-laboratory frame).

The CDF and D0 collaborations have found comparable values in the $t\bar{t}$-rest frame: $A_{FB}^{t\bar{t},{\rm CDF}}=(15.8\pm7.5)\%$~\cite{cdf2} and $A_{FB}^{t\bar{t},{\rm D0}}=(19.6\pm6.5)\%$~\cite{d0}. The asymmetry in the $p\bar{p}$-laboratory frame measured by CDF is $A_{FB}^{p\bar{p},{\rm CDF}} = (15.0\pm5.5)\%$~\cite{cdf2}. The CDF collaboration has also measured the dependence of $A^{t\bar{t}}_{FB}$ with respect to the $t\bar{t}$-invariant mass $M_{t\bar{t}}$: the asymmetry increases with $M_{t\bar{t}}$, and $A^{t\bar{t}}_{FB}(M_{t\bar{t}}>450\; {\rm GeV})=(47.5\pm11.4)\%$~\cite{cdf2}. The measured asymmetries are much larger than the usual SM estimates for the top quark forward-backward asymmetries. The NLO QCD contributions to the asymmetric $t\bar{t}$-production yield $A_{FB}^{t\bar{t}}\simeq 7\%$ and $A^{p\bar{p}}_{FB}\simeq 5\%$~\cite{zgsi,nnloasy,nnloasy2,sterman,nki2}, which are about $2\sigma$-deviation from the above measurements. For the case of $M_{t\bar{t}}>450\; {\rm GeV}$, using the MCFM program~\cite{mcfm}, one obtains $A^{t\bar{t}}_{FB} (M_{t\bar{t}}>450\; {\rm GeV})\sim 8.8\%$ which is about $3.4\sigma$-deviation from the data. A recent reevaluation of the electroweak correction raises the QCD asymmetries by at most $20\%$: i.e. $A_{FB}^{t\bar{t}} (A^{p\bar{p}}_{FB}) \sim 9\%\;(7\%)$~\cite{qedc1,qedc2} and $A^{t\bar{t}}_{FB} (M_{t\bar{t}}>450\; {\rm GeV}) \sim 12.8\%$~\cite{qedc2}. The large discrepancies between the SM estimates and the data have aroused interest, because of the possibility for probing new physics beyond the SM. However, these comparisons are based on the conventional scale setting for choosing the renormalization scale. In the following, we will show how the use of PMC can greatly improve our estimates within the SM.

\subsubsection{\it Basic Formulas}

Numerical results for the top-quark pair production at the Tevatron and LHC have been presented in Sec.\ref{topquarkCS}. We will compare the total cross-sections derived from PMC scale setting and the conventional scale setting. We emphasize two points in addition to the ones listed in the above subsection:

\begin{itemize}

\item At the Tevatron, the top-quark pair production is dominated by the $(q\bar{q})$-channel which provides about $85\%$ contribution to the total cross-section. The $(q\bar{q})$-channel due to interference of the one gluon and two gluon intermediate states is asymmetric at the NLO level, which leads to sizable top-quark forward-backward asymmetry at the Tevatron. In addition, emission of real gluons gives an asymmetry. In contrast, one finds that the dominant channel at the LHC is the symmetric $(gg)$-channel, so the top-quark forward-backward asymmetry from other channels will be greatly diluted at the LHC; i.e., this asymmetry becomes small which agrees with the CMS and ATLAS measurements~\cite{cms2,atlas2}.

\item More specifically, at the lowest order, the two channels $q\bar{q}\to t\bar{t}$ and $gg\to t\bar{t}$ do not discriminate the final top-quark and top-antiquark, so their differential distributions are symmetric for the hadronic production. At the NLO level, either the virtual or real gluon emission will cause sizable differences between the differential top-quark and top-antiquark production, thus leading to an observable top-quark forward-backward asymmetry. At the Tevatron, the asymmetric channels are $(q\bar{q})$-, $(gq)$- and $(g\bar{q})$- channels accordingly. Table \ref{tab:tev} shows the total cross-sections of the $(gq)$ and $(g\bar{q})$ channels are quite small, less than $1\%$ of that of $(q\bar{q})$-channel, so their contributions to the asymmetry can be safely neglected.

\end{itemize}

Writing the numerator and the denominator of the two asymmetries $A_{FB}$ defined by Eqs.(\ref{Afbtt},\ref{Afblab}) in powers of $\alpha_s$, we formally obtain
\begin{eqnarray}
A_{FB} &=& \frac{\alpha_s^{3} N_{1}+\alpha_s^{4} N_{2}+ {\cal O}(\alpha_s^5)}{\alpha_s^{2} D_{0} +\alpha_s^{3} D_{1}+\alpha_s^{4} D_{2} +{\cal O}(\alpha_s^5)} \nonumber\\
&=& \frac{\alpha_s}{D_{0}}\left[N_{1}+ \alpha_s\left(N_{2}-\frac{D_{1} N_{1}}{D_{0}}\right)+\alpha_s^2 \left(\frac{D_1^2 N_1}{D_0^2} -\frac{D_1 N_2}{D_0} -\frac{D_2 N_1}{D_0}\right) +\cdots \right] ,\label{afbfull}
\end{eqnarray}
where the $D_i$-terms stand for the total cross-sections at certain $\alpha_s$-order and the $N_i$-terms stand for the asymmetric cross-sections at certain $\alpha_s$-order. The terms up to NLO ($D_{0},D_{1},N_{1}$) have been calculated, whereas only parts of $D_{2}$ and $N_{2}$ are currently known~\cite{nason1,nason2,nason3,beenakker1,beenakker2,czakon1,czakon2,moch1,moch2,moch3,beneke1,beneke2,andrea,vogt,hathor}.

As shown in Table~\ref{tab:tev}, using conventional scale setting, the relative importance of the denominator terms is $\left[{\alpha_s^2} D_{0} : {\alpha_s^3} D_{1} : {\alpha_s^4} D_{2} \sim 1: 18\% : 12\% \right]$, and the numerator terms for the asymmetric $(q\bar{q})$-channel satisfy $\left[\alpha_s^{3} N_{1} : \alpha_s^{4} N_{2} \sim 1 : 50\% \right]$. Since at present the NNLO numerator term $N_2$ is not available, as a first approximation, we treat these asymmetric terms to have the same relative importance as their total cross-sections; i.e. $(\alpha_s^{3} N_{1})_{q\bar{q}} : (\alpha_s^{4} N_{2})_{q\bar{q}} \sim (\alpha_s^{3} D_{1})_{q\bar{q}} : (\alpha_s^{4} D_{2})_{q\bar{q}}$. Thus, the $N_{1}D_{1}/D_{0}$ term and the $N_{2}$ term have the same importance. Then, to be consistent, one has to keep only the first term in Eq.(\ref{afbfull}); i.e. dealing with only the so-called LO asymmetry~\cite{Kuhn,zgsi,qedc1,qedc2}: $A_{FB}=\frac{N_{1}}{D_{0}} \alpha_s$.

On the other hand, after PMC scale setting, we have $\left[{\alpha_s^2} D_{0} : {\alpha_s^3} D_{1} : {\alpha_s^4} D_{2} \sim 1: 41\% : 2\%\right]$ and the numerators for the asymmetric $(q\bar{q})$-channel becomes $\left[\alpha_s^{3} N_{1} : \alpha_s^{4} N_{2} \sim 1 : 3\% \right]$. It shows that, after PMC scale setting, the NNLO corrections for both the total cross-sections and the asymmetric part are lowered by about one order of magnitude. Therefore, the NNLO-terms $N_2$ and $D_2$ can be safely neglected in the calculation, and we can obtain an accurate asymmetry at the NNLO level:
\begin{displaymath}
\quad\quad A_{FB}=\frac{\alpha_s}{D_{0}}\left[N_{1}-\alpha_s\left(\frac{D_{1} N_{1}}{D_{0}}\right)+\alpha_s^2 \left(\frac{D_1^2 N_1}{D_0^2}\right) \right] .
\end{displaymath}
Furthermore, it is natural to assume that those higher-order terms $N_i$ and $D_i$ with $i>2$ after PMC scale setting will also give negligible contribution; the above asymmetry can thus be resummed to a more convenient form:
\begin{equation}
A_{FB}= \frac{\alpha_s^{3} N_{1}}{\alpha_s^{2} D_{0} +\alpha_s^{3} D_{1}} \;.
\end{equation}
Furthermore, as shown in Refs.~\cite{Kuhn,qedc1,qedc2}, the electromagnetic and weak contributions $\tilde{N}_{0,1}$ provides an extra $\sim 20\%$ increment for the asymmetry; thus the electromagnetic contribution provides a non-negligible fraction of the QCD-based antisymmetric cross-section with the same overall sign. Then, our final formula to calculate the asymmetry changes to
\begin{equation}\label{final}
A_{FB}= \frac{\alpha_s^{3} N_{1}+\alpha_s^{2}\alpha \tilde{N}_{1}+\alpha^2 \tilde{N}_0}{\alpha_s^{2} D_{0} +\alpha_s^{3} D_{1}} \;.
\end{equation}

Based on the above considerations, the top-quark forward-backward asymmetry after PMC scale setting can be written as
\begin{eqnarray}
A_{FB}^{t\bar{t},{\rm PMC}} &=& \frac{1} {\sigma^{\rm tot, PMC}_{H_1 H_2 \to t\bar{t}+X}(\mu^{\rm PMC}_r)} \left[\sigma_{asy}^{(q\bar{q})}\left(\mu^{\rm PMC}_r; y^{t\bar{t}}_t > 0\right)- \sigma_{asy}^{(q\bar{q})}\left(\mu^{\rm PMC}_r; y^{t\bar{t}}_t < 0\right) \right] , \\
A_{FB}^{p\bar{p},{\rm PMC}} &=& \frac{1}{\sigma^{\rm tot, PMC}_{H_1 H_2 \to t\bar{t}+X}(\mu^{\rm PMC}_r)} \left[\sigma_{asy}^{(q\bar{q})}\left(\mu^{\rm PMC}_r; y^{p\bar{p}}_{t} > 0\right) - \sigma_{asy}^{(q\bar{q})}\left(\mu^{\rm PMC}_r; y^{p\bar{p}}_{t} < 0\right) \right] ,
\end{eqnarray}
where according to Eq.(\ref{final}): $\sigma^{\rm tot}_{H_1 H_2 \to t\bar{t}+X}$ is the total hadronic cross-section up to NLO; $\sigma_{asy}^{(q\bar{q})}$ stands for the asymmetric cross-section of the $(q\bar{q})$-channel which includes the above mentioned ${\cal O}(\alpha_s^3)$, ${\cal O}(\alpha_s^2 \alpha)$ and ${\cal O}(\alpha^2)$ terms. In the denominator for the total cross-section up to NLO, for each production channel, we need to introduce two LO PMC scales which are for the Coulomb part and non-Coulomb part accordingly, and one NLO PMC scale for the non-Coulomb part. While in the numerator, we only need to know the NLO PMC scale $\mu^{\rm PMC, NLO}_r$ for the $(q\bar{q})$-channel, since it is the only asymmetric component.

It is interesting to observe that there is a dip for the NLO scale $\mu^{\rm PMC, NLO}_{r}$ of the $(q\bar{q})$-channel when $\sqrt{s} \simeq [\sqrt{2}\exp(5/6)]m_t \sim 563$ GeV, which, as shown in Sec.\ref{topquarkCS}, is caused by the correlation among the PMC coefficients for NLO and NNLO terms. More specifically, it is found that
\begin{eqnarray}
{\mu^{\rm PMC, NLO}_r} = \exp\left(\frac{\tilde{B}_{2q\bar{q}}} {\tilde{A}_{1q\bar{q}}}\right) {\mu^{\rm PMC, LO}_r} =\exp\left(\frac{\tilde{B}_{2q\bar{q}}}{\tilde{A}_{1q\bar{q}}}\right) \exp\left(\frac{3B_{1q\bar{q}}}{2A_{0q\bar{q}}}+{\cal O}(\alpha_s) \right) {\mu^{\rm init}_r},
\end{eqnarray}
where the coefficients are defined through the standard PMC scale setting, cf. Eq.(\ref{eq:pmctopstart}). As shown in Fig.(\ref{qqcoe}), the value of $\tilde{B}_{2q\bar{q}}$ is always negative and $\tilde{A}_{1q\bar{q}}$ has a minimum value at small $\sqrt{s}$. Quantitatively, the NLO PMC scale $\mu^{\rm PMC, NLO}_r$ for the $(q\bar{q})$-channel is considerably smaller than $m_t$ in small $\sqrt{s}$-region. The NLO cross-section of the $(q\bar{q})$-channel will thus be greatly increased; it is a factor of two times larger than its value derived under conventional scale setting, as shown by Table~\ref{tab:tev}.

\subsubsection{\it Numerical Analysis for the Backward-Forward Asymmetry}

The PMC asymmetries $A_{FB}^{t\bar{t},{\rm PMC}}$ and $A_{FB}^{p\bar{p},{\rm PMC}}$ can be compared with the asymmetries calculated using conventional scale setting. For definiteness, we apply PMC scale setting to improve Hollik and Pagani's results~\cite{qedc2}, and we obtain
\begin{eqnarray} \label{pmcr1}
A_{FB}^{t\bar{t},{\rm PMC}}&=& \left\{\frac{\sigma^{\rm tot, HP}_{H_1 H_2 \to t\bar{t}X} } {\sigma^{\rm tot, PMC}_{H_1 H_2 \to t\bar{t}X}} \right\} \left\{ \frac{{\overline{\alpha}_s}^3\left(\overline{\mu}^{\rm PMC, NLO}_r\right)} {{\alpha^{HP}_s}^3 \left(\mu^{\rm conv}_r\right)} A_{FB}^{t\bar{t},{\rm HP}}|_{{\cal O}(\alpha_s^3)} + \frac{{\overline{\alpha}_s}^2\left(\overline{\mu}^{\rm PMC, NLO}_r\right)} {{\alpha^{HP}_s}^2 \left(\mu^{\rm conv}_r\right)} A_{FB}^{t\bar{t},{\rm HP}}|_{{\cal O}(\alpha_s^2\alpha)}+ A_{FB}^{t\bar{t},{\rm HP}}|_{{\cal O}(\alpha^2)} \right\} \\
A_{FB}^{p\bar{p},{\rm PMC}}&=& \left\{\frac{\sigma^{\rm tot, HP}_{H_1 H_2 \to t\bar{t}X} } {\sigma^{\rm tot, PMC}_{H_1 H_2 \to t\bar{t}X}} \right\} \left\{ \frac{{\overline{\alpha}_s}^3\left(\overline{\mu}^{\rm PMC, NLO}_r\right)} {{\alpha^{HP}_s}^3 \left(\mu^{\rm conv}_r\right)} A_{FB}^{p\bar{p},{\rm HP}}|_{{\cal O}(\alpha_s^3)} + \frac{{\overline{\alpha}_s}^2\left(\overline{\mu}^{\rm PMC, NLO}_r\right)} {{\alpha^{HP}_s}^2 \left(\mu^{\rm conv}_r\right)} A_{FB}^{p\bar{p},{\rm HP}}|_{{\cal O}(\alpha_s^2\alpha)}+ A_{FB}^{p\bar{p},{\rm HP}}|_{{\cal O}(\alpha^2)} \right\} \label{pmcr2}
\end{eqnarray}
Here $\mu^{\rm conv}_r$ stands for the scale set by conventional scale setting and the symbol HP stands for the corresponding values of Ref.~\cite{qedc2}; i.e. for $\mu^{\rm conv}_r =m_t$, it shows~\cite{qedc2}: $\sigma^{\rm tot, HP}_{H_1 H_2 \to t\bar{t}X} = 5.621 \;{\rm pb}$ and
\begin{eqnarray}
&& A_{FB}^{t\bar{t},{\rm HP}}|_{{\cal O}(\alpha_s^3)}=7.32\% \;\;\; A_{FB}^{t\bar{t},{\rm HP}}|_{{\cal O}(\alpha_s^2\alpha)}=1.36\% \;\;\; A_{FB}^{t\bar{t},{\rm HP}}|_{{\cal O}(\alpha^2)}=0.26\% \nonumber\\
&& A_{FB}^{p\bar{p},{\rm HP}}|_{{\cal O}(\alpha_s^3)}=4.85\% \;\;\; A_{FB}^{p\bar{p},{\rm HP}}|_{{\cal O}(\alpha_s^2\alpha)}=0.90\% \;\;\; A_{FB}^{p\bar{p},{\rm HP}}|_{{\cal O}(\alpha^2)}=0.16\% \nonumber
\end{eqnarray}
where
\begin{itemize}
\item $A_{FB}^{t\bar{t},{\rm HP}}|_{{\cal O}(\alpha_s^3)}$ and $A_{FB}^{p\bar{p},{\rm HP}}|_{{\cal O}(\alpha_s^3)}$ stand for the pure QCD asymmetry at the $\alpha_s^3$-order under the $t\bar{t}$-rest frame and the $p\bar{p}$ lab frame, respectively.
\item $A_{FB}^{t\bar{t},{\rm HP}}|_{{\cal O}(\alpha_s^2\alpha)}$ and $A_{FB}^{p\bar{p},{\rm HP}}|_{{\cal O}(\alpha_s^2\alpha)}$ stand for the combined QED and weak with the QCD asymmetry at the $\alpha_s^2 \alpha$-order under the $t\bar{t}$-rest frame and the $p\bar{p}$ lab frame, respectively.
\item $A_{FB}^{t\bar{t},{\rm HP}}|_{{\cal O}(\alpha^2)}$ and $A_{FB}^{p\bar{p},{\rm HP}}|_{{\cal O}(\alpha^2)}$ stand for the pure electroweak asymmetry at the $\alpha^2$-order under the $t\bar{t}$-rest frame and the $p\bar{p}$ lab frame, respectively.
\end{itemize}

\begin{table}
\begin{center}
\begin{minipage}[tb]{16.5 cm}
\caption{Total cross-sections (in unit: pb) for the top-quark pair production at the Tevatron with $p\bar{p}$-collision energy $\sqrt{s}=1.96$ TeV. For conventional scale setting, we set the scale $\mu_r\equiv Q$. For PMC scale setting, we set the initial scale $\mu^{\rm init}_r=Q$ and then apply the PMC procedure. Here we take $Q=m_t=172.9$ GeV and use the MSRT 2004-QED parton distributions~\cite{mrst} as the PDF. }
\label{tab:newtev}
\end{minipage}
\begin{tabular}{|c||c|c|c|c||c|c|c|c|}
\hline
& \multicolumn{4}{c||}{Conventional scale setting} & \multicolumn{4}{c|}{PMC scale setting} \\
\hline
~ ~ &~LO~ &~NLO~ &~NNLO~ &~ {\it total} ~&~LO~ &~NLO~ &~NNLO~ &~ {\it total} ~\\
\hline
$(q\bar{q})$-channel & 4.890 & 0.963 & 0.483 & 6.336 & 4.748 & 1.727 & -0.058 & 6.417 \\
$(gg)$-channel & 0.526 & 0.440 & 0.166 & 1.132 & 0.524 & 0.525 & 0.160 & 1.208 \\
$(gq)$-channel & 0.000 &-0.0381 & 0.0049& -0.0332 & 0.000 & -0.0381 & 0.0049 & -0.0332 \\
$(g\bar{q})$-channel & 0.000 &-0.0381 & 0.0049& -0.0332 & 0.000 & -0.0381 & 0.0049 & -0.0332 \\
sum  & 5.416 & 0.985 & 0.659 & 7.402 & 5.272 & 2.176 & 0.112 & 7.559 \\
\hline
\end{tabular}
\end{center}
\end{table}

Total cross-sections for the top-quark pair production at the Tevatron with $p\bar{p}$-collision energy $\sqrt{s}=1.96$ TeV and with the same parameters of Ref.~\cite{qedc2} are given in Table~\ref{tab:newtev}. In the formulas (\ref{pmcr1},\ref{pmcr2}), we have defined an effective coupling ${\overline{\alpha}_s} \left(\overline{\mu}^{\rm PMC, NLO}_r \right)$ for the asymmetric part, which is the weighted average of the QCD coupling for the $(q\bar{q})$-channel; i.e. in using the effective coupling ${\overline{\alpha}_s} \left(\overline{\mu}^{\rm PMC, NLO}_r \right)$, one obtains the same $(q\bar{q})$-channel NLO cross-section as that of ${\alpha}_s (\mu^{\rm PMC, NLO}_r )$\footnote{In principle, one could divide the cross-sections into symmetric and asymmetric components and find PMC scales for each of them. For this purpose, one needs to identify the $n_f$-terms or the $n_f^2$-terms for both the symmetric and asymmetric parts at the NNLO level separately. }.

\begin{figure}[tb]
\begin{center}
\begin{minipage}[t]{10 cm}
\epsfig{file=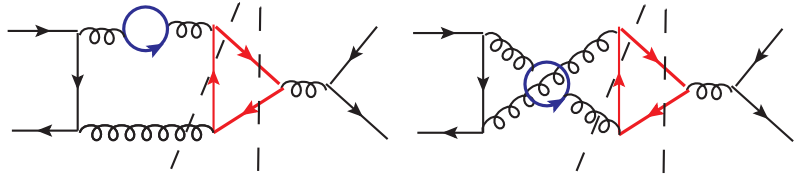,scale=0.8}
\end{minipage}
\begin{minipage}[t]{16.5 cm}
\caption{Dominant cut diagrams for the $n_f$-terms at the $\alpha^4$-order of the $(q\bar{q})$-channel, which are responsible for the smaller effective NLO PMC scale $\overline{\mu}^{\rm PMC, NLO}_r$, where the solid circles stand for the light-quark loops. \label{qqnloscale}}
\end{minipage}
\end{center}
\end{figure}

\begin{figure}[tb]
\begin{center}
\begin{minipage}[t]{14 cm}
\epsfig{file=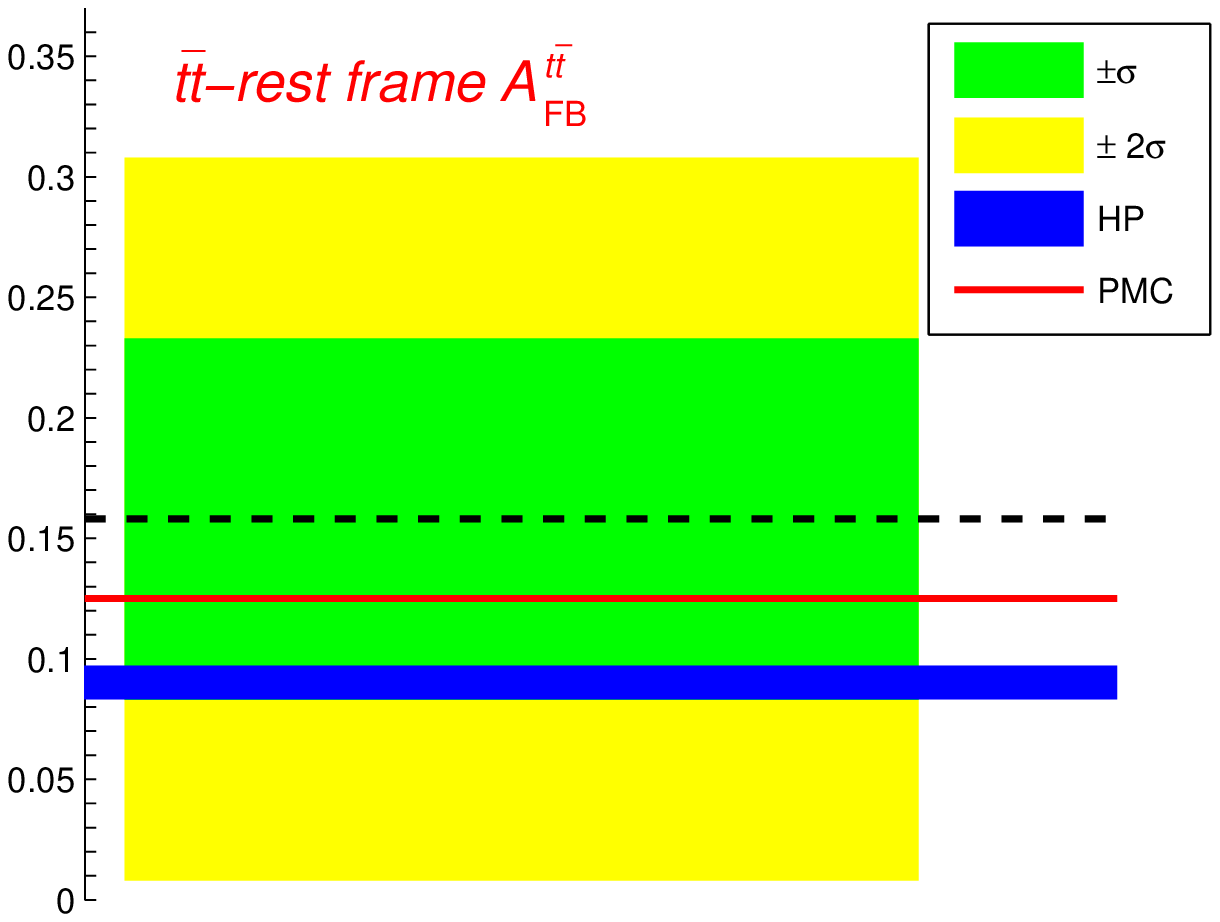,scale=0.5}
\epsfig{file=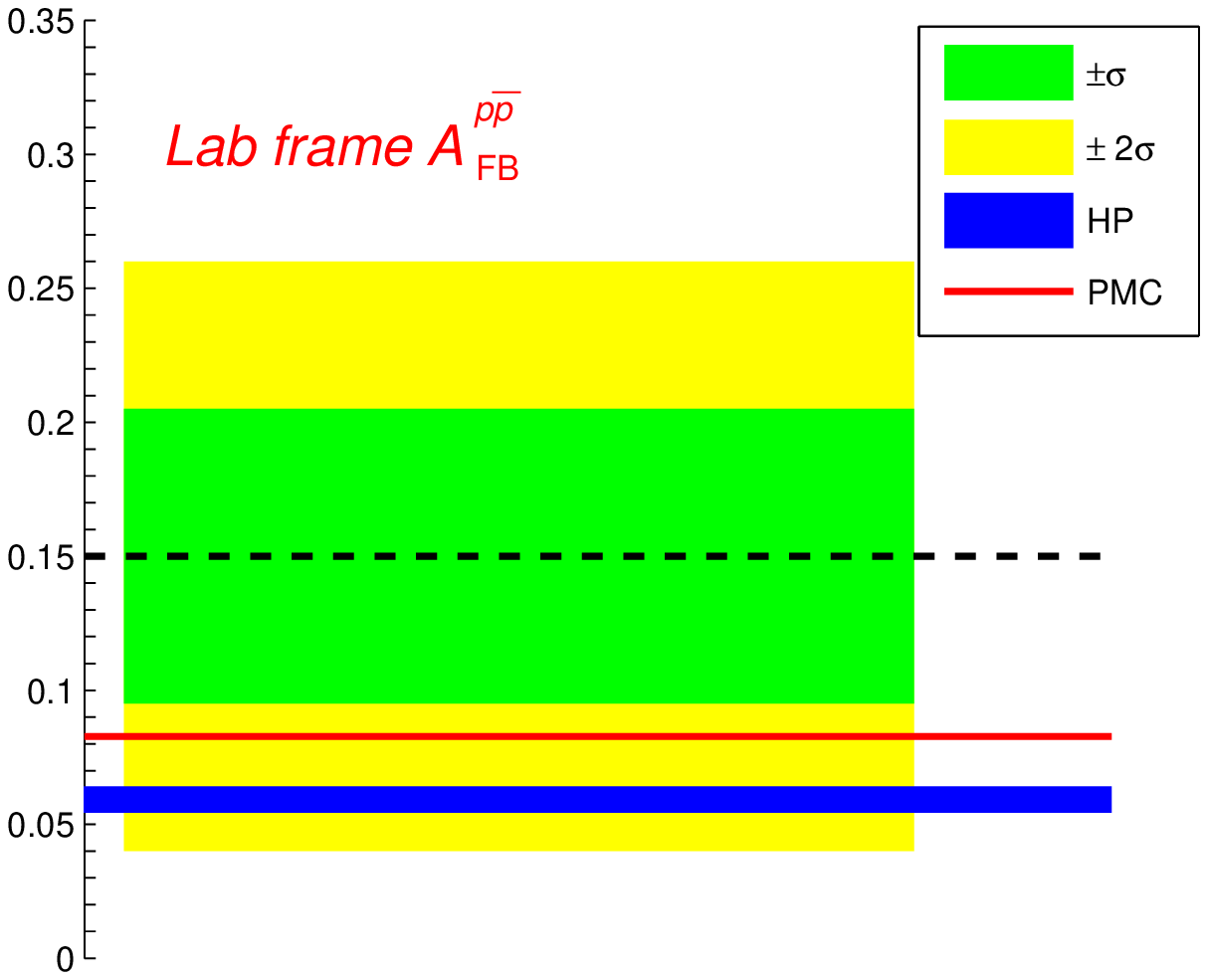,scale=0.5}
\end{minipage}
\begin{minipage}[t]{16.5 cm}
\caption{Comparison of the PMC prediction with the CDF data~\cite{cdf2} for the $t\bar{t}$-pair forward-backward asymmetry for the whole phase-space. The Hollik and Pagani's results (HP)~\cite{qedc2} using conventional scale setting are presented for a comparison. The result for D0 data~\cite{d0} shows a similar behavior. \label{pmcasy}}
\end{minipage}
\end{center}
\end{figure}

It is noted that the NLO-level asymmetric part for $(q\bar{q})$-channel only involves the NLO PMC scale for the non-Coulomb part, so the effective coupling ${\overline{\alpha}_s}\left(\overline{\mu}^{\rm PMC, NLO}_r\right)$ can be unambiguously determined. We obtain a smaller effective NLO PMC scale $
\overline{\mu}^{\rm PMC, effective}_r \simeq \exp(-9/10)m_t\sim 70$ GeV, which corresponds to ${\overline{\alpha}_s}\left(\overline{\mu}^{\rm PMC, NLO}_r\right)=0.1228$. It is larger than ${\alpha^{HP}_s}\left(m_t\right)\simeq 0.098$~\cite{qedc1,qedc2}. This effective NLO PMC scale is dominated by the non-Coulomb $n_f$-terms at the $\alpha_s^4$-order, which are shown in Fig.(\ref{qqnloscale}). In these diagrams, the momentum flow in the virtual gluons possess a large range of virtualities. This effect for NLO PMC scale $\overline{\mu}^{\rm PMC, effective}_r$ can be regarded as a weighted average of these different momentum flows in the gluons, which can be softer than the nominal scale, $m_t$. Finally, we obtain
\begin{equation}
A_{FB}^{t\bar{t},{\rm PMC}} \simeq 12.7\% \; ;\;\; A_{FB}^{p\bar{p},{\rm PMC}} \simeq 8.39\% \;.
\end{equation}
Thus after PMC scale setting, the top-quark asymmetry under the conventional scale setting is increased by $\sim 42\%$ for both the $t\bar{t}$-rest frame and the $p\bar{p}$-laboratory frame. This large improvement is explicitly shown in Fig.(\ref{pmcasy}), where Hollik and Pagani's results which are derived under conventional scale setting~\cite{qedc2} are presented for comparison.

Another possible effect from QCD can be the lensing effect of the final state interactions of the $t$ and $\bar{t}$ with the beam spectators. The same diagrams causes Sivers single-spin asymmetry and diffractive deep inelastic scattering\footnote{We thanks Benjamin von Harling and Yue Zhao for conversions on this possibility.}.

\begin{figure}[htb]
\begin{center}
\begin{minipage}[t]{9 cm}
\epsfig{file=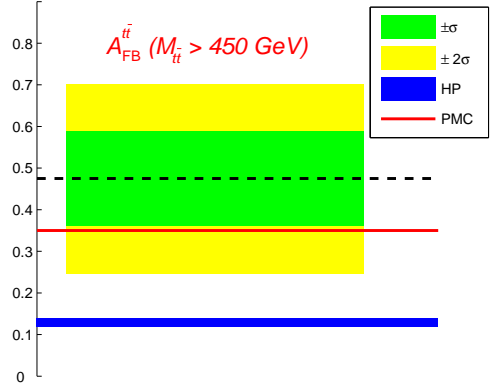,scale=0.55}
\end{minipage}
\begin{minipage}[t]{16.5 cm}
\caption{The PMC prediction of $A_{FB}^{t\bar{t}}(M_{t\bar{t}}>450\; {\rm GeV})$ and the corresponding CDF data~\cite{cdf2} for the $t\bar{t}$-pair forward-backward asymmetry for $M_{t\bar{t}}>450$ GeV. The Hollik and Pagani's results (HP)~\cite{qedc2} using conventional scale setting are presented for a comparison. \label{asycut}}
\end{minipage}
\end{center}
\end{figure}

The CDF collaboration has found that when the $t\bar{t}$-invariant mass, $M_{t\bar{t}}>450$ GeV, the top-quark forward-backward asymmetry $A^{t\bar{t}}_{FB}(M_{t\bar{t}}>450 \;{\rm GeV})$ is about $3.4$ standard deviations above the SM asymmetry prediction under the conventional scale setting~\cite{mcfm}. After applying PMC scale setting, we have $\sigma^{\rm tot, PMC}_{H_1 H_2 \to t\bar{t}X}(M_{t\bar{t}}>450\; {\rm GeV})=2.406$ pb and $ {\overline{\alpha}_s}\left(\overline{\mu}^{\rm PMC, NLO}_r\right)=0.1460$ with $\overline{\mu}^{\rm PMC, NLO}_r \sim\exp(-19/10)m_t \simeq 26$ GeV. Then, we obtain
\begin{equation}
 A^{t\bar{t},PMC}_{FB}(M_{t\bar{t}}>450\; {\rm GeV})\simeq 35.0\% \;,
\end{equation}
which is increased by about $1.7$ times of the previous one $A_{FB}^{t\bar{t},{\rm HP}}(M_{t\bar{t}}>450\; {\rm GeV})=12.8\%$~\cite{qedc2}. Our present prediction is only about $1\sigma$-deviation from the CDF data, which is shown in Fig.(\ref{asycut}). This shows that, after PMC scale setting, the discrepancies between the SM estimate and the present CDF and D0 data are greatly reduced.

\subsection{\it Sum rules for special moments of the deep-inelastic structure functions}

Deep-inelastic structure functions obey a series of sum rules for special
moments, such as the Adler sum rule~\cite{Adler:1965ty}, the unpolarized Bjorken sum rule~\cite{Bjorken:1967px}, the Gross-Llewellyn Smith (GLS) sum rule~\cite{Gross:1969jf}, the polarized Bjorken sum rule~\cite{Bjorken:1969mm}, the Gerasimov-Drell-Hearn sum rule~\cite{Gerasimov:1965et,Drell:1966jv}, the Burkhardt-Cottingham sum rule~\cite{Burkhardt:1970ti}, the Efremov-Teryaev-Leader sum rule~\cite{Efremov:1996hd}, the Ellis-Jaffe sum rule~\cite{Ellis:1973kp}, the Gottfried sum rule~\cite{Gottfried:1967kk} and etc., all of which are of interest for experimental tests. A brief description of those sum rules can be found in Ref.\cite{mainsumrules}. The Bjorken sum rule and the GLS sum rule obey the well-known Crewther relations \cite{crewther1,crewther2,crewther3,crewther4,Baikov:2010je, Baikov:2012zn,crewther5,crewther6,crewther7}, which through the Adler function can be used to expose their conformal parts. As has been shown in Sec.\ref{csrQCD}, one can obtain the generalized Crewther relation by using the commensurate scale relation among them. It is shown~\cite{BMW2} that both sum rules after PMC scale setting have perturbative expansions that match exactly the inverse of the anomalous dimension, $\gamma^{-1}$, and is what one expects in a conformal field theory.

The Bjorken sum rule expresses the integral over the spin distributions of quarks inside of the nucleon in terms of its axial charge times a coefficient function ${C}^{Bjp}$:
\begin{eqnarray}
\Gamma_1^{p-n}(Q^2) = \int_0^1 [g_1^{ep}(x,Q^2)-g_1^{en}(x,Q^2)]dx =\frac{g_A}{6}
C^{Bjp}(a) + \sum_{i=2}^{\infty}\frac{\mu_{2i}^{p-n}(Q^2)}{Q^{2i-2}} \ ,
\label{gBSR}
\end{eqnarray}
where $g_1^{ep}$ and $g_1^{en}$ are the spin-dependent proton and neutron structure functions, $g_A$ is the nucleon axial charge as measured in neutron $\beta$ decays. The sum in the second line of Eq.(\ref{gBSR}) describes the nonperturbative power corrections (higher twists) which are inaccessible for pQCD. Focusing on the perturbative part, we define
\begin{equation}
{C}^{Bjp}(Q^2) = 1  - 3 \,C_F\, a(Q^2) + \sum_{n=2}^{\infty} \  {\tilde{C}}_n\, a(Q^2)^n \ .
\end{equation}

The GLS sum rule,
\begin{equation}
\frac{1}{2}\int_0^1 F_3^{\nu p + \bar\nu p}(x,Q^2) dx = 3\, n_f C^{GLS}(a) \ ,
\end{equation}
relates the lowest moment of the isospin singlet structure function $F_3^{\nu p + \bar\nu p}(x,Q^2)$ to a coefficient $C^{CLS}(a_s)$, which appears in the operator product expansion of the axial and vector non-singlet currents.
We are again only considering the perturbative contribution and define
\begin{equation}
{C}^{GLS}(Q^2) = 1 - 3 \,C_F\, a(Q^2) +\sum_{n=2}^{\infty}{C}_n a(Q^2)^n .
\end{equation}

The Adler function can be written in terms of the vector field anomalous dimension, $\gamma$, and the vacuum polarization function, $\Pi$, as follows \cite{Chetyrkin:1996ia,Baikov:2012zm}
\begin{equation}
\label{Adler}
\bar{D}(Q^2) = \kappa^{-1} D(Q^2) =  \gamma(a) - \beta(a) \frac{d}{da} \Pi(Q^2,a) \ .
\end{equation}
where $\beta(a)$ is the $\beta$-function of the running coupling and we defined the normalized Adler function $\bar{D}$ with $\kappa = d_F \sum_f Q_f^2$ and $d_F$ is the dimension of the quark color representation. The (generalized) Crewther relation \cite{crewther4,crewther5,crewther6,crewther7} states that there exist a relation between the two sum rules through the Adler function $D(Q^2)$ as follows :
\begin{eqnarray}
\tilde{\bar{D}}(Q^2)\, C^{Bjp}(a) &=&  1+ \frac{\beta(a)}{a}\, \tilde{K}(a) ~,
\label{gCrewtherNS} \\
\tilde{K}(a) &=& a\,\tilde{K}_1 + a^2\,\tilde{K}_2 +a^3\,\tilde{K}_3 + \dots
\end{eqnarray}
and
\begin{eqnarray}
{ \displaystyle \bar{D}(Q^2)\,  C^{GLS}(a)} \label{gCrewtherFull}
&=&  1 + \frac{\beta(a)}{a}\, K(a) \ , \\
K(a) &=& a\,K_1 + a^2\,K_2 +a^3\,K_3 + \dots
\end{eqnarray}
The tilde on $\bar{D}$ and $K$ indicates the expressions without the light-by-light type terms, and the term proportional to the $\beta$-function describes the deviation from the limit of exact conformal invariance, with the deviations starting at order $a^2$.

Both sum rules have been explicitly computed to four loops and shown to obey the generalized Crewther relations~\cite{Baikov:2010je, Baikov:2012zn} \footnote{There is a recent claim \cite{Larin:2013yba} that the existing four-loop coefficient of the Bjorken sum rule \cite{Baikov:2010je, Baikov:2012zn} is missing some singlet-diagram contributions. This is relevant only for the explicit evaluation of $\tilde{K}_3$, and does not change the results of this section.}. We can use the Crewther relations to extract the conformal and non-conformal parts of $C^{Bjp}$ and $C^{GLS}$~\cite{BMW2}. Denoting the power expansion of $\bar{D}$ by
\begin{equation}
\label{expansion}
\bar{D}(Q^2) = 1 +  \sum_{n=1}^\infty d_n a(Q^2)^n \ ,
\end{equation}
and expanding its inverse perturbatively gives us
\begin{eqnarray}
&& C^{GLS}(a) = 1-d_1 a+ a^2 \left[d_1^2-d_2-\beta _0 K_1\right] + a^3 \left[2d_1 d_2 -d_1^3-d_3+\beta _0 \left(d_1 K_1 -K_2\right) -\beta _1 K_1\right] \nonumber\\
&& +a^4 \left[d_1^4+d_2^2-d_4-3 d_1^2d_2 +2d_1 d_3+\beta _1 \left(d_1 K_1 -K_2\right) +\beta _0 \left(-d_1^2 K_1+d_1 K_2 +d_2 K_1-K_3\right) -\beta _2 K_1 \right]
\end{eqnarray}
The expression for $C^{Bjp}$ is the same after putting tildes on the coefficients. The $d_i$ are given in terms of $\gamma_i$, $\Pi_i$ and $\beta_i$ as follows:
\begin{eqnarray}
d_1 &=& \gamma_1 = 3 C_F \\
d_{i\geq 2} &=& \gamma_i + \sum_{k=0}^{i-2} (i-1-k) \beta_k \Pi_{i-1-k} \ .
\end{eqnarray}
We use this to find the degenerate $r_{i,j}$ coefficients of Eq.(\ref{betapattern}).
\begin{eqnarray}
r_{2,1} &=& - K_1 - \Pi_1 \ , \\
r_{3,1} &=& - \frac{K_2}{2} - \Pi_2+ \left ( \frac{K_1}{2} + \Pi_1\right) \gamma_1  \ , \\
r_{4,1} &=& -\frac{K_3}{3} -\Pi _3 +(K_2 + 4 \Pi_2)\frac{\gamma_1}{3}
-\left ( \frac{K_1}{3} + \Pi_1 \right ) \gamma_1^2+( K_1+2 \Pi_1 ) \frac{\gamma_2}{3} \\
r_{4,2} &=& \frac{1}{3} ( K_1 \Pi_1 + \Pi_1^2)  \\
r_{3,2} &=& 0 \ , \quad r_{4,3} = 0
\end{eqnarray}
The degeneracy allows us to resum the series as described earlier. The final result is:
\begin{eqnarray}
&C^{GLS}(a) = 1-a(Q_1) \gamma _1+a(Q_2)^2 \left(\gamma _1^2-\gamma _2\right)
+a(Q_3)^3 \left(-\gamma _1^3+2 \gamma _2 \gamma _1-\gamma _3\right) \nonumber \\
&+a(Q_4)^4 \left(\gamma _1^4-3 \gamma _2 \gamma _1^2+2 \gamma _3
  \gamma _1+\gamma _2^2-\gamma _4\right) +{\cal O}\left(a^5\right) \ ,
\end{eqnarray}
exposing the $r_{i,0}$ coefficients. This expression is simply the inverse of the anomalous dimension:
\begin{equation}
C^{GLS}(a) = \gamma^{-1}(Q_1, Q_2, Q_3,  \ldots) \ ,
\end{equation}
where we used that $\gamma_0 = 1$. The argument of $\gamma^{-1}$ on the right hand side indicate the effective scales at each order in perturbation theory, once the inverse is Taylor expanded. All the above expressions apply to the Bjorken sum rules, with the coefficients replaced by the ones with tilde. In particular,
\begin{displaymath}
C^{Bjp}(a) = \tilde{\gamma}^{-1}(\tilde{Q}_1, \tilde{Q}_2, \tilde{Q}_3, \ldots) .
\end{displaymath}

Since, the Adler function itself is after PMC scale setting simply given by the anomalous dimension:
\begin{equation}
D(Q) = \gamma(Q_1, Q_2,Q_3,  \ldots )
\end{equation}
and correspondingly for $\tilde{D}$, the Crewther relations can be expressed as \begin{eqnarray}
\tilde{\bar{D}}(\tilde{Q}) C^{Bjp}(\mu) &= \frac{\tilde{\gamma}(\tilde{Q}_1, \tilde{Q}_2, \ldots )}{\tilde{\gamma}(\tilde{\mu}_1, \tilde{\mu}_2, \ldots)} = 1 \\
{\bar{D}}(Q) C^{GLS}(\mu) &= \frac{\gamma({Q}_1, {Q}_2, \ldots )}{{\gamma}(\mu_1, \mu_2, \ldots)} = 1
\end{eqnarray}
where the last equality follows due to conformality.

\section{Summary}
\label{secVI}

Because of the RG invariance (\ref{inv-scale},\ref{inv-sch}), the predictions for a physical observable must be independent of the renormalization scheme and the initial scale. The results cannot depend on which scheme the theorist chooses; e.g. $\overline{MS}$-scheme, MOM-scheme, etc. Note that the conventional $\overline{MS}$-scheme is somewhat artificial. One can introduce a more general $MS$-like renormalization scheme, ${\cal R}_\delta$-scheme, by further absorbing an arbitrary constant $\delta$ into ${1}/{\epsilon}$ pole, i.e. $\frac{1}{\epsilon} +\ln(4 \pi) - \gamma_E-\delta$. Physical results cannot depend on the choice of $\delta$.

At a fixed-order the dependence on the renormalization scheme and initial scale choice leads to large uncertainties for perturbative QCD predictions. The problem is compounded in multi-scale processes.
The conventional scale setting procedure assigns an arbitrary range and an arbitrary systematic error to fixed-order pQCD predictions. As we have discussed in this review, this {\it ad hoc} assignment of the range and associated systematic error is unnecessary and can be eliminated by a proper scale setting as the PMC.

The extended RG equations, which includes the dependence on the scheme parameters, provide a convenient way for estimating both the scheme and scale dependence of the perturbative predictions for a physical process. It provides a way for the running coupling to run reliably either in scale or in scheme. With the help of the extended RG equations, we have presented a general demonstration for the RG invariance. Furthermore, this formalism provides a platform for a reliable error analysis, and it also provides a precise definition for the QCD asymptotic scale under any renormalization ${\cal R}$-scheme, $\Lambda^{'tH-{\cal R}}_{QCD}$, which is defined as the pole of the strong coupling in the 't Hooft scheme associated with ${\cal R}$-scheme.

Several scale setting methods have been proposed in the literature: FAC, PMS, BLM and PMC. The FAC (Fastest Apparent Convergence) use the scale to contract the prediction to one term. The PMS (Principle of Minimum Sensitivity) chooses the scale at the point of minimum variation. The BLM (Brodsky-Lepage-Mackenzie) and PMC (Principle of Maximum Conformality) procedures shift all $\{\beta_i\}$-terms into the argument of the running coupling. Based on the extended RG equation, we have discussed the self-consistency conditions for a scale setting method, which include the {\it existence} and {\it uniqueness} of the renormalization scale, {\it reflexivity}, {\it symmetry}, and {\it transitivity}. These properties are natural requirements of RG invariance. We have shown that the FAC and BLM/PMC satisfy these requirements, whereas the PMS does not. The PMS is designed to be renormalization-scheme independent; however it violates the {\it symmetry} and {\it transitivity} properties of the renormalization group, and does not reproduce the {\rm Gell Mann-Low} scale for QED observables. In addition, the application of PMS to jet production from the $e^+e^- \to q \bar{q} g$ gives unphysical results; i.e. its PMS scale rises without bound for small jet energy, since it sums physics into the running coupling not associated with renormalization. This implies the necessity of further careful studies of the theoretical principles underlying PMS.

Among these scale setting methods, the advantages of PMC are clear. In PMC, the same procedure is valid for both space-like and time-like arguments; in particular, this leads to a well-behaved perturbative expansion, since all the large $\{\beta_i\}$-dependent terms on the time-like side involving $\pi^{2}$-terms are fully absorbed into the running coupling. Through the PMC - BLM correspondence, the PMC and the BLM are equivalent to each other. Thus the features of BLM scale setting are also adaptable to PMC.

For convenience, we summarize the dominant features of PMC in the following:
\begin{itemize}
\item It keeps the information of the higher order corrections but in a more convergent perturbative series. After PMC scale setting, the divergent ``renormalon" series with $n!$-growth disappear, so that a more convergent perturbative series is obtained.

\item Its estimation is renormalization-scheme independent, because after PMC scale setting,
\begin{itemize}
\item the resulting expressions are conformally invariant and thus do not depend on the choice of renormalization scheme;
\item one obtains the proper scale-displacements among the PMC scales derived for different schemes or conventions;
\item one also obtains CSRs connecting observables and schemes. For example, by using the PMC procedure, one can obtain the well-known one-loop displacement between the argument of the coupling in the ${\overline{MS}}$ scheme relative to the {\rm GM-L} scheme, $\alpha^{GM-L}_{em}(t)=\alpha^{\overline{MS}}_{em}(e^{-5/3}t)$~\cite{conLam1}, which ensures the estimates under the $\overline{MS}$-scheme and GM-L scheme are the same.
\end{itemize}

    These features become clear using the ${\cal R}_\delta$-scheme. The $\delta$-terms always accompany nonconformal $\beta$-dependent terms, and thus the elimination of $\delta$-terms by shifting the scale of the running coupling is equivalent to the elimination of $\{\beta_i\}$-terms. The PMC estimate can therefore also be achieved through a proper treatment of $\delta$-terms. This new way for PMC scale setting can be readily programmed for automatically setting the PMC scales to all orders~\cite{BMW,BMW2}.

\item The PMC provides a systematic way to set the optimized renormalization scale for a fixed-order calculation. In principle, the PMC needs an initial value to initialize renormalization scale. It is found that the estimates of PMC are to high accuracy independent of the initial renormalization scale; even the PMC scales themselves are in effect independent of the initial renormalization scale and are `physical' at any fixed order. This is because the PMC scale itself is a perturbative series and those unknown higher-order $\{\beta_i\}$-terms will be absorbed into the higher-order terms of the PMC scale, which is strongly power suppressed.

\item By applying the PMC scale setting to the known Gross-Llewellyn Smith sum rule and the polarized Bjorken sum rule up to four-loop level, one can improve the precision of these two sum rules, and the perturbative convergence is greatly improved. Furthermore, one can obtain the generalized Crewther relation by using the commensurate scale relation between them.

\item The Gross-Llewellyn Smith and the polarized Bjorken sum rules provide ideal platforms to check the conformal properties of the series after the PMC scale setting. More explicitly, by using the $R_\delta$-scheme, we have found that by using the results for these two sum rules up to four-loop level, both sum rules after PMC scale setting have perturbative expansions that exactly match the inverse of the anomalous dimension, $\gamma^{-1}$, which is in accordance with what one expects in a conformal field theory~\cite{BMW2}.

\item By applying PMC to the top-quark pair hadroproduction up to NNLO level, it has been found that the PMC scales and the resulting finite-order total cross-sections are both to high accuracy independent of the choice of an initial scale~\cite{pmc3,pmc5,pmc4}. After PMC scale setting, the top quark forward-backward asymmetries at the Tevatron are : $A_{FB}^{t\bar{t}} \simeq 12.5\%$, $A_{FB}^{p\bar{p}} \simeq 8.28\%$ and $A_{FB}^{t\bar{t}}(M_{t\bar{t}}>450 \;{\rm GeV}) \simeq 35.0\%$~\cite{pmc4}. These predictions deviate approximately $1\,\sigma$ from the CDF and D0 measurements. The large discrepancy of the top quark forward-backward asymmetry between the standard model estimate and the data is thus greatly reduced.

\item A PMC analysis for the charmonium production processes, $e^{+}+e^{-}\to J/ \psi (\psi') + \chi_{cJ}$ with $(J=0,1,2)$, has been given in Ref.~\cite{wangpmc}; it shows that the scale uncertainty for both the polarized and the unpolarized cross sections are greatly suppressed even at the NLO level.

\item The PMC is adoptable for the QED case. In the Abelian limit $N_C \to 0$ at fixed $\alpha=C_F \alpha_s$ with $C_F=(N_c^2-1)/2N_c$~\cite{qed1,qed2}, the PMC also agrees with the standard {\rm Gell Mann-Low} procedure for setting the renormalization scale in QED. Any method used in perturbative QCD must be applicable to QED in the $N_C \to 0$ limit, since they share the same Yang-Mills Lagrangian.
\end{itemize}

The elimination of the renormalization scale ambiguity and the scheme dependence using the PMC procedure will not only increase the precision of QCD tests, but it will also increase the sensitivity of collider experiments to new physics beyond the SM. The PMC procedure can be advantageously applied to the entire range of perturbatively-calculable QCD and Standard Model processes, eliminating an unnecessary systematic error.

\hspace{2cm}

{\bf Acknowledgements}: We thank Leonardo di Giustino, George Sterman, Andrei L. Kataev, Sergey V. Mikhailov, Konstantin Chetyrkin, Zvi Bern and Dmitry V. Shirkov for helpful conversations. This work is dedicated to Dmitry V. Shirkov on the occasion of his $85$th birthday. This work was supported in part by the Program for New Century Excellent Talents in University under Grant No.NCET-10-0882, the Fundamental Research Funds for the Central Universities under Grant No.CQDXWL-2012-Z002, Natural Science Foundation of China under Grant No.11075225 and No.11275280, the Department of Energy contract DE-AC02-76SF00515, and the Danish National Research Foundation, Grant No. DNRF90. SLAC-PUB-15282. CP${}^3$-Origins-2013-001. DIAS-2013-1.

\end{document}